\documentclass[12pt]{article}
\usepackage{epsfig, epsf, graphicx, subfigure, color}
\usepackage{pstricks, pst-node, psfrag}
\usepackage{amssymb,amsmath}
\usepackage{verbatim,enumerate}
\usepackage{rotating, lscape}
\usepackage{setspace}
\usepackage{bbm}
\usepackage{natbib}
\usepackage{amsthm}
\usepackage{url}

\usepackage{gensymb}

\usepackage{appendix}

\usepackage{caption}
\usepackage{hyperref}

\hypersetup{
  colorlinks   = true, %Colours links instead of ugly boxes
  urlcolor     = blue, %Colour for external hyperlinks
  linkcolor    = blue, %Colour of internal links
  citecolor   = blue %Colour of citations
}%\usepackage{subcaption}

\theoremstyle{definition}

\newcommand{\bi}{\begin{itemize}}
	\newcommand{\ei}{\end{itemize}}

\begin{document}
	\thispagestyle{empty} \baselineskip=28pt \vskip 5mm
	\begin{center} {\Large{\bf Estimation of Spatial Deformation for Nonstationary Processes via Variogram Alignment}}
	\end{center}
	
	\baselineskip=12pt \vskip 5mm
	
	\begin{center}\large
		Ghulam A. Qadir\footnote[1]{
			\baselineskip=10pt CEMSE Division, King Abdullah University of Science and Technology, Thuwal 23955-6900, Saudi Arabia.
			E-mail:  ghulam.qadir@kaust.edu.sa; ying.sun@kaust.edu.sa
		}, Ying Sun$^1$ and Sebastian Kurtek\footnote[2]{
			\baselineskip=10pt Department of Statistics, The Ohio State University, Columbus, OH 43210, USA.
			E-mail: kurtek.1@stat.osu.edu}\end{center}
	
	\baselineskip=16pt \vskip 1mm \centerline{\today} \vskip 8mm
	
	%%%%%%%%%%%%%%%%%%%%%%%%%%%%%%%%%%%%%%%%%%%%%%%%%%%%%%%%%%%%%%%%%%%%%%%%
	\begin{center}
		{\large{\bf Abstract}}\end{center}
        In modeling spatial processes, a second-order stationarity assumption is often made. However, for spatial data observed on a vast domain, the covariance function often varies over space, leading to a heterogeneous spatial dependence structure, therefore requiring nonstationary modeling. Spatial deformation is one of the main methods for modeling nonstationary processes, assuming the nonstationary process has a stationary counterpart in the deformed space. The estimation of the deformation function poses severe challenges. Here, we introduce a novel approach for nonstationary geostatistical modeling, using space deformation, when a single realization of the spatial process is observed. Our method is based, at a fundamental level, on aligning regional variograms, where warping variability of the distance from each subregion explains the spatial nonstationarity. We propose to use multi-dimensional scaling to map the warped distances to spatial locations. We asses the performance of our new method using multiple simulation studies. Additionally, we illustrate our methodology on precipitation data to estimate the heterogeneous spatial dependence and to perform spatial predictions.
	
	\baselineskip=17pt

	\begin{doublespace}
		
		\par\vfill\noindent
	\par\medskip\noindent
		{\bf Short title}: Estimation of Spatial Deformation.
	\end{doublespace}
	
	\clearpage\pagebreak\newpage \pagenumbering{arabic}
	\baselineskip=26.5pt
	
	%\begin{doublespace}

\section{Introduction}\label{sec:intro}

Spatial Statistics methods are widely used in various disciplines such as meteorology, hydrology and earth science, to model environmental processes for a better understanding of the latent dependence structure, and for making predictions at unobserved locations. Statistical analysis of spatial processes generally involves a second-order stationarity assumption stating that, for a random process $\{X(\textbf{s}):\textbf{s}\in \mathbb{R}^d,d\geq 1\}$, the mean is a constant, i.e., ${E}\big(X(\textbf{s})\big)=c$ for some $c\in \mathbb {R}$, and the covariance between any two locations depends only on the lag vector between those two locations, i.e., $\text{Cov}\big(X(\textbf{s}),X(\textbf{s}+\textbf{h})\big)=\text{C}(\textbf{h})$. Isotropic and anisotropic processes are two special cases of a second-order stationary process. The former implies that the covariance function depends only on the $\mathbb{L}^2$ norm of a lag vector, i.e.,  $\|\textbf{h}\|$, whereas the latter is a minute generalization that incorporates both length and direction into the covariance function through a linear transformation of the lag vector as $\|\textbf{A}^{-\frac{1}{2}}\textbf{h}\|$, where $\textbf{A}$ is a $d \times d$ positive definite matrix known as the anisotropy matrix. Modeling spatial processes by assuming a translation-invariant spatial dependence is a convenient, but non-viable approach, especially when the spatial domain is large and statistical features of the process vary in space; in this case, such an assumption is a misspecification of the process.

In recent decades, considerable research has been directed toward developing methods to model nonstationary processes. \cite{risser} and \cite{Fouedjio2017} extensively reviewed the existing literature on this topic and published a comprehensive summary of nonstationary modeling approaches for univariate geostatistical data. \cite{Higdon1998} proposed a process-convolution approach with a spatially varying convolution kernel to model the nonstationary dependence structure. Further adaptation of this approach in~\cite*{Swall},~\cite{Paciorek:2006aa} and~\cite{ENV:ENV852} resulted in a covariance function with spatially varying parameters. Subsequently, \cite*{FOUEDJIOconvo} generalized the idea of the process-convolution model by introducing a convolution with a spatially varying random weighting function. Recent work by \cite{NYCHKA201821} introduced a computationally efficient method to model convolution type nonstationarity for large spatial datasets.~\cite{10.2307/4140567} constructed a nonstationary process through convolution of locally stationary processes, which was later used by~\cite{Reich:2011aa} to introduce a novel spatio-temporal covariance function, addressing nonstationarity by using covariate information.  \cite{10.2307/2290458} published one of the first studies on nonstationary spatial modeling by pioneering the method of spatial deformation; their work served as the fundamental idea for further studies by \cite*{damian2001bayesian}, \cite{M.:2003aa}, \cite{doi:10.1198/1061860043100}, \cite{anderes2008}, \cite{anderes2009}, and \cite*{FOUEDJIO201545}. Some other popular nonstationary spatial modeling approaches include basis function methods (\citealp{Nychka1998,holland1999spatial}; \citealp*{doi:10.1191/1471082x02st037oa}; %\citealp{pintore,Stephenson2005}
\citealp{Stephenson2005}), stochastic partial differential equations (SPDE's) approaches (\citealp*{rue2011}: \citealp{fuglstad2015exploring}), and moving window methods \citep{haas1990a,haas1990b,lloyd2000,lloyd2002non}. 

The prominent approach to model nonstationarity, using the method of spatial deformation introduced by \cite{10.2307/2290458}, involves mapping the locations in a geographic space ($\mathcal{G}$) to transformed locations in a deformed space ($\mathcal{D}$), where the process is expected to be stationary and isotropic. This original concept provides an invaluable direction for modeling nonstationarity, but it fundamentally requires multiple independent realizations of the spatial process which, in practice, are often not observed. In addition, one major drawback of their method is the folding of space. This occurs if the estimated deformation function that maps geographical locations to the deformed space is not injective. In the presence of spatial data replicates, \cite*{damian2001bayesian} and \cite{M.:2003aa}  attempted to address the folding of space in a Bayesian framework, whereas \cite*{born} addressed this issue in a frequentist framework by retaining the original locations of the geographic space and by adding extra dimensions to embed a nonstationary field of lower dimensions to a higher dimensional stationary field. The problem of estimation of a spatial deformation by using only one realization of the spatial process was first addressed by \cite{anderes2008} and \cite{anderes2009}. However, their proposed quasi-conformal mappings-based methodology requires very dense spatial data and its application on a real dataset has not yet been illustrated. \cite*{FOUEDJIO201545} developed a method for estimating the deformation function, using a single realization of the spatial field that avoids the problem of folding of space, but their method relies heavily on many tuning parameters and subjective selection of anchor points.

Here, we propose a metric-based nonparametric method for estimating a spatial deformation by applying the functional data registration method, proposed by ~\cite{anuj}, to spatial variograms. Our method extricates the strong assumption of replicates of spatial data and allows us to estimate the deformed space in higher dimensions, consequently avoiding the problem of folding of space. The key concept underlying the proposed method is based on aligning regional variograms belonging to different subregions of the entire spatial domain to estimate the warping variability in inter-point distances. The principal tools used in the proposed method are: (1) kernel smoothing, (2) classical (metric) multi-dimensional scaling (CMDS) \citep*{torgerson1958theory,1979} and (3) a functional data registration algorithm \citep{anuj}; we use these tools to obtain a one-to-one mapping of locations in a geographic space ($\mathcal{G}$) to transformed locations in a deformed space ($\mathcal{D}$). Our method does not require the use of thin-plate splines (a key component in the methods of \cite*{10.2307/2290458,born} and \cite*{FOUEDJIO201545}) to estimate the deformed coordinates of unobserved locations. Both observed and unobserved locations can be mapped to their corresponding deformed coordinates in a single step, and hence can be used directly to obtain kriging estimates. Besides the estimation of a heterogeneous spatial dependence structure for spatial predictions, the proposed method also serves as a useful exploratory tool to visualize the degree of nonstationarity in spatial data. We illustrate the proposed method with a simulated example. We also apply it to precipitation data from the state of Colorado in the United States. 

Section~\ref{sec:method} describes the proposed spatial deformation estimation procedure, including a brief discussion of the functional data registration algorithm used in the proposed method. In Section~\ref{sec:simulation}, we illustrate our methodology on a simulated example. Section~\ref{sec:app} discusses an application to the precipitation dataset, followed by a discussion in Section~\ref{sec:disc} highlighting the main contributions of this work and some directions for the future.

\section{Methodology}\label{sec:method}
According to \cite*{born}, ``Environmental systems might exhibit behavior that looks locally stationary, yet when considered over large and heterogeneous domain they very often exhibit nonstationarity''. Our method is motivated by such locally stationary behavior of environmental processes that can be well approximated by piecewise or regionwise stationary models. It involves a partitioning of the entire spatial domain into smaller subregions such that the process shows homogeneous spatial dependence within each subregion, but may exhibit heterogeneous spatial dependence across subregions. One common way to quantify homogeneous spatial dependence is by using a stationary variogram that measures the variability in observations, depending on the distance between them. Therefore, heterogeneous spatial dependence across subregions implies that the regional variogram, as a function of distance, may vary across subregions. We treat these regional variograms as functional data. However, unlike the traditional functional data registration problem where functional data are directly observed, the regional variograms need to be estimated from spatial observations prior to alignment. In this section, we give a brief introduction to the functional data registration algorithm (Section \ref{fda}), followed by a detailed discussion of the proposed method for estimating spatial deformations. The estimation procedure can be broadly classified into two steps: (1) an ``alignment step", and (2) a ``construction step"; the steps are described in detail in Section~\ref{astep} and Section~\ref{cstep}, respectively.
\subsection{Functional Data Registration}\label{fda}

We first introduce the functional data registration algorithm developed by \cite{anuj}; \cite*{kurt6} and \cite{srivastava2016functional} that we use in our work for variogram alignment. In those works, they defined the notion of ``elastic functions'', i.e.,  functions with warping or phase variability, and proposed a framework for separation of amplitude ($y$-axis) and phase ($x$-axis) in these elastic functions by warping the $x$-axis. They considered the following representation:\begin{equation} \label{eq:1}
f_i=c_i(g\text{\hspace{0.08cm}}\circ \text{\hspace{0.08cm}}\phi_i)+e_i, \text{\hspace{0.2cm}} i=1,2,\dots,n,
\end{equation}where $f_i$ denote the observed functions (assumed to be absolutely continuous), $c_i \in \mathbb{R}^+$ are the individual scalings, $e_i \in \mathbb{R}$ are the vertical translations, and $g$ is an underlying template. Each function $f_i$ represents an observation of the template $g$ under a random warping of the $x$-axis $\phi_i$, and a random scaling and translation, $c_i$ and $e_i$, respectively. %The algorithm considers diffeomorphisms of the function domain as the class of warping functions $\Phi$, and $\phi_i\in\Phi,\;i=1,2,\dots,n$. 
For a given sample of functions $\{f_i\}$, the main task is to obtain a consistent estimator of the template $g$; this additionally results in estimates of the optimal warping functions $\phi^*_i$ (phase component of $f_i$), and the set of optimally registered functions $f^*_i=f_i\circ\phi_i^{*-1}$ (amplitude component of $f_i$). Standard solutions to the warping problem based on the $\mathbb{L}^2$ Hilbert space framework are known to have theoretical and practical issues, such as the lack of isometry of the $\mathbb{L}^2$ metric under the action of the warping group. This, in turn, results in degenerate warping solutions and the so-called pinching effect \citep{marron2015}. To overcome these problems, \cite{anuj} proposed an approach based on the extended Fisher-Rao metric and the \emph{square-root velocity function} (SRVF) representation of observed functional data. The SRVF allows for efficient computation of the optimal warping functions via Dynamic Programming \citep{DynPro}. Their registration algorithm (available in the R-package \text{\tt{fdasrvf}}\footnote{\url{https://cran.r-project.org/web/packages/fdasrvf/fdasrvf.pdf}}
) has been extensively studied to demonstrate theoretical guarantees for the consistent estimation of the unknown template $g$ \citep*{kurt6,kurt7,srivastava2016functional}. Furthermore, its practical efficiency has been explored in various applied contexts (\citealp{kurt2}; \citealp*{kurt5}; \citealp{kurt4,kurt3}). However, its application in Spatial Statistics has not yet been considered. For brevity, we skip the complete discussion of the registration algorithm, and instead refer the interested readers to \cite{anuj} and \cite{srivastava2016functional} for details. In this work, we adapt this algorithm to the spatial setting for the registration of regional variograms.%; this, in turn, allows us to deform the geographic space to achieve stationarity.% (see Section~\ref{esti} for details).

\subsection{Estimation of Spatial Deformation}\label{esti}
Let $\{X(\textbf{s}): \textbf{s}\in \mathcal{G}\subset \mathbb{R}^{d^{\mathcal{G}}}\}$ be a zero-mean nonstationary random field defined on the geographic space $\mathcal{G}$ of dimensionality ${d^{\mathcal{G}}}$, and $\{Y(\textbf{u}): \textbf{u}\in \mathcal{D}\subset \mathbb{R}^{d^\mathcal{D}}\}$ be the corresponding zero-mean stationary random field defined on the deformed space $\mathcal{D}$ of dimensionality $d^\mathcal{D}$. Here, $d^{\mathcal{D}}$ is not necessarily equal to $d^{\mathcal{G}}$, and in fact, $\{d^{\mathcal{D}}=d^{\mathcal{G}}+\psi,\: \psi \in \{0,1,2,\dots\} \}$, i.e., the domain of the stationary process $Y$ can have a higher dimensionality relative to the nonstationary process $X$. The primary objective is to estimate a deformation $\theta:\mathcal{G}\rightarrow\mathcal{D}$ such that $\{X(\textbf{s})=Y(\theta(\textbf{s})), \:\forall\: \textbf{s} \in \mathcal{G}\}$ and $\{Y(\textbf{u})=X(\theta^{-1}(\textbf{u})),\: \forall \: \textbf{u} \in \mathcal{D}\}$. This allows us to model the nonstationary covariance of $X$ as:\begin{equation} \label{eq:model}
\text{Cov}^{NS}(\textbf{s},\textbf{s}')=\text{C}_\mathcal{D}(\|\theta(\textbf{s})-\theta(\textbf{s}')\|),\: \forall \: (\textbf{s,s}')\in \mathcal{G}\times \mathcal{G},\end{equation}where $\text{C}_\mathcal{D}(\|\cdot\|)$ represents any valid stationary and isotropic covariance function that depends only on the $\mathbb{L}^2$ distance between points in the deformed space. The corresponding nonstationary semivariogram (simply called variogram hereafter) of $X$ is then given by $\gamma^{NS}(\textbf{s},\textbf{s}')=\gamma_{\mathcal{D}}(\|\theta(\textbf{s})-\theta(\textbf{s}')\|),$ where $\gamma_\mathcal{D}(\|\cdot\|)$ is a valid stationary and isotropic variogram model which is related to $\text{C}_\mathcal{D}(\|\cdot\|)$ via $\gamma_\mathcal{D}(\|\textbf{h}\|)=\text{C}_\mathcal{D}(\|\textbf{0}\|)-\text{C}_\mathcal{D}(\|\textbf{h}\|).$

Our method is based on a mild assumption of regional stationarity of the process $\{X(\textbf{s}): \textbf{s}\in \mathcal{G}\subset \mathbb{R}^{d^{\mathcal{G}}}\}$, which  implies that $\mathcal{G}$ can be partitioned into $k$ mutually exclusive subregions $\mathcal{G}_1,\mathcal{G}_2,\dots,\mathcal{G}_k$ ($\{\mathcal{G}=\cup_{i=1}^k\mathcal{G}_i\}$) such that for each $i=1,2,\dots,k$, $\{X(\textbf{s}): \textbf{s}\in \mathcal{G}_i\}$ is a stationary process with spatial dependence described by the stationary and isotropic variogram model $\gamma_i(\|\textbf{h}\|)$. The variogram models may differ from each other through various features such as smoothness, autocorrelation range, variance (sill) and nugget, making the process $X(\textbf{s})$ nonstationary over the domain $\mathcal{G}$. For each subregion $\mathcal{G}_i,\ i=1,2,\dots,k,$ %of the geographic space $(\mathcal{G})$, 
we have a corresponding subregion $\mathcal{D}_i,\ i=1,2,\dots,k,$ in the deformed space $\mathcal{D}$ such that $\mathcal{D}=\cup_{i=1}^k\mathcal{D}_i$.

Considering the regional variograms $\gamma_i(\|\textbf{h}\|)$ as elastic functions results in the following representation (adaptation of Equation \ref{eq:1}): \begin{equation} \label{eq:2}
\gamma_i(\|\textbf{h}\|)=c_i(\gamma\:\circ\: \phi_i)(\|\textbf{h}\|)+e_i, \: i=1,2,\dots,k.
\end{equation}
In the spatial context, Equation~\ref{eq:2} leads to the following interpretation: each regional variogram is an observation from the global stationary variogram model $\gamma$, under a regional distance warping function $\phi_i$, with a scaling $c_i\in \mathbb{R}^+$ and a vertical translation $e_i \in \mathbb{R}^+$ (note that $e_i$ is non-negative because variogram values are always non-negative). For instance, if we assume that the features of the global variogram model $\gamma$ such as nugget and variance are 0 and 1 respectively, then $c_i$ and $e_i$ can be interpreted as the variance and nugget for the regional variogram $\gamma_i$. The application of functional data registration %\citep{srivastava2016functional}
to Equation~\ref{eq:2} allows us to estimate the $k$ regional distance warping functions that are of paramount importance in estimating the deformation $\theta$. Specifically, they inform us about the inter-point distances in different subregions $\mathcal{D}_i,\ i=1,2,\dots,k,$ of the deformed space. Consequently, $\theta$ can be defined locally for the $i^{th}$ subregion $(i=1,2,\dots,k)$ as $\theta:\mathcal{G}_i\rightarrow\mathcal{D}_i,$ and the following condition drives its estimation.\\\textbf{Condition 1}: \emph{For any two arbitrary locations} $\textbf{s}_1,\textbf{s}_2\in \mathcal{G}_i$, \emph{the distance between their corresponding locations in the deformed space is given by warping the distance between them in the geographic space with a warping function $\phi_i,$ i.e.,} $\|\theta(\textbf{s}_1)-\theta(\textbf{s}_2)\|=\phi_i(\|\textbf{s}_1-\textbf{s}_2\|).$

Following the interpretation of Equation ~\ref{eq:2} and imposing Condition 1 in the estimation of $\theta$ implies that the variogram models describing the spatial dependence for the processes $\{Y(\textbf{u})=X(\theta^{-1}(\textbf{u})), \: \forall \: \textbf{u}\in \mathcal{D}_i,\; i=1,2,\dots,k.\}$  share the same features, such as smoothness and autocorrelation range; this indicates that the nonstationarity in smoothness and autocorrelation range can be addressed by variogram registration. However, the processes might have varying regional variances and nuggets. The functional data registration algorithm used in our method is invariant to scalings and vertical translations, and therefore cannot deal with the nonstationarity in those components. More specifically, the proposed method addresses the nonstationarity only in the correlation function to introduce nonstationarity in the covariance function. The components of the covariance function other than the correlation function, namely the variance and nugget, can be made to be nonstationary straightforwardly by allowing them to be spatially varying as discussed later in Section~\ref{sec:disc}. 

For a complete specification of the deformation $\theta$, we need to define it globally, i.e., $\theta:\mathcal{G}\rightarrow\mathcal{D}$, and therefore, a global distance function $\phi$, which governs the inter-point distances in the deformed space $\mathcal{D}$, is required: $\|\theta(\textbf{s})-\theta(\textbf{s}')\|=\phi(\textbf{s},\textbf{s}'),\: \forall\: (\textbf{s},\textbf{s}') \in \mathcal{G}\times \mathcal{G}$; this global distance function should also be consistent with Condition 1. We propose to define a global distance function $\phi:\mathcal{G}\times\mathcal{G}\rightarrow\mathbb{R}^+\cup\{0\}$ as a weighted linear combination of the regional distance warping functions as follows:\begin{equation} \label{eq:3}\phi(\textbf{s},\textbf{s}')=\sum_{\mathcal{G}_i\in \mathcal{L}(\textbf{s},\textbf{s}')}\mathcal{W}_i(\textbf{s},\textbf{s}')\phi_i(\|\textbf{s}-\textbf{s}'\|),\: \forall\: (\textbf{s},\textbf{s}') \in \mathcal{G}\times\mathcal{G},\end{equation}
where $\mathcal{L}(\textbf{s},\textbf{s}')$ is the set of subregions $\mathcal{G}_i$ such that the line segment joining the locations $\textbf{s}$ and $\textbf{s}'$ passes through all of the subregions in this set, and $\mathcal{W}_i(\textbf{s},\textbf{s}')$ are the location-dependent weights for the $i^{th}$ regional distance warping function. We define the weights as ${\mathcal{W}_i(\textbf{s},\textbf{s}')=\frac{\mathcal{P}(i,\textbf{s},\textbf{s}')}{\|\textbf{s}-\textbf{s}'\|}}$, where $\mathcal{P}(i,\textbf{s},\textbf{s}')$ is the length of the line segment joining $\textbf{s}$ and $\textbf{s}'$ that lies in the subregion $\mathcal{G}_i$. This special choice of weights %used to define the global distance function 
used in Equation \ref{eq:3} 
imparts robustness to our method under different subdivisions of the spatial domain (see Supplementary Material Section S2), and are specifically chosen to satisfy the following two properties that are crucial to our approach. %The proofs for the following properties are included in the Supplementary Material (Section 1). 
\\\textbf{Property 1}: \emph{The global distance function  $\phi:\mathcal{G}\times\mathcal{G}\rightarrow\mathbb{R}^+\cup\{0\}$ is consistent with Condition 1, i.e.,} $\phi(\textbf{s},\textbf{s'})=\phi_i(\|\textbf{s}-\textbf{s}'\|)\text{ }\forall\text{ } (\textbf{s},\textbf{s}')\in \mathcal{G}_i\times\mathcal{G}_i,\ i=1,2,\dots,k.$  \\\textbf{Property 2}: \emph{If the process} $\{X(\textbf{s}):\textbf{s}\in \mathcal{G} \subset \mathbb{R}^{d^{\mathcal{G}}}\}$ \emph{is second-order stationary, then} $\phi(\textbf{s},\textbf{s}')=\|\textbf{s}-\textbf{s}'\|$, \emph{implying that the geographic and deformed spaces are identical, up to a rotation and/or translation.}

Due to the global distance function, the deformed space $\mathcal{D}$ is now known through the inter-point distances between different locations. Therefore, we propose to map these distances to deformed coordinates ($\theta(\textbf{s}),\textbf{s}\in \mathcal{G}$) using the CMDS algorithm \citep*{torgerson1958theory,1979}. For a given distance matrix, the application of CMDS seeks to find the coordinates in a
space of a specified dimensionality, such that the associated distance matrix is as close as
possible to the given distance matrix \citep{cmdsapp1,cmdsapp2}. The distance matrix for the deformed space with $n$
locations denoted by $\Delta_{(n\times n)}=\{\phi(\textbf{s}_i,\textbf{s}_j)\}_{i,j=1}^{n}$ is supplied to the CMDS algorithm for a given dimension $d^{\mathcal{D}}=d^{\mathcal{G}}+\psi$ to estimate the deformation $\hat{\theta}$. %Note that any other tool with similar functionality can be used in place of CMDS for the estimation of $\hat{\theta}$ in a straightforward manner. 
The estimation procedure is described in more detail in Sections \ref{astep} and \ref{cstep}.

\subsubsection{Variogram Estimation and Registration}\label{astep}

As already outlined in Section~\ref{esti}, our method is based on the regionwise stationary behavior of the spatial process on a vast domain, and hence requires the identification of homogeneous subregions. Some informative covariates, or prior knowledge of the underlying physical process, can be instrumental in making this decision. In case such information is not available, we can divide the entire region into any number of subregions, provided that each subregion has enough observations to describe the local dependence structure in the corresponding process.

The variogram representation in Equation~\ref{eq:2} requires the true regional variogram models $\{\gamma_i(\|\textbf{h}\|),\ i=1,2,\dots,k\}$ that are often unknown in practical situations. Therefore, we fit a valid stationary and isotropic variogram model for each subregion, and use the estimated variogram models $\{\hat{\gamma}_i(\|\textbf{h}\|),\ i=1,2,\dots,k\}$ to redefine Equation~\ref{eq:2} as
%\begin{equation} \label{eq:4}
$\hat{\gamma}_i(\|\textbf{h}\|)=c_i (\gamma\:\circ\: \phi_i)(\|\textbf{h}\|)+e_i,\; i=1,2,\dots,k.$
%\end{equation}

The choice of the number of subregions $k$ is critical, as it controls the trade-off between flexibility of the model and efficiency of our method. Higher values of $k$ allow us to introduce a higher degree of nonstationarity, but potentially lead to inaccurate parameter estimates of the regional variogram models, due to a reduced number of observations per subregion. Similarly, lower values of $k$ lead to better estimates of the regional variograms, but render a lower degree of nonstationarity in the model. Depending on the size of the data, an appropriate value of $k$ should be chosen to maintain a balance between flexibility and estimation accuracy.

Once we have determined the appropriate value of $k$, the next step is to divide the geographic space $\mathcal{G}$ into well-defined $k$ subregions $\mathcal{G}_1,\dots,\mathcal{G}_k$, and to fit a valid stationary and isotropic variogram model for each subregion. %Among the various available parametric variogram models, the most popular choice among geostatisticians is Mat{\'e}rn \citep{matern,guttorp}. It is described by three parameters: the variance ($\sigma^2>0$), the spatial range ($\alpha>0$) and the smoothness ($\nu>0$). 
In our study, we use the Mat{\'e}rn variogram model \citep{matern,guttorp} which is described by three parameters: the variance ($\sigma^2>0$), the spatial range ($\alpha>0$) and the smoothness ($\nu>0$). %Thus, fitting $k$ regional variogram models implies an estimation of $3k$ parameters $\{\sigma_i^2,\alpha_i,\nu_i,\ i=1,2,\dots,k\}$, where the subscript $i$ denotes parameter estimates for the $i^{th}$ subregion. 
To represent the regional variograms as functions for registration, we evaluate the fitted variograms at a sequence of equally spaced points in the interval $[0, \|\textbf{h}_{t}\|]$, where $\|\textbf{h}_{t}\|$ is the distance at which all of the estimated variograms become numerically constant. We then apply the functional data registration algorithm to estimate the $k$ regional distance warping functions $\hat{\phi}_i^{fdr},\ i=1,2,\dots,k,$ which are defined over the domain $[0, \|\textbf{h}_{t}\|]$. In order to define the regional distance warping functions over the domain $[0,\infty]$, we augment identity warping to $\hat{\phi}_i^{fdr},\ i=1,2,\dots,k$ for any distance $\|\textbf{h}\|>\|\textbf{h}_t\|$ so that $\hat{\phi}_i,\ i=1,2,\dots,k,$ is now given by: \begin{equation}\label{neweq}\hat{\phi}_i(\|\textbf{h}\|)=\begin{cases}\hat{\phi}_i^{fdr}(\|\textbf{h}\|), \; \|\textbf{h}\| \leq \|\textbf{h}_{t}\| \\
  \|\textbf{h}\|, \;  \|\textbf{h}\| > \|\textbf{h}_{t}\|.
  \end{cases}\end{equation}The identity warping for large distances $(\|\textbf{h}\|>\|\textbf{h}_t\|)$ ensures that they remain unchanged, as beyond those distances all subregions exhibit spatial homogeneity (constant and identical spatial variograms)..%do we need to add scaling comment also?%

The regional distance warping functions serve as a valuable exploratory tool to visualize the degree of nonstationarity in the spatial data. Since the 45 degree line represents the identity warping, a larger deviation of regional distance warping functions from the identity warping indicates a higher degree of nonstationarity across subregions. %A crude numerical measure for the degree of nonstationarity can also be given by computing the spread of the regional distance warping functions around the identity warp.  
Additionally, the warping functions tell us about the amount of stretching and compression required for different subregions to achieve stationarity in the deformed space. Note that the estimated regional distance warping functions represent the warped pre-determined distances when evaluating the regional variograms. Thus, to be able to warp any distance in the interval $[0, \|\textbf{h}_{t}\|]$, we use kernel smoothing on the warped distances using a Gaussian kernel with a fairly low bandwidth. From the estimated regional distance warping functions given in Equation \ref{neweq}, we estimate a global distance function using Equation \ref{eq:3}: $  \label{eq:5}\hat{\phi}(\textbf{s},\textbf{s}')=\sum_{\mathcal{G}_i\in \mathcal{L}(\textbf{s},\textbf{s}')}\mathcal{W}_i(\textbf{s},\textbf{s}')\hat{\phi}_i(\|\textbf{s}-\textbf{s}'\|),\: \forall\: (\textbf{s},\textbf{s}') \in \mathcal{G}\times\mathcal{G}$. The next step is to use this global distance function to obtain a distance matrix for the deformed space $\mathcal{D}$ and to estimate the deformation $\theta$.

\subsubsection{Estimation of the Coordinates in Deformed Space}\label{cstep}
The global distance function is defined for any arbitrary pair of locations in the geographic space $\mathcal{G}$, i.e., for any observed or unobserved pair of locations. We can then compute their corresponding pairwise distance in the deformed space. Let $\textbf{s}_1,\textbf{s}_2,\dots,\textbf{s}_n\in \mathcal{G}$ be the $n$ observed locations, and let $\textbf{s}_{n+1},\textbf{s}_{n+2},\dots,\textbf{s}_{n+m}\in \mathcal{G}$ be $m$ unobserved locations. The aim is to estimate $\theta$ such that the approximation $\|\hat{\theta}(\textbf{s}_i)-\hat{\theta}(\textbf{s}_j)\|\approx \hat{\phi}(\textbf{s}_i,\textbf{s}_j)$ holds true for all $i,j=1,2,\dots,n+m$.  %\begin{equation} \label{eq:6}\|\hat{\theta}(\textbf{s}_i)-\hat{\theta}(\textbf{s}_j)\|\approx \hat{\phi}(\textbf{s}_i,\textbf{s}_j),\: \forall\: i,j=1,2,\dots,n+m.\end{equation}

To achieve this goal, we first compute the transformed distance matrix $\Delta_{(n+m)\times(n+m)}=\{\hat{\phi}(\textbf{s}_i,\textbf{s}_j)\}_{i,j=1}^{n+m}$ and then apply CMDS to $\Delta_{(n+m)\times(n+m)}$ to estimate $\hat{\theta}$ for a space of dimensionality $d^{\mathcal{D}}$. As we increase $d^{\mathcal{D}}$, the approximation improves, but an exact distance match is not guaranteed, even for a large $d^{\mathcal{D}}$. We know that $d^{\mathcal{D}}=d^{\mathcal{G}}+\psi,\ \psi\in \{0,1,\dots\}$, and thus, the value of $\psi$ needs to be chosen appropriately.  The value of $\psi$ can be increased to $\psi+1$ in the case of co-located deformed coordinates to tackle the space-folding problem.

\section{Simulation Study}\label{sec:simulation}
To assess the performance of our method, we apply it to a two-dimensional Gaussian process which has a regionally stationary dependence structure. Specifically, we consider a zero-mean Gaussian process $X$ over a domain $\mathcal{G}=[0,2]^2$, with a spatial dependence described by the following nonstationary Mat{\'e}rn covariance function \citep{Paciorek:2006aa}:
\begin{equation}\label{eq:7}\text{C}^{NS}(\textbf{s}_i,\textbf{s}_j:\tilde{\eta})=\sigma(\textbf{s}_i)\sigma(\textbf{s}_j)\frac{|\Sigma(\textbf{s}_i)|^{1/4}|\Sigma(\textbf{s}_j)|^{1/4}}{2^{\nu-1}\Gamma(\nu)}\Big|\frac{\Sigma(\textbf{s}_i)+\Sigma(\textbf{s}_j)}{2}\Big|^{-1/2}(2\sqrt{\nu Q_{ij}})^{\nu}K_{\nu}(2\sqrt{\nu Q_{ij}}),\end{equation}where $\tilde{\eta}$ represents the vector of parameters, $\sigma(\textbf{s})$ is a location-dependent standard deviation, $\nu$ is the smoothness parameter, $Q_{ij}$ is the Mahalanobis distance between a pair of locations $\textbf{s}_i=(x_i,y_i)$ and $\textbf{s}_j=(x_j,y_j)$, $K_\nu$ is a modified Bessel function of second order, and $\Sigma(\textbf{s})$ is a spatially varying kernel matrix that supervises the range and direction of spatial dependence.

We simulate $X$ at $30\times 30$ points on a regular grid, on $\mathcal{G}$, with a smoothness $\nu=0.6$, standard deviation $\{\sigma(\textbf{s})=1,\:\forall \:\textbf{s}\in \mathcal{G} \}$ and a regionally varying kernel matrix such that $\Sigma(\textbf{s})=\text{diag}(0.0400,0.0400)$ for the subregion where $x\leq1$ and $\Sigma(\textbf{s})=\text{diag}(0.1849,0.1849)$ for the subregion where $x>1$. This allows us to simulate a realization from a regionally stationary process with nonstationarity only in the spatial range.
In this setting, we already know the homogeneous subregions of $\mathcal{G}$; therefore, we divide $\mathcal{G}$ into two subregions $\mathcal{G}_1=[0,1]\times[0,2]$ and $\mathcal{G}_2=(1,2]\times[0,2]$. We fit the isotropic Mat{\'e}rn variogram model for both of the subregions via Maximum Likelihood Estimation (MLE), and register the two estimated regional variograms.
\begin{figure}[!t]
\centering     %%% not \center
\subfigure[]{\label{fig:2a}\includegraphics[width=60mm]{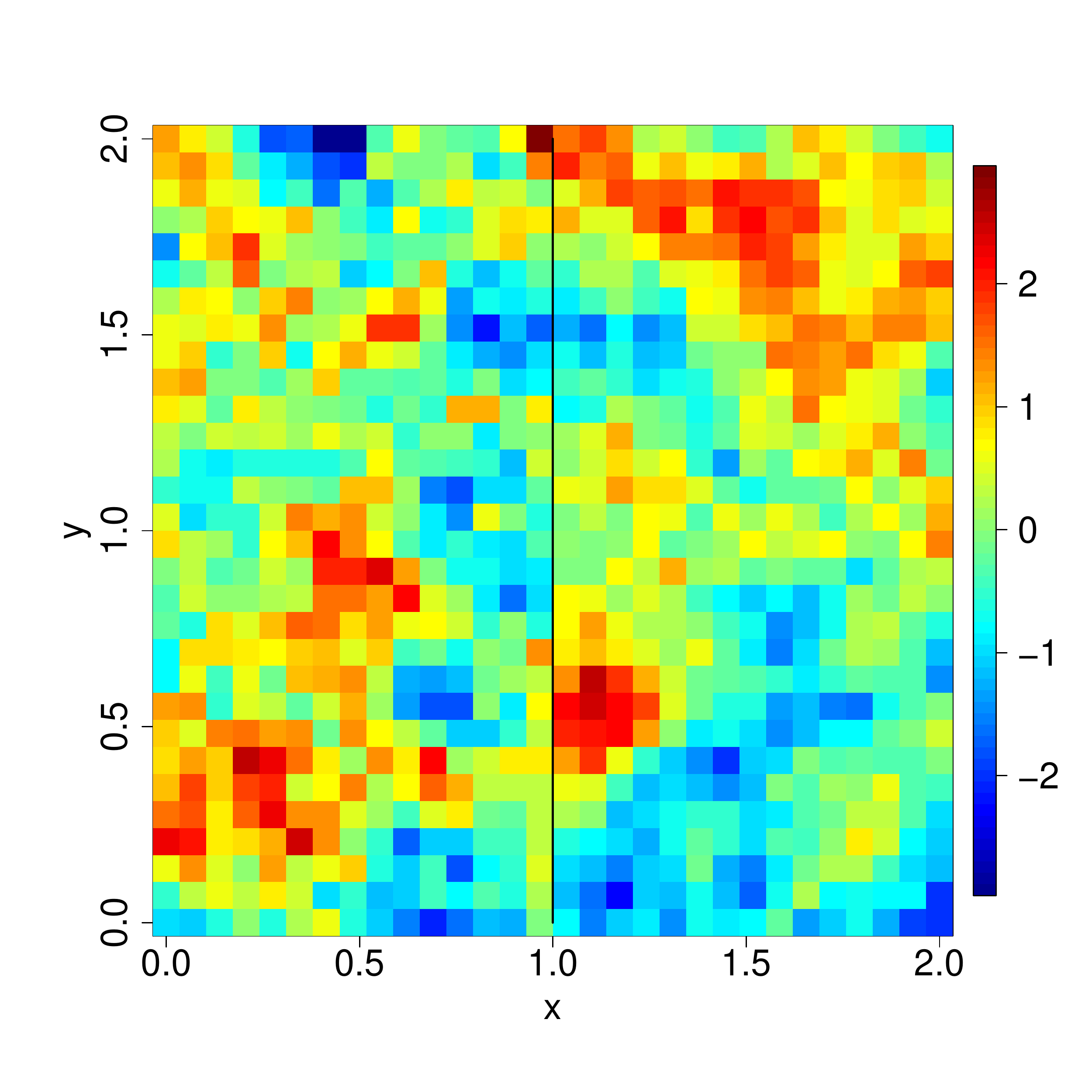}}
\subfigure[]{\label{fig:2b}\includegraphics[width=60mm]{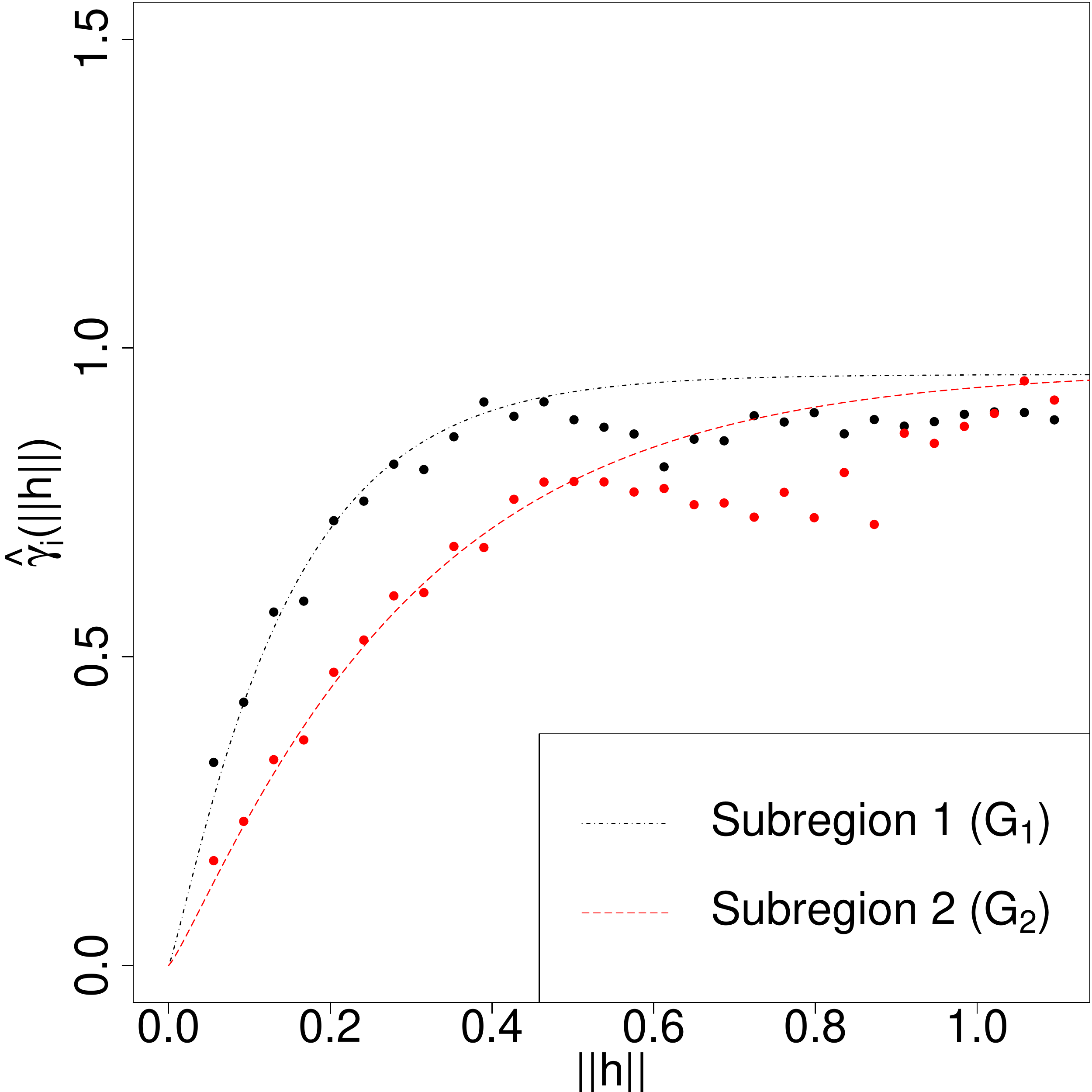}}
\subfigure[]{\label{fig:2c}\includegraphics[width=60mm]{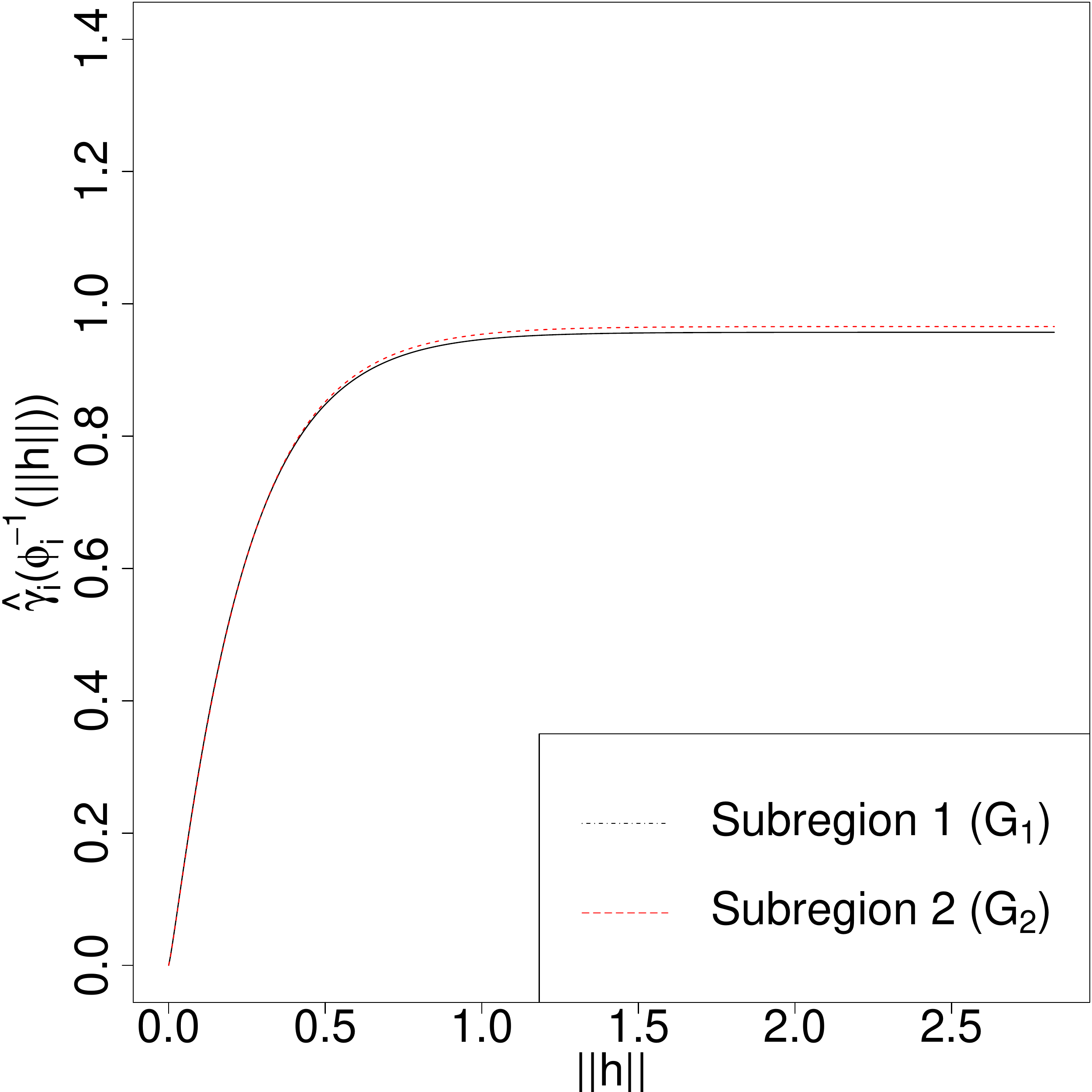}}
\subfigure[]{\label{fig:2d}\includegraphics[width=60mm]{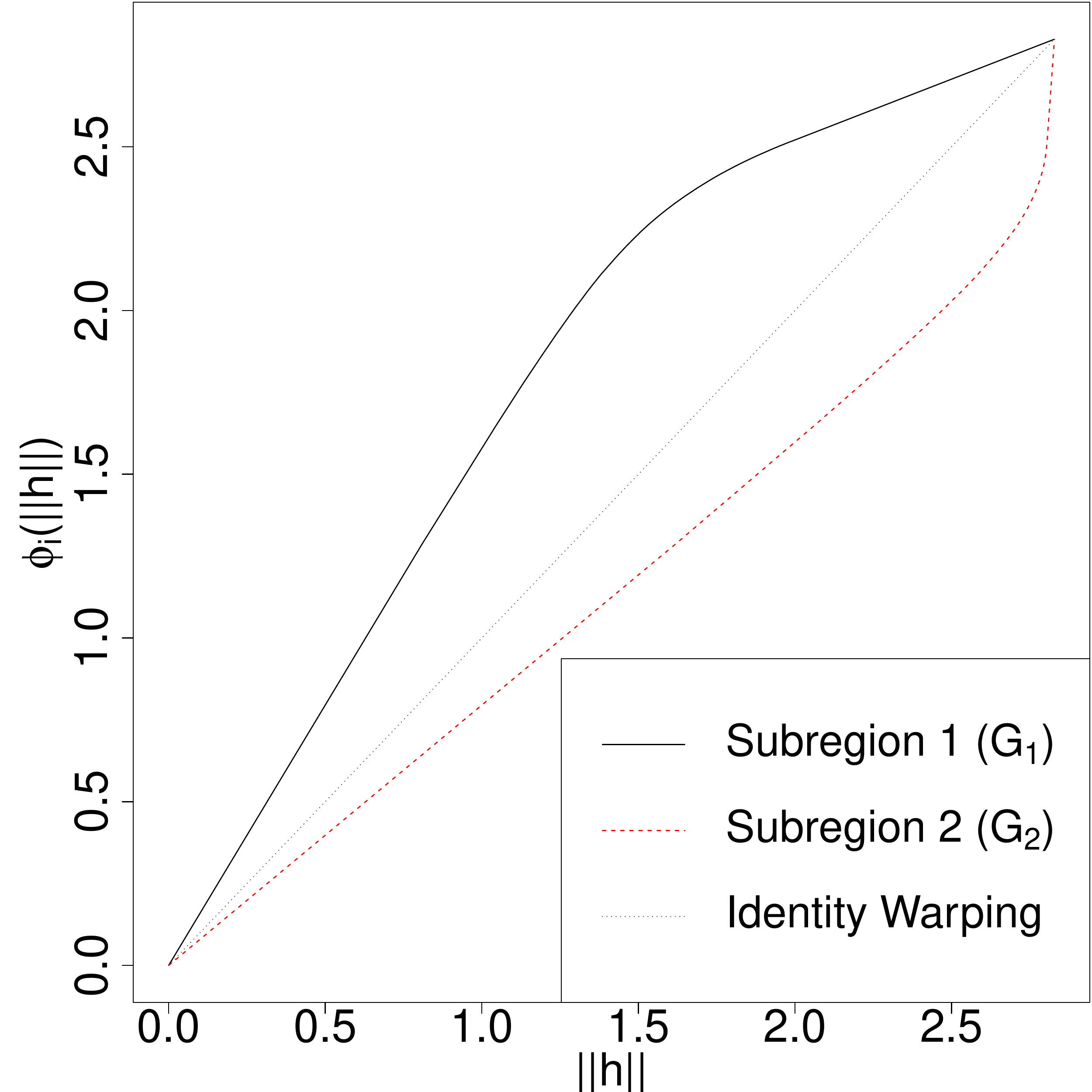}}

\caption{(a) Example of a realization of the zero-mean regionally stationary Gaussian process, with a solid black line indicating the partitioning of the geographic space. (b) Estimated regional variograms overlaid on the regional empirical variograms. (c) Registered variograms. (d) Regional distance warping functions.}
\label{fig:2}
\end{figure}

%The results of the registration step on this simulated dataset are shown in Figure~\ref{fig:2}. 
Figure~\ref{fig:2a} shows a realization of the simulated process in the geographic space, with a solid black line depicting the chosen partitioning. The estimated regional variograms in Figure~\ref{fig:2b} exhibit varying spatial range for the two subregions, and their registration eliminates this variability (Figure~\ref{fig:2c}). The estimated regional distance warping functions are shown in Figure~\ref{fig:2d}. The extent of nonstationarity in the simulated data can be assessed visually by looking at this figure, where the large deviation of both warping functions from the identity means a high degree of nonstationarity. It also suggests that stretching in subregion 1 ($\mathcal{G}_1$) and compression in subregion 2 ($\mathcal{G}_2$) are required to achieve stationarity.

\begin{figure}[!t]
\centering     %%% not \center
\subfigure[]{\label{fig:3a}\includegraphics[width=60mm]{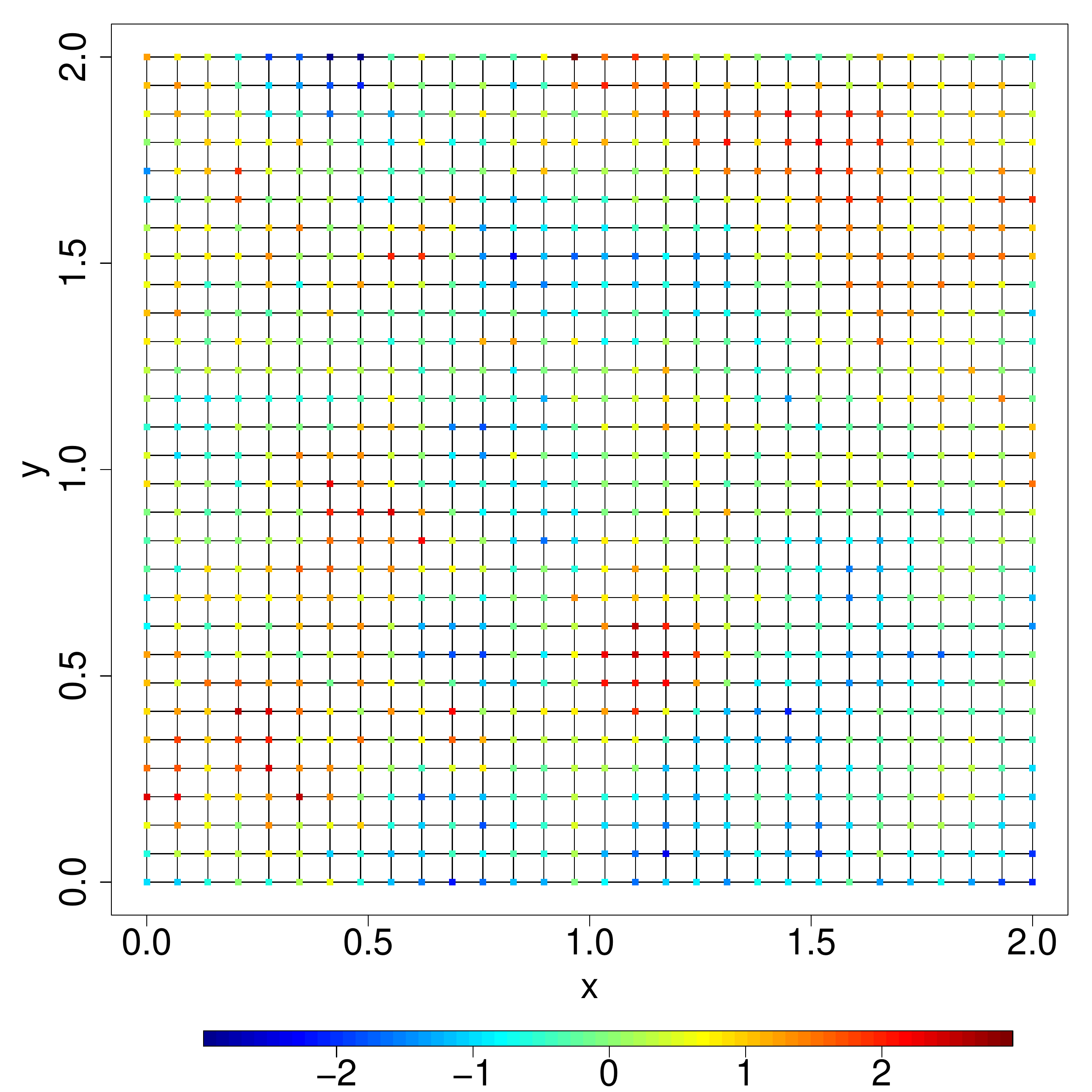}}
\subfigure[]{\label{fig:3c}\includegraphics[width=90mm]{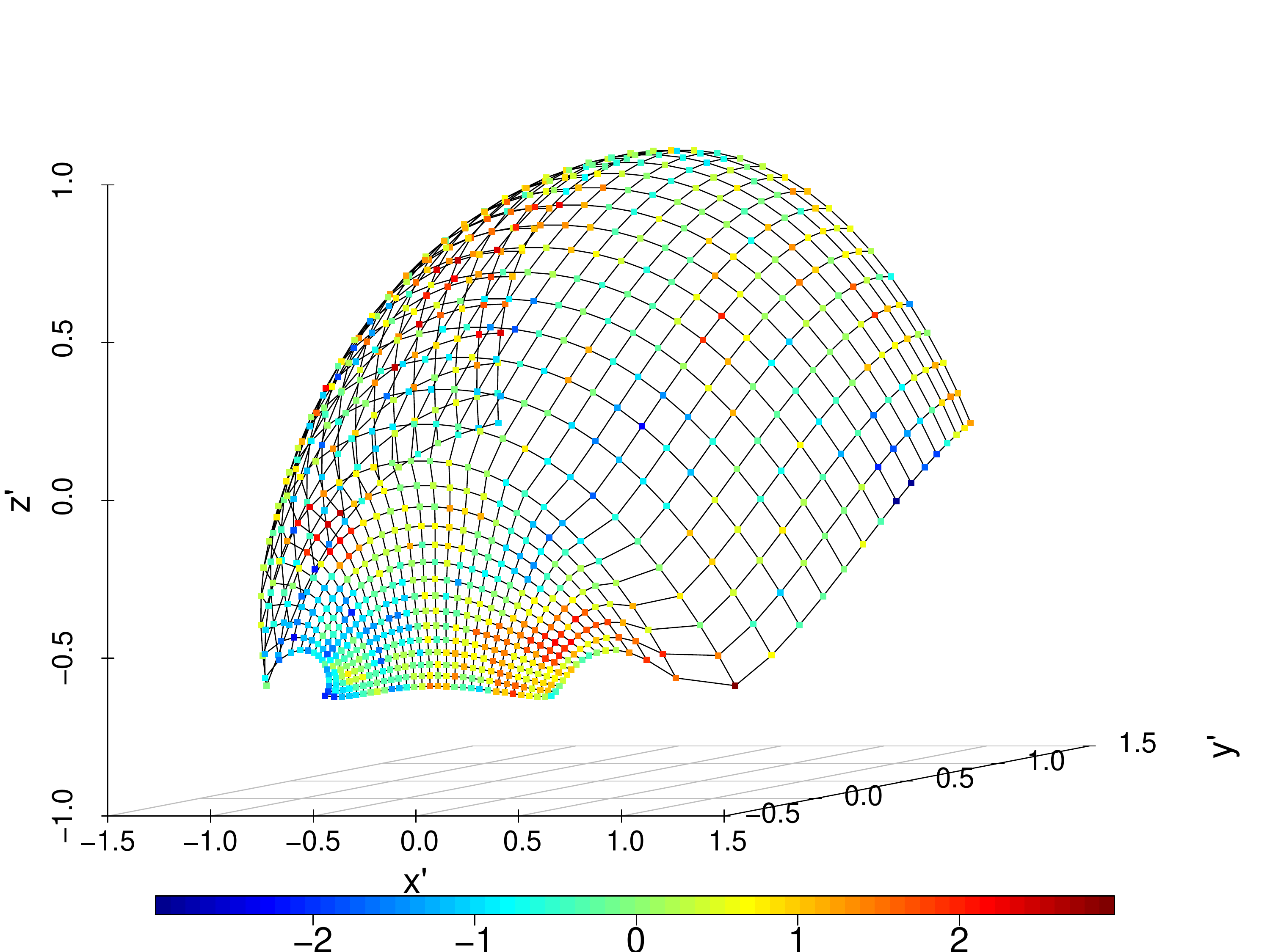}}

\caption{(a) Geographic space. (b) Estimated deformed space in 3-D.}
\label{fig:3}
\end{figure}

We estimate the deformation $\theta$ for $\psi=1$ (3-D) and the corresponding deformed space is shown in Figure~\ref{fig:3c}. The highly correlated observations corresponding to $\mathcal{G}_1$ are placed at higher inter-point distances in the deformed space, leading to a higher spatial range relative to $\mathcal{G}_1$. On the other hand, the spatial range is lowered in the deformed space for the observations corresponding to $\mathcal{G}_2$, due to compression. The compression and stretching bring the spatial range of both subregions to nearly the same level, allowing the spatial dependence structure to be adequately modeled with a stationary variogram model in the deformed space.

%\begin{figure}[!t]
%\centering     %%% not \center
%\subfigure[]{\label{fig:4a}\includegraphics[width=137mm]{sim1corr1o.pdf}}\\
%\subfigure[]{\label{fig:4b}\includegraphics[width=137mm]{sim1corr2o.pdf}}\\
%\subfigure[]{\label{fig:4c}\includegraphics[width=137mm]{sim1corr3o.pdf}}

%\caption{(a) True correlations. (b) Estimated correlations in 2-D deformed space. (c) Estimated correlations in 3-D deformed space.}
%\label{fig:4}
%\end{figure}

\begin{figure}[!t]
    \centering
    \includegraphics[scale=0.30]{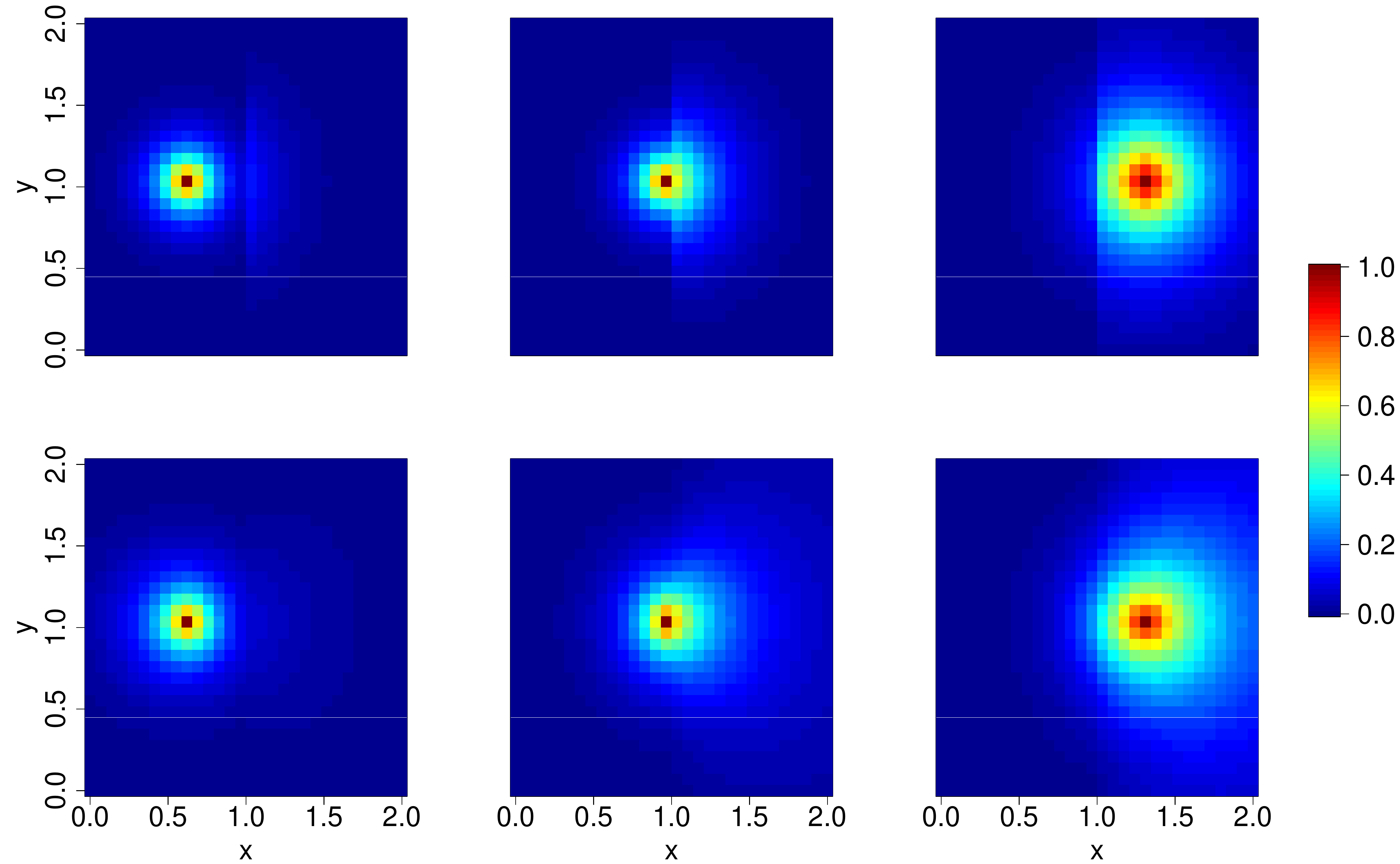}
    \caption{True correlations (top row) and estimated nonstationary correlations (bottom row) for three different locations.}
    \label{fig:nscor}
\end{figure}

To evaluate the performance of our method, we fit the nonstationary covariance model in Equation~\ref{eq:model} by choosing $\text{C}_\mathcal{D}(\cdot)$ to be an isotropic Mat{\'e}rn covariance function. We select three locations and plot their correlations with every other location on the simulation grid. The resulting correlation maps are displayed in Figure~\ref{fig:nscor}. The true correlation function underlying this simulation is used to produce the map in the top row of Figure~\ref{fig:nscor}, whereas the bottom row shows the correlation map from the estimated nonstationary covariance model. The similarities between the true correlations and the estimated nonstationary correlations in Figure~\ref{fig:nscor} demonstrate the effectiveness of our approach in capturing the nonstationary spatial dependence as the deformation-based model satisfactorily recovers the varying spatial range for the two subregions. 

\begin{figure}[!t]
\centering     %%% not \center
\subfigure[]{\label{fig:19a}\includegraphics[width=44mm]{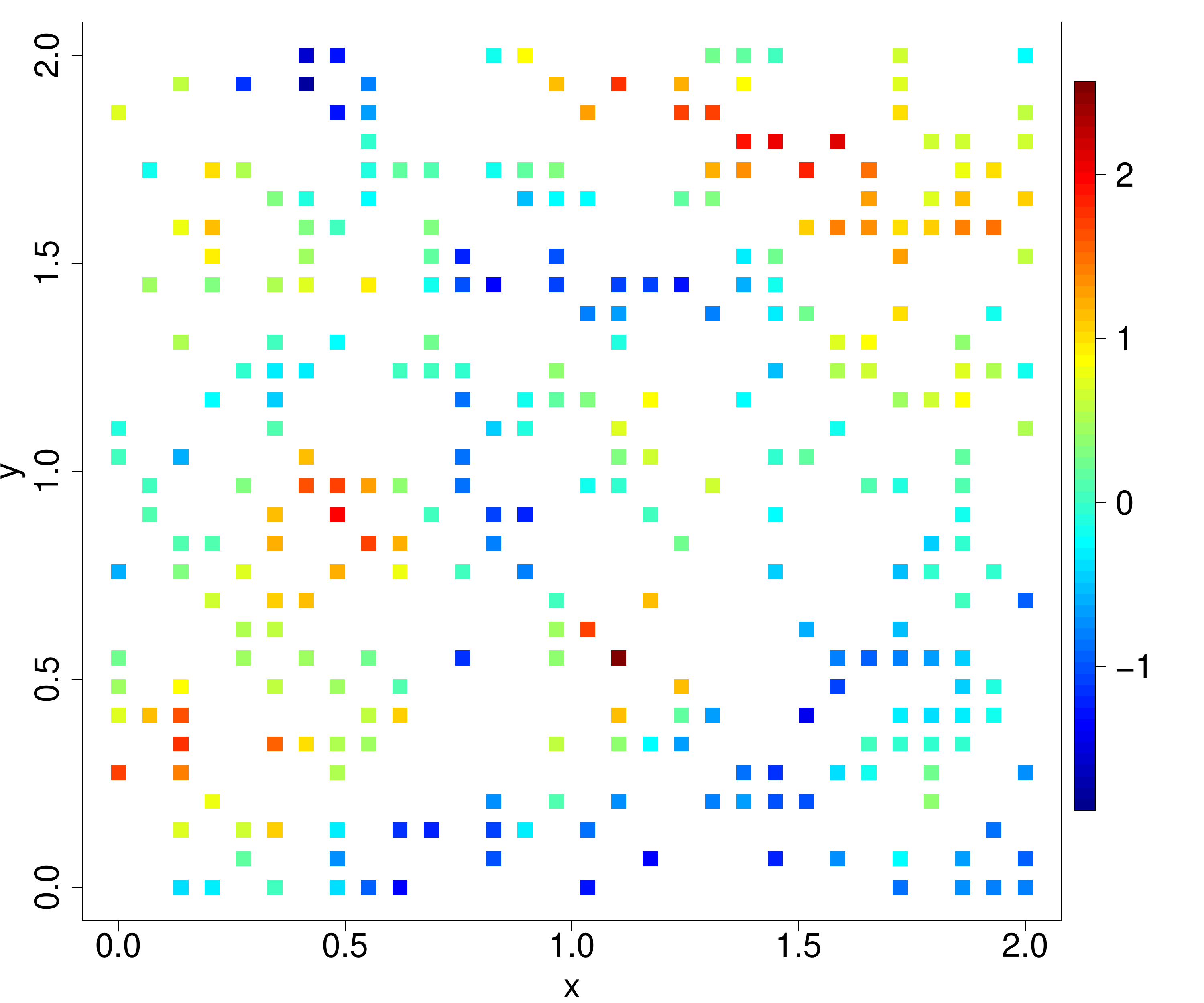}}
\subfigure[]{\label{fig:19b}\includegraphics[width=44mm]{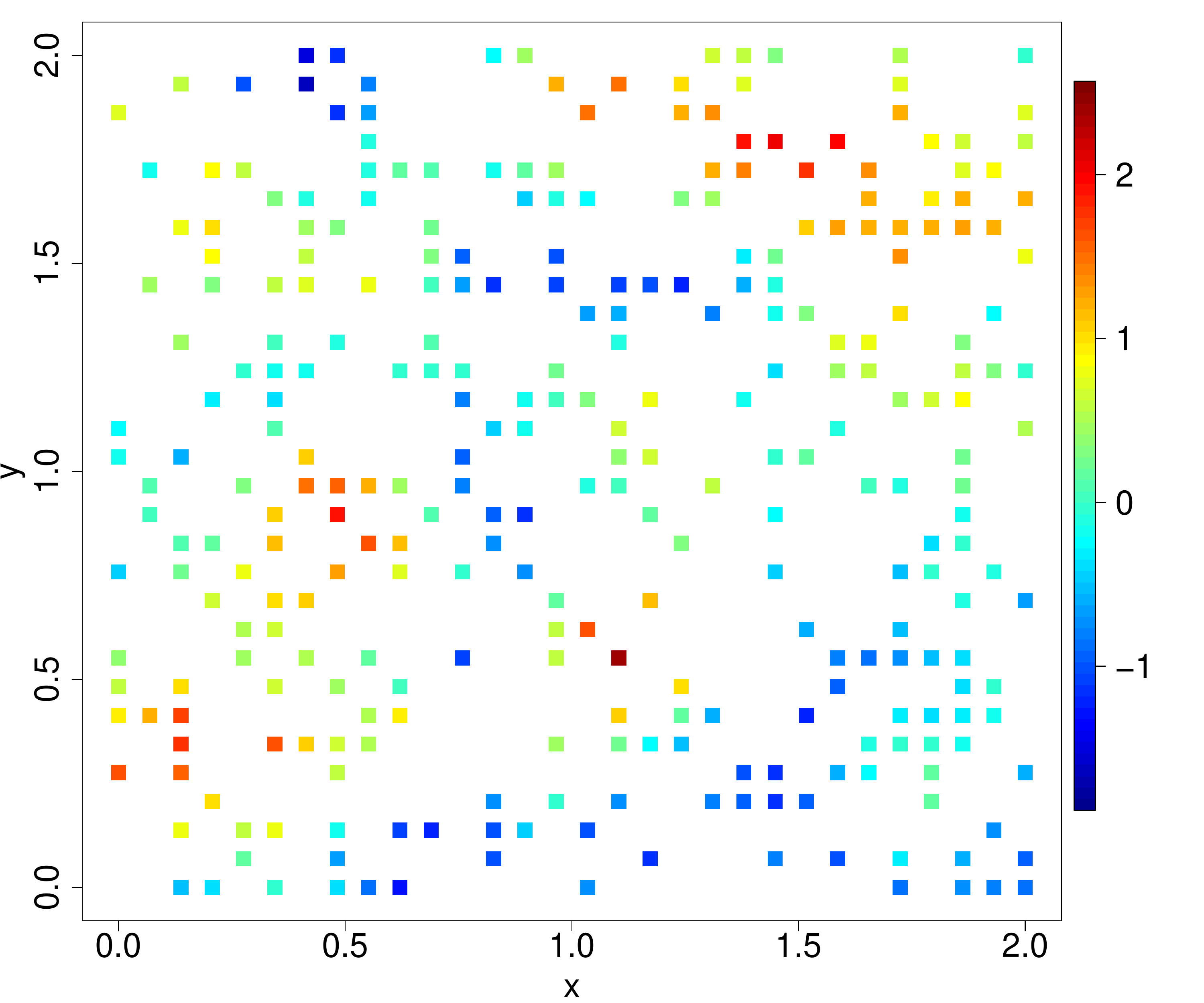}}
\subfigure[]{\label{fig:19c}\includegraphics[width=44mm]{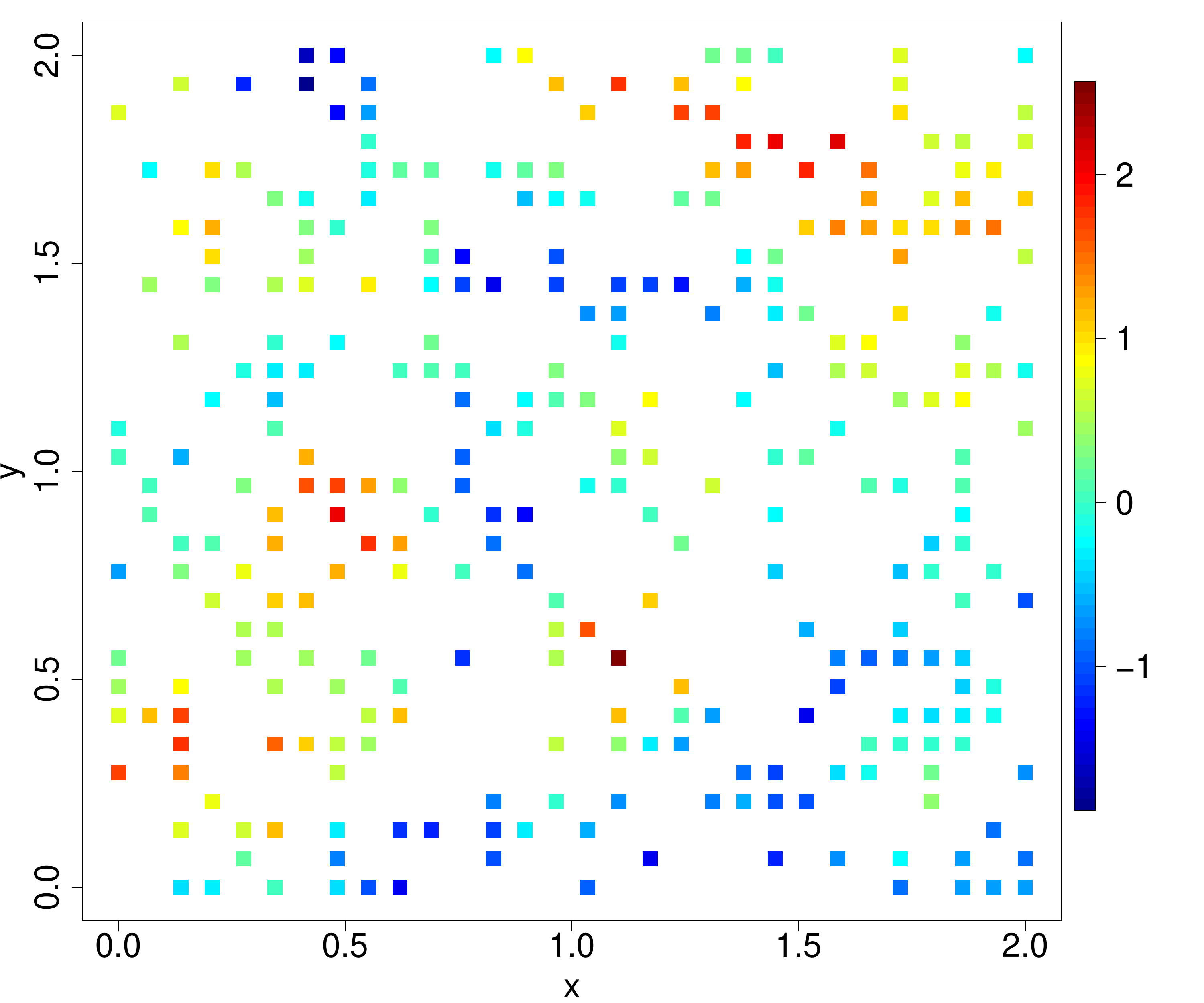}}\\
\subfigure[]{\label{fig:19d}\includegraphics[width=44mm]{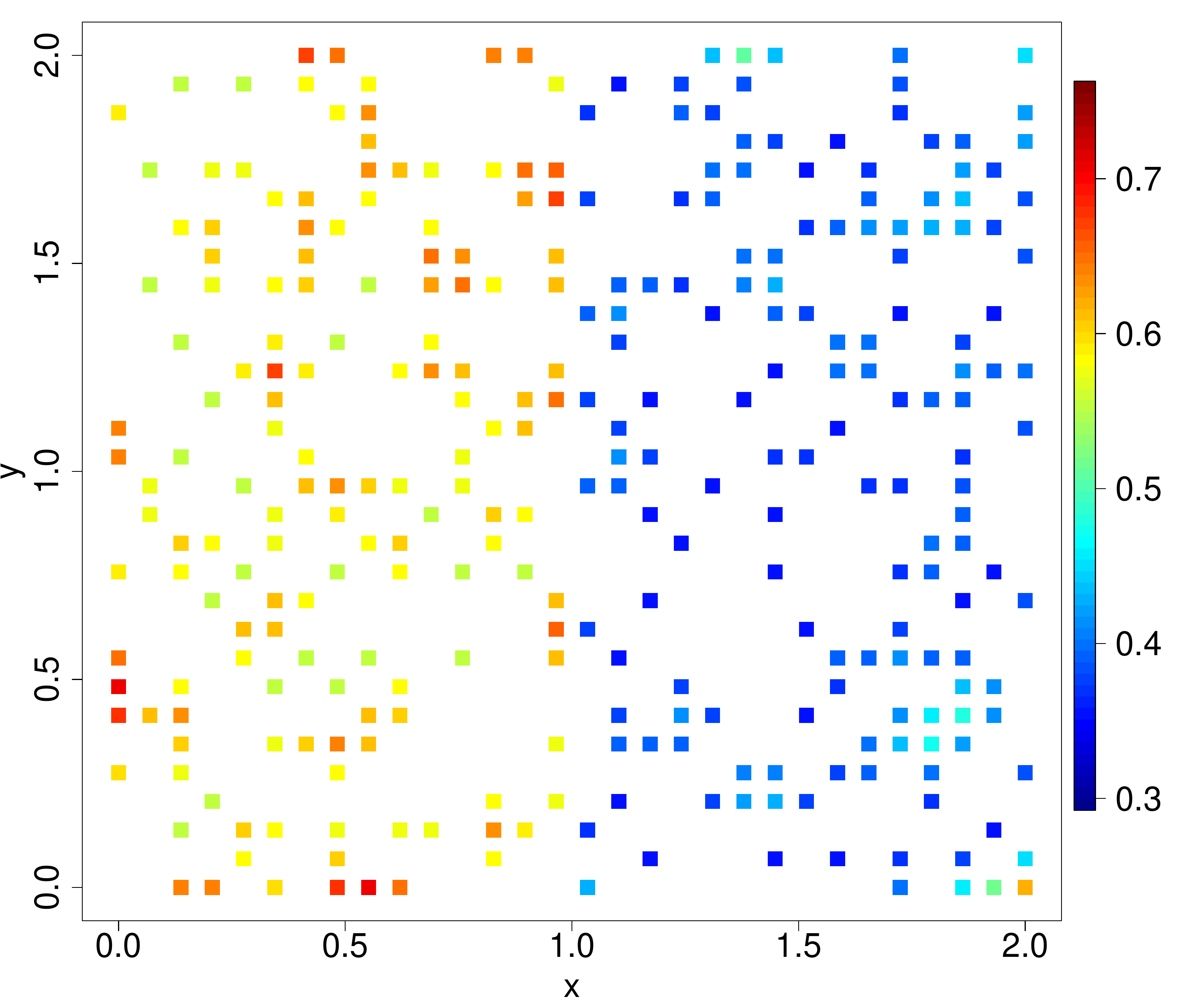}}
\subfigure[]{\label{fig:19e}\includegraphics[width=44mm]{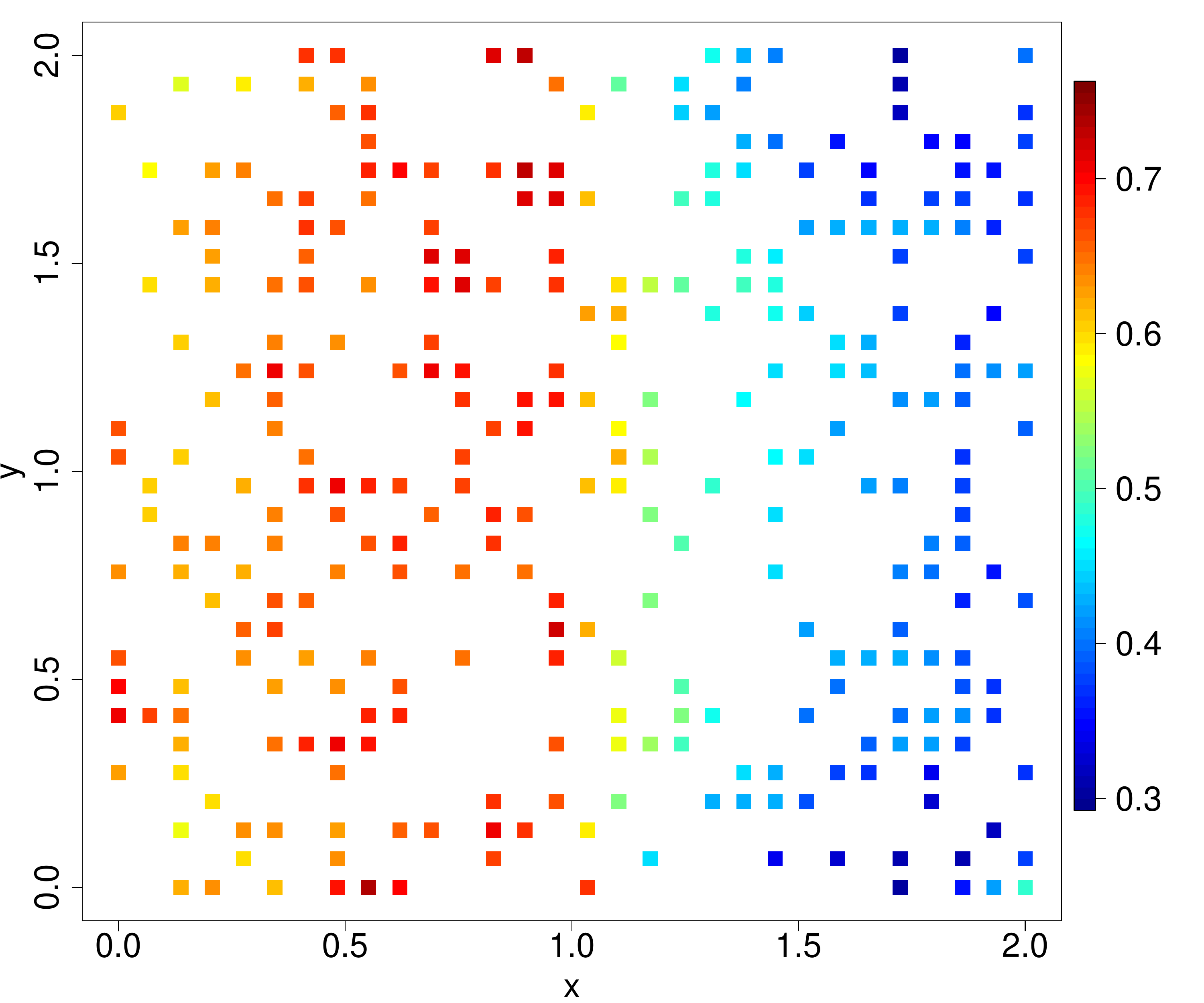}}
\subfigure[]{\label{fig:19f}\includegraphics[width=44mm]{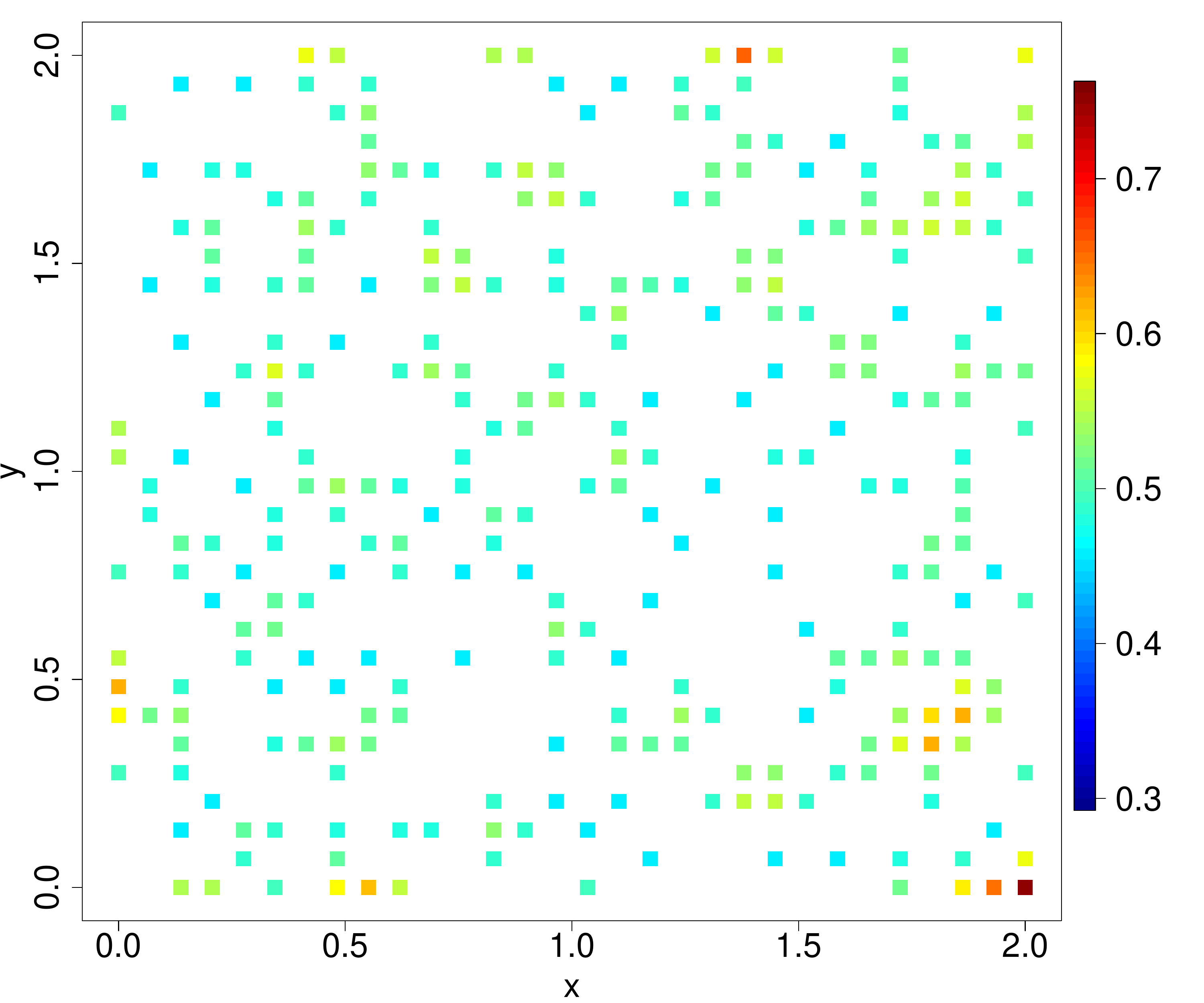}}

\caption{Kriged values with (a) true covariance function, (b) an isotropic Mat{\'e}rn covariance function in the deformed space and (c) an isotropic Mat{\'e}rn covariance function in the geographic space. Kriging standard deviation with (d) true covariance function, (e) an isotropic Mat{\'e}rn covariance function in the deformed space and (f) an isotropic Mat{\'e}rn covariance function in the geographic space. }
\label{fig:19}
\end{figure}

To assess the gains in prediction using our method, we randomly divide the simulated data into a training set of 600 points, and a validation set of 300 points, and re-estimate the deformation $\theta$ based on training data. We perform kriging on the 300 test locations with (1) the true covariance function, (2) an isotropic Mat{\'e}rn covariance function estimated in the geographic space, and (3) the nonstationary covariance model in Equation~\ref{eq:model}, with $\text{C}_\mathcal{D}(\cdot)$ being an isotropic Mat{\'e}rn covariance function estimated in 3-D deformed space. Kriged values and kriging standard deviations for the true model, stationary approach and the nonstationary model are shown in Figure~\ref{fig:19}. The close resemblance of the predicted values, shown in Figures~\ref{fig:19a},~\ref{fig:19b} and~\ref{fig:19c} highlights qualitative equivalence in the performance of point prediction, using  both the stationary approach and our method. The similarity between Figures~\ref{fig:19d} and~\ref{fig:19e}, in contrast with the lack of similarity between Figures~\ref{fig:19d} and~\ref{fig:19f}, conveys a significant improvement in estimating prediction uncertainties (kriging standard deviations) using our method in comparison with the stationary approach. Our method significantly outperforms the stationary approach in estimating the regional variations of the prediction uncertainties, and therefore provides us with more reliable prediction estimates and prediction intervals.

\section{Application to Precipitation Data}\label{sec:app}
\begin{figure}[t]
\centering     %%% not \center
\subfigure[]{\label{fig:9a}\includegraphics[width=60mm]{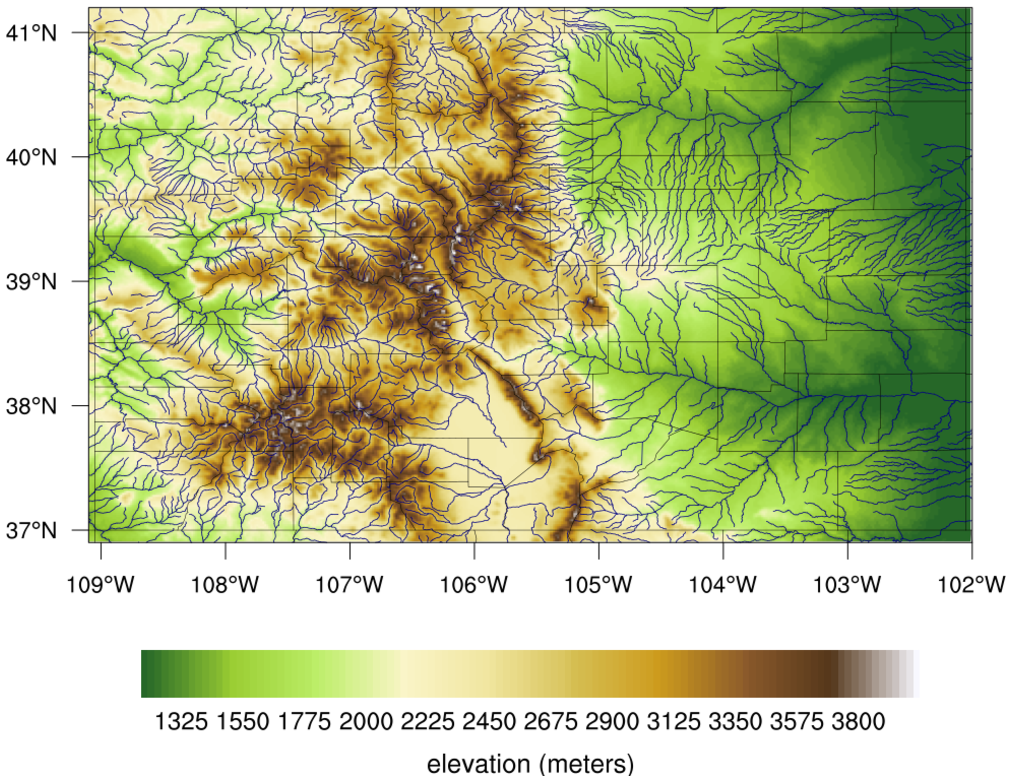}}
\subfigure[]{\label{fig:9b}\includegraphics[width=60mm]{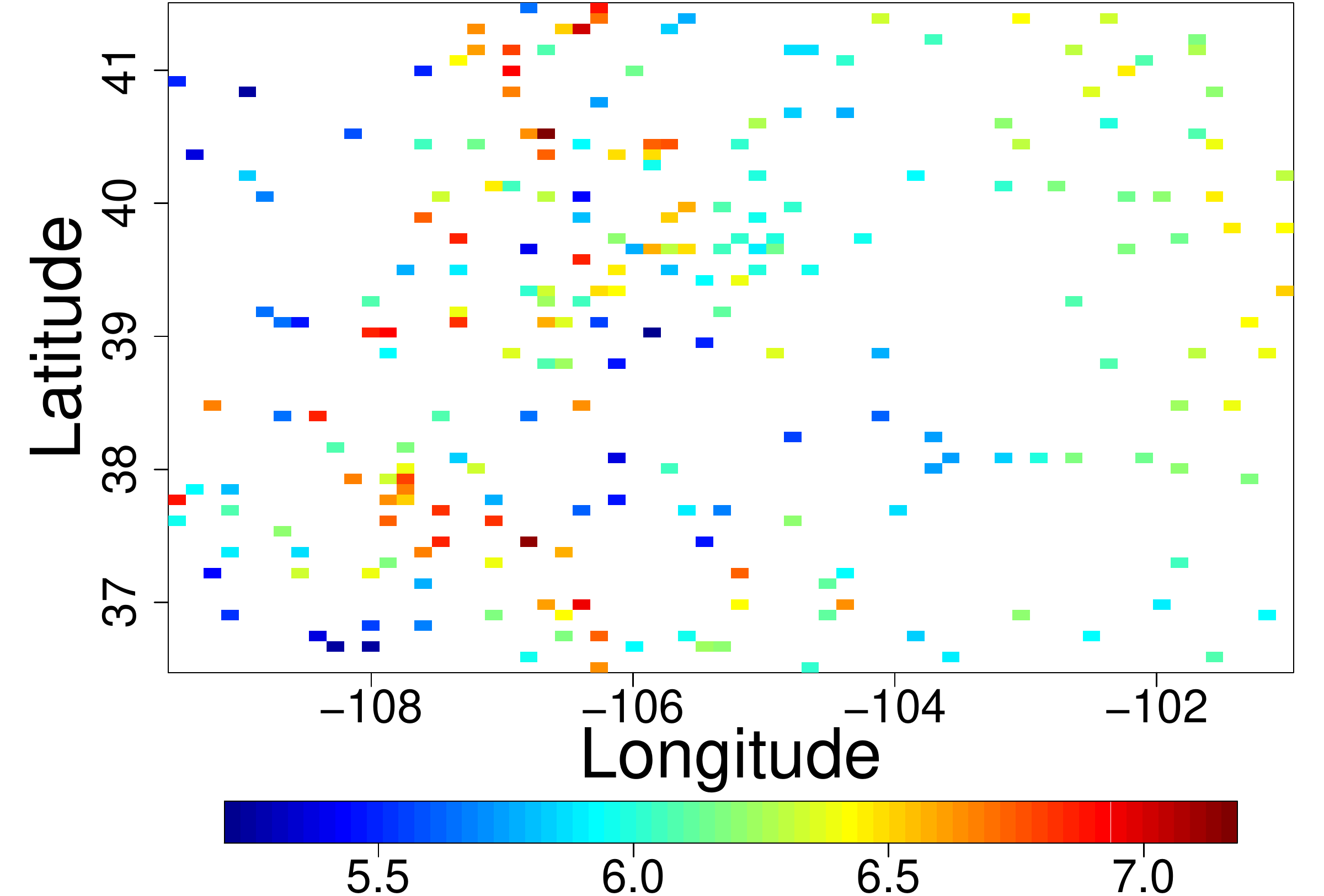}}\\
\caption{(a) Topography of Colorado showing mountainous Western region and Eastern flat plains. (Image source: \cite{ncl}). % (\url{https://www.ncl.ucar.edu/Applications/Images}). 
 (b) Observed log-transformed total annual precipitation for the year 1992.}
\label{fig:9}
\end{figure}

In this section, we illustrate the application of our method to the total annual precipitation data for the state of Colorado in the United States. The data came from Colorado's climate record provided by the Geophysical Statistics Project at the National Center for Atmospheric Research (NCAR) (\url{http://www.image.ucar.edu/GSP/Data/US.monthly.met/CO.html}). It contains monthly total precipitation (in millimeters) recorded from a network of weather stations located across the state of Colorado over the period of 1895-1997. The spatial domain of interest has a varied topography with noticeable distinction between the mountainous region in the West and the flat plains in the East, as shown in Figure \ref{fig:9a}. Furthermore, the topographical variability in Western Colorado is much higher than in Eastern Colorado. The diverse topography induces landform driven nonstationarity in the precipitation data which has been studied previously by \cite{Paciorek:2006aa}. For our analysis, we consider the log-transformed total annual precipitation data for the year 1992 (shown in Figure \ref{fig:9b}), since the number of weather stations (254) having non-missing recordings for total annual precipitation is highest for this year. The distribution of log-transformed precipitation data is approximately Gaussian, which makes it suitable for modeling as a Gaussian process. We apply the proposed deformation-based approach to model the nonstationary spatial dependence in the data by assuming regional stationarity within the mountainous subregion in the West and flat plains in the East. Based on the estimated nonstationary model, we interpolate the sparsely observed data to a fine resolution of $0.29^\circ$ longitude $\times$ of $0.17^\circ$ latitude by kriging. We also compare the proposed method to the stationary approach based on prediction performance.

\begin{figure}[t]
\centering     %%%% not \center
\subfigure[]{\label{fig:10a}\includegraphics[width=60mm]{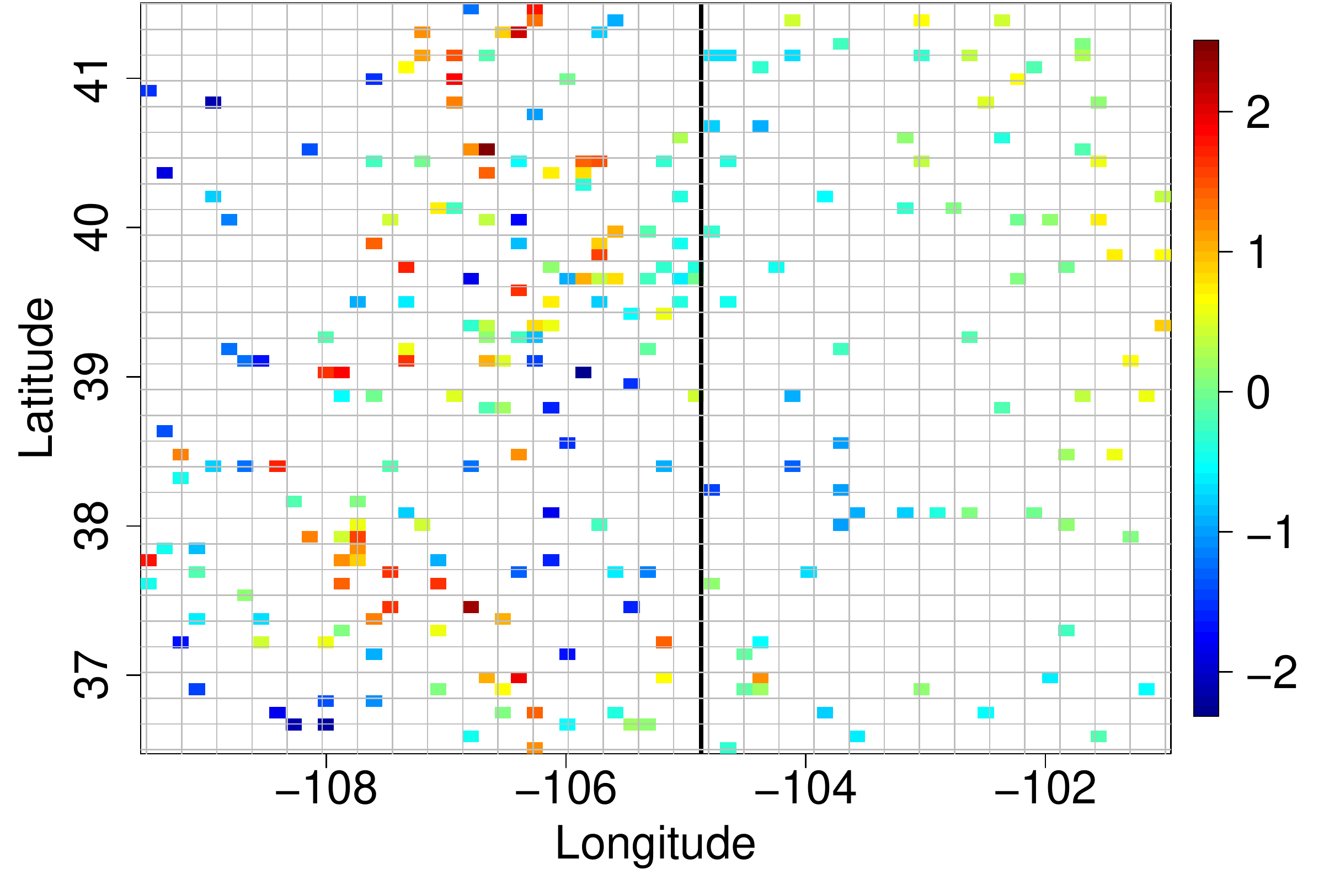}}
\subfigure[]{\label{fig:10b}\includegraphics[width=60mm]{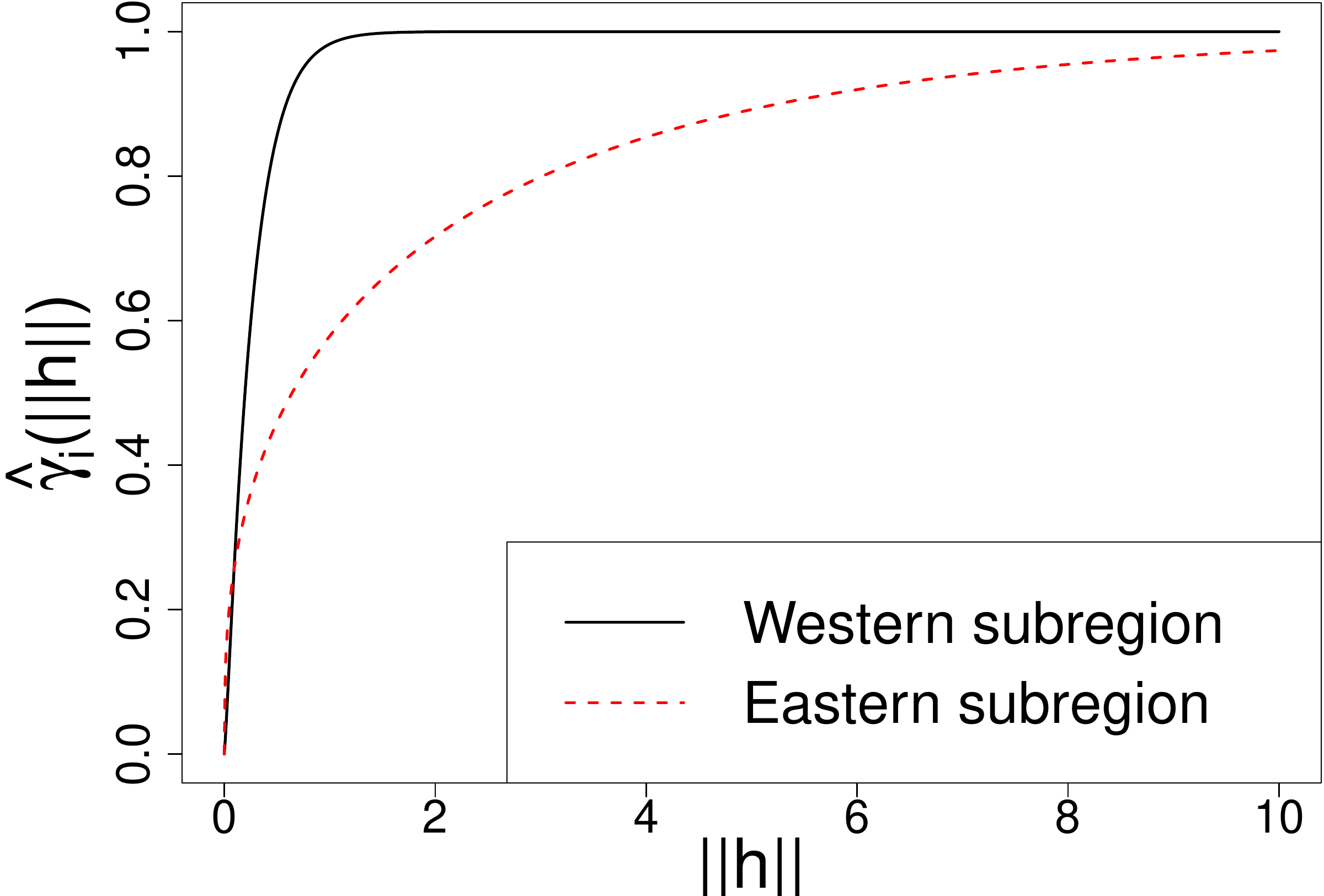}}
\subfigure[]{\label{fig:10c}\includegraphics[width=60mm]{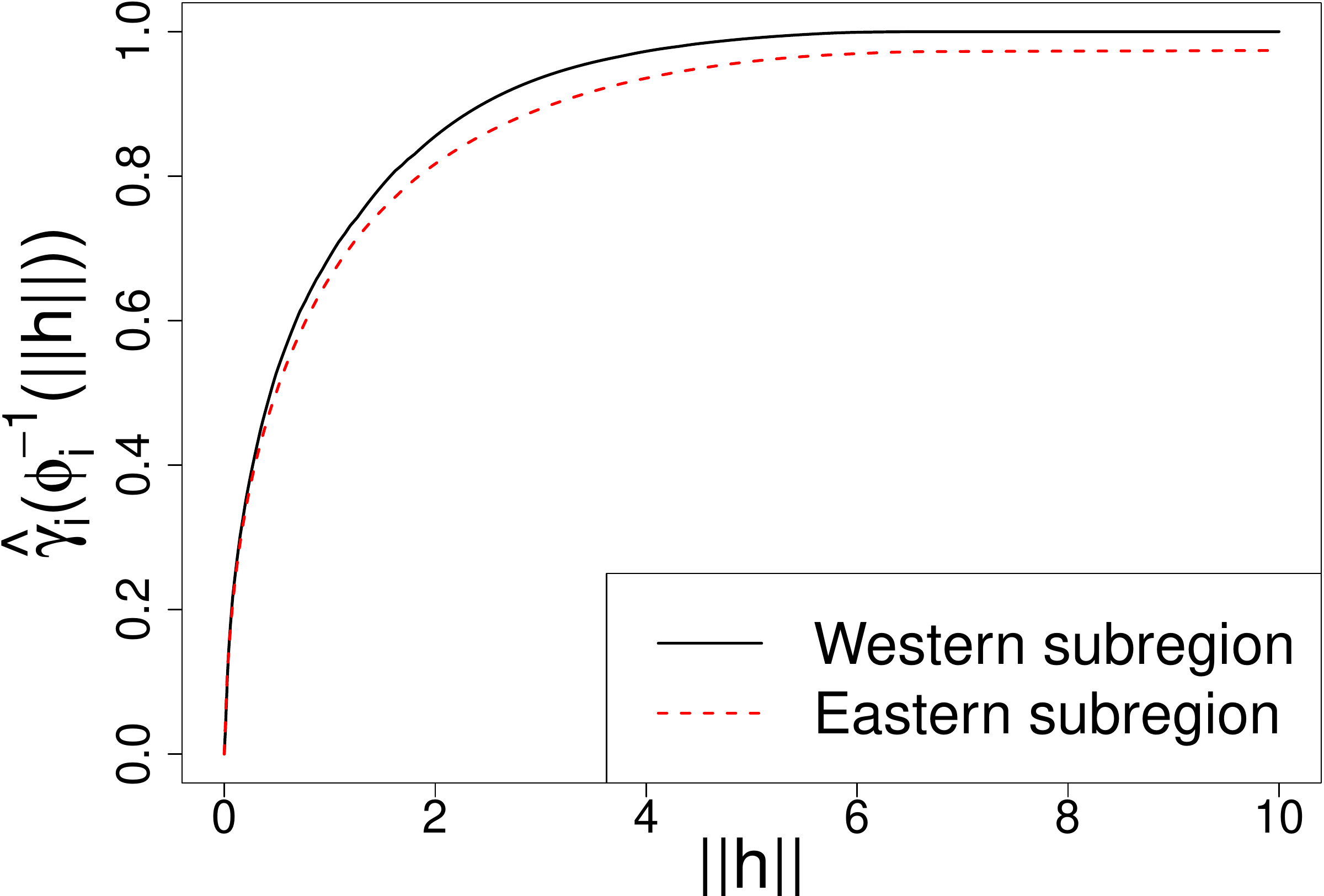}}
\subfigure[]{\label{fig:10d}\includegraphics[width=60mm]{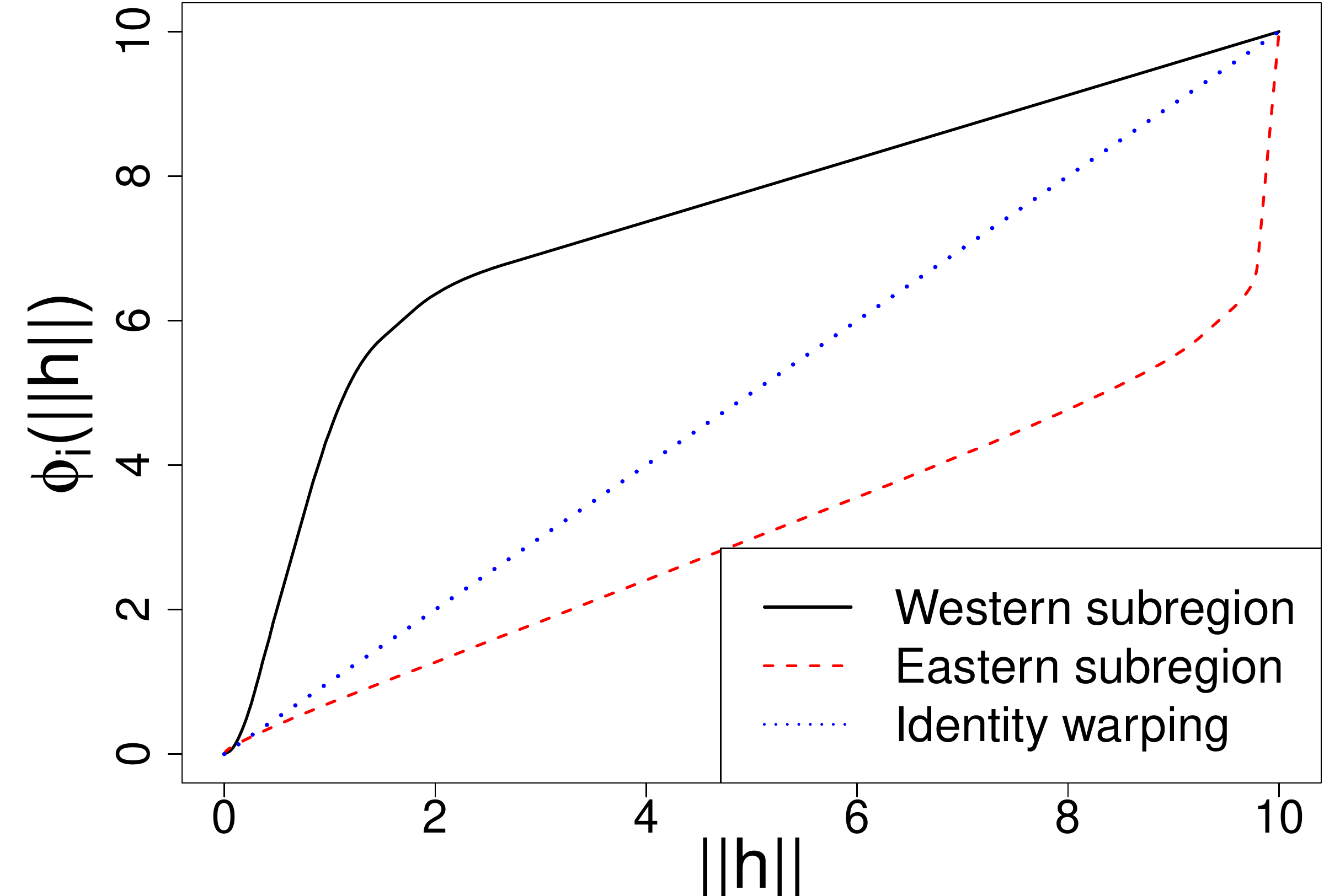}}
\caption{(a) Standardized log-transformed total annual precipitation data with solid black line depicting the chosen partitioning. The grey-colored lines depict the additional grid locations for the interpolation. (b) Estimated standardized regional variograms. (c) Registered variograms. (d) Regional distance warping functions. }
\label{fig:10}
\end{figure}

We begin by standardizing the data and splitting the entire region into Western and Eastern subregions demarcated by the longitude 104.873$^\circ$ W, as shown in Figure~\ref{fig:10a}. The chosen partitioning, which was also considered by \cite{Paciorek:2006aa} in their analysis, is motivated by the fact that the resulting two subregions differ significantly in their topographic features. Additionally, due to the Eastern and Western subregions being mostly flat plains and mountainous, respectively, the process can be reasonably assumed to be regionally stationary. The 254 observed locations are then randomly divided into a test set of 30 locations and a training set of 224 locations. The training set contains 153 observed locations in the Western subregion and 71 observed locations in the Eastern subregion. An isotropic Mat{\'e}rn variogram model, estimated using MLE, is considered for describing the spatial dependence structure for each subregion. Figure~\ref{fig:10b} shows the estimated regional Mat{\'e}rn variogram models, standardized using their respective regional variances. The estimated variance parameter varies slightly for the two subregions. However, due to their negligible difference, we choose to ignore such insignificant nonstationarity in variance. The estimated regional variograms show long range spatial dependence for the Eastern subregion and relatively shorter range spatial dependence for the Western subregion. Figure~\ref{fig:10d} shows the two estimated regional distance warping functions obtained by registration of the standardized estimated regional variograms. The large deviation of regional distance warping functions from the identity warping shows the prevalence of a high degree of nonstationarity when the entire region is considered. Based on these regional distance warping function, we can infer that modeling the two subregions with a common stationary variogram in the geographic space is clearly an imprecise approach to describe the spatial dependence of this process.

\begin{figure}[h]
\centering     %%% not \center
\includegraphics[scale=0.25]{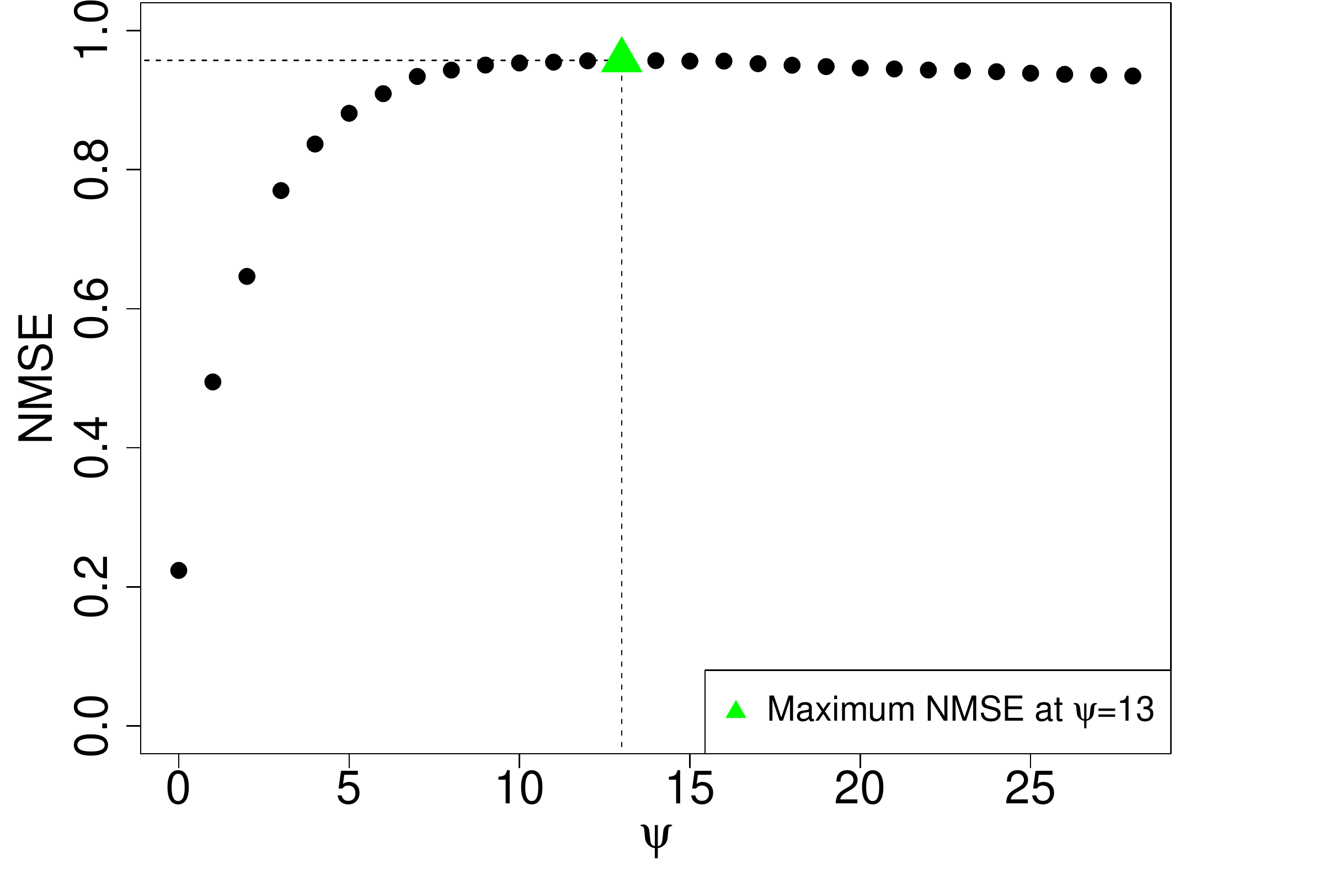}
\caption{NMSE between the transformed distance matrix $\Delta$ and the distance matrix for the estimated deformed space of dimensionality $d^{\mathcal{G}}+\psi$, for different values of $\psi$ ranging from 0 to 28. $d^{\mathcal{G}}$ in this case is 2.}
\label{fig:dim}
\end{figure}

\begin{figure}[!t]
\centering
\subfigure[]{\label{fig:11a}\includegraphics[width=60mm]{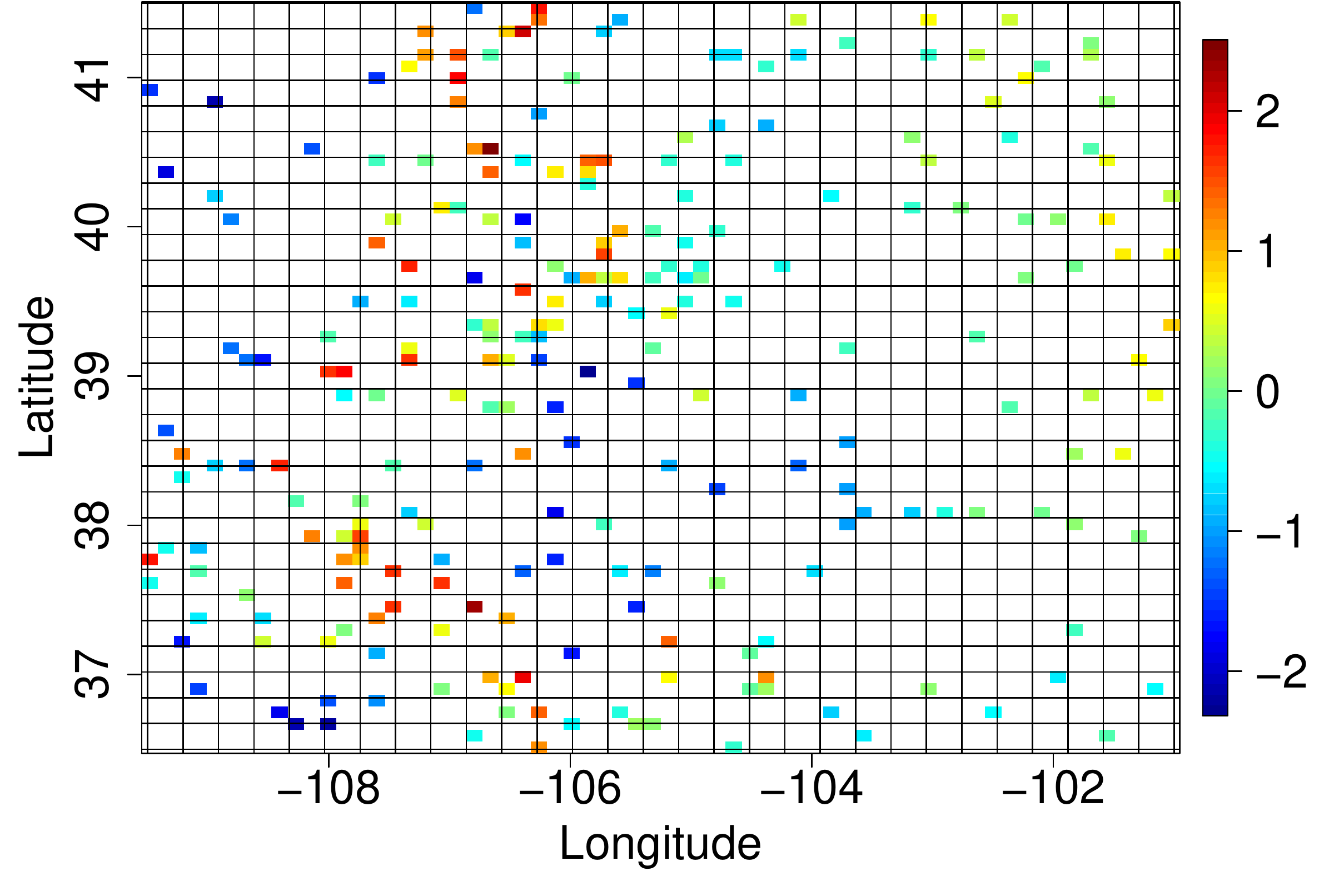}}
\subfigure[]{\label{fig:11b}\includegraphics[width=60mm]{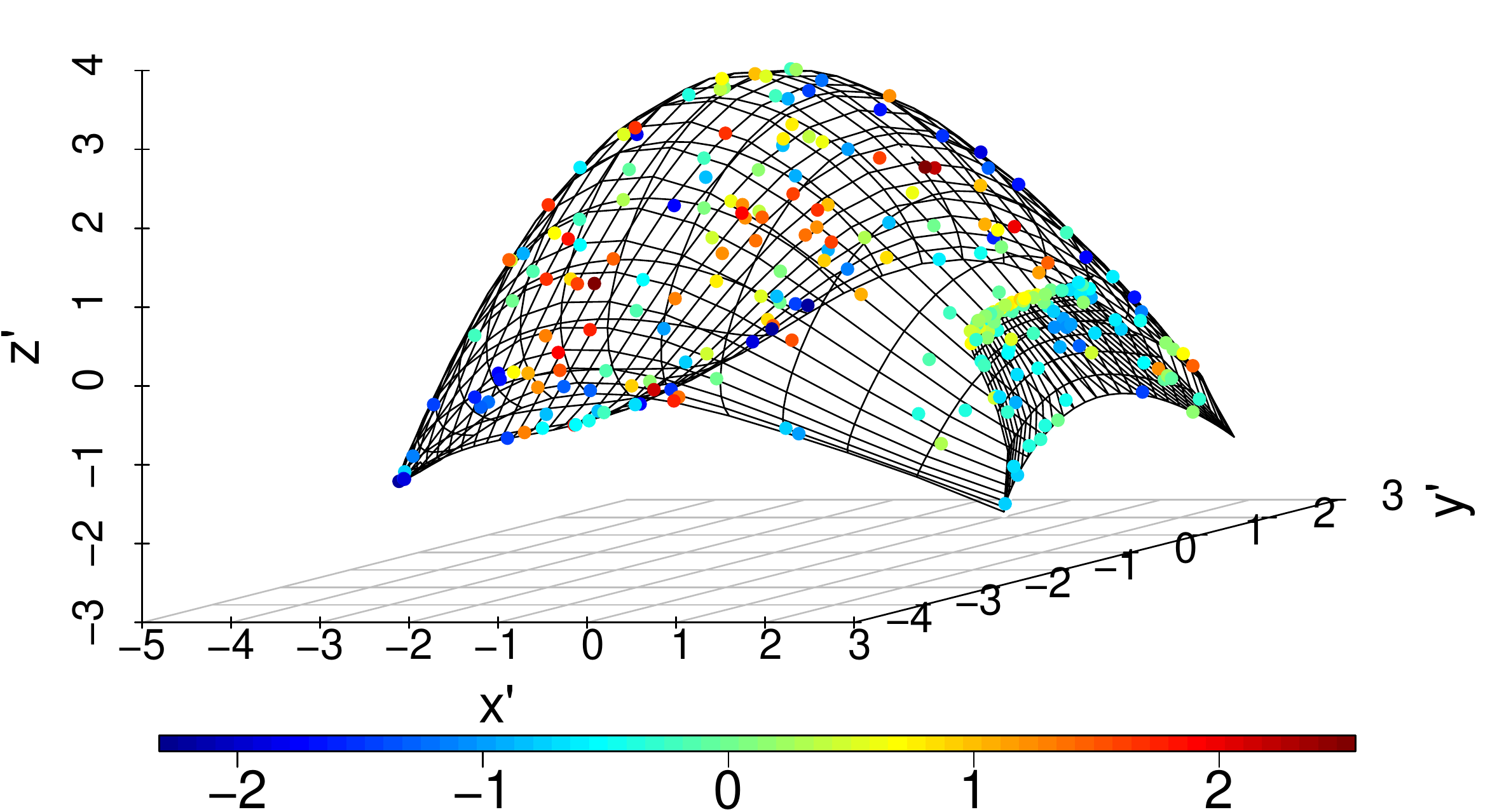}}
\caption{Standardized log-transformed total annual precipitation data in (a)  geographic space and (b) estimated deformed space in first three dimensions of maximum variation.}
\label{fig:11}
\end{figure}

We proceed to estimate the deformed space for locations corresponding to the training locations, test locations and an additional grid of locations (shown in Figure~\ref{fig:10a}) chosen for the interpolation. The estimation of the deformed space requires the specification of dimensionality $d^{\mathcal{D}}=d^{\mathcal{G}}+\psi,$ which should be based on the accuracy of the CMDS approximation. For an optimal choice of $d^\mathcal{D}$, or equivalently $\psi$, we estimate the deformed space for $\psi=0,1,...,28$, and compute the normalized mean squared error (NMSE) between the transformed distance matrix $\Delta$ and the distance matrix for the estimated deformed space for each value of $\psi$. Figure \ref{fig:dim} shows the computed NMSE versus different values of $\psi$. The value of NMSE closest to one indicates least discrepancy between the two distance matrices, which in this case indicates highest accuracy of the CMDS approximation. %Since $\psi=13$ achieves the NMSE closest to one, we consider the deformed space in 15-D $(\psi=13)$ as as our final estimate of deformation $\theta$.
We choose $\psi=13$ resulting in a 15-D deformed space and an associated final estimate of the deformation $\theta$. Figure~\ref{fig:11b} shows the estimated deformed space in the first three dimensions of maximum variation. We observe that the deformation leads to a very tight configuration of highly correlated points corresponding to the Eastern subregion and a highly sparse configuration of points corresponding to the Western subregion, resulting in an approximately constant spatial range and smoothness over the entire deformed space. This makes it suitable for modeling using a stationary variogram model.

\begin{figure}[h]
\centering     %%% not \center
\includegraphics[scale=0.20]{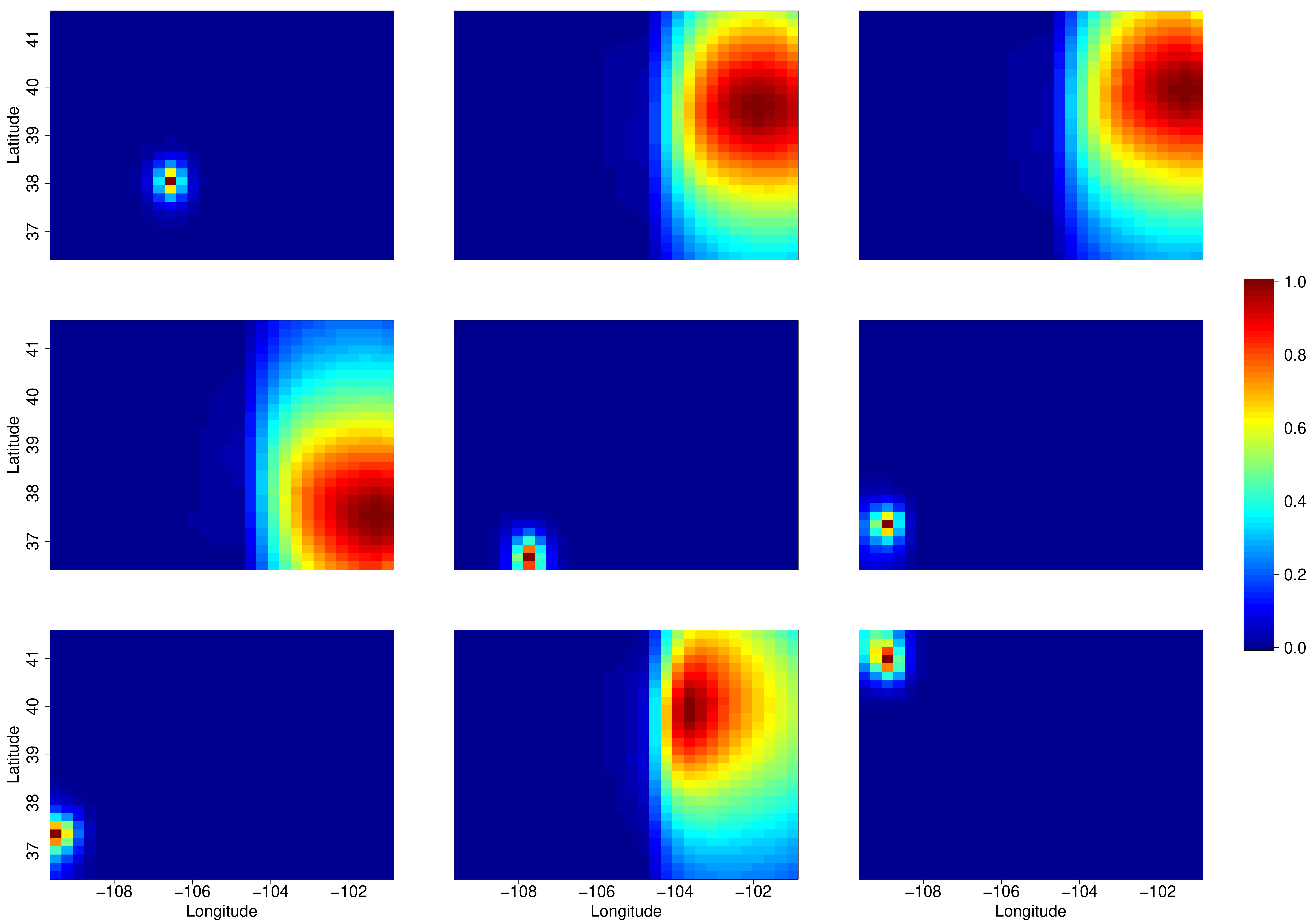}
\caption{Correlation map for nine randomly selected prediction locations.}
\label{fig:12}
\end{figure}

We now demonstrate the kriging predictions over the test locations and an additional grid of locations with our deformation-based nonstationary covariance function% in Equation ~\ref{eq:model}
, and compare prediction performance to the stationary approach. We fit an isotropic Mat{\'e}rn+Nugget covariance function in the geographic space and our nonstationary covariance model in Equation~\ref{eq:model}, with $\text{C}_\mathcal{D}(\cdot)$ also being an isotropic Mat{\'e}rn+Nugget covariance function in the 15-D deformed space. A visualization of the spatial correlations at nine randomly selected prediction locations for the estimated nonstationary covariance model %~\ref{eq:model}
is shown in Figure~\ref{fig:12}. We can see that our model perfectly captures the regionally varying spatial dependence structure, with strong spatial correlations in the Eastern subregion and relatively weaker spatial correlations in the Western subregion.

\begin{figure}[!t]
\centering     %%% not \center
\subfigure[]{\label{fig:13a}\includegraphics[width=60mm]{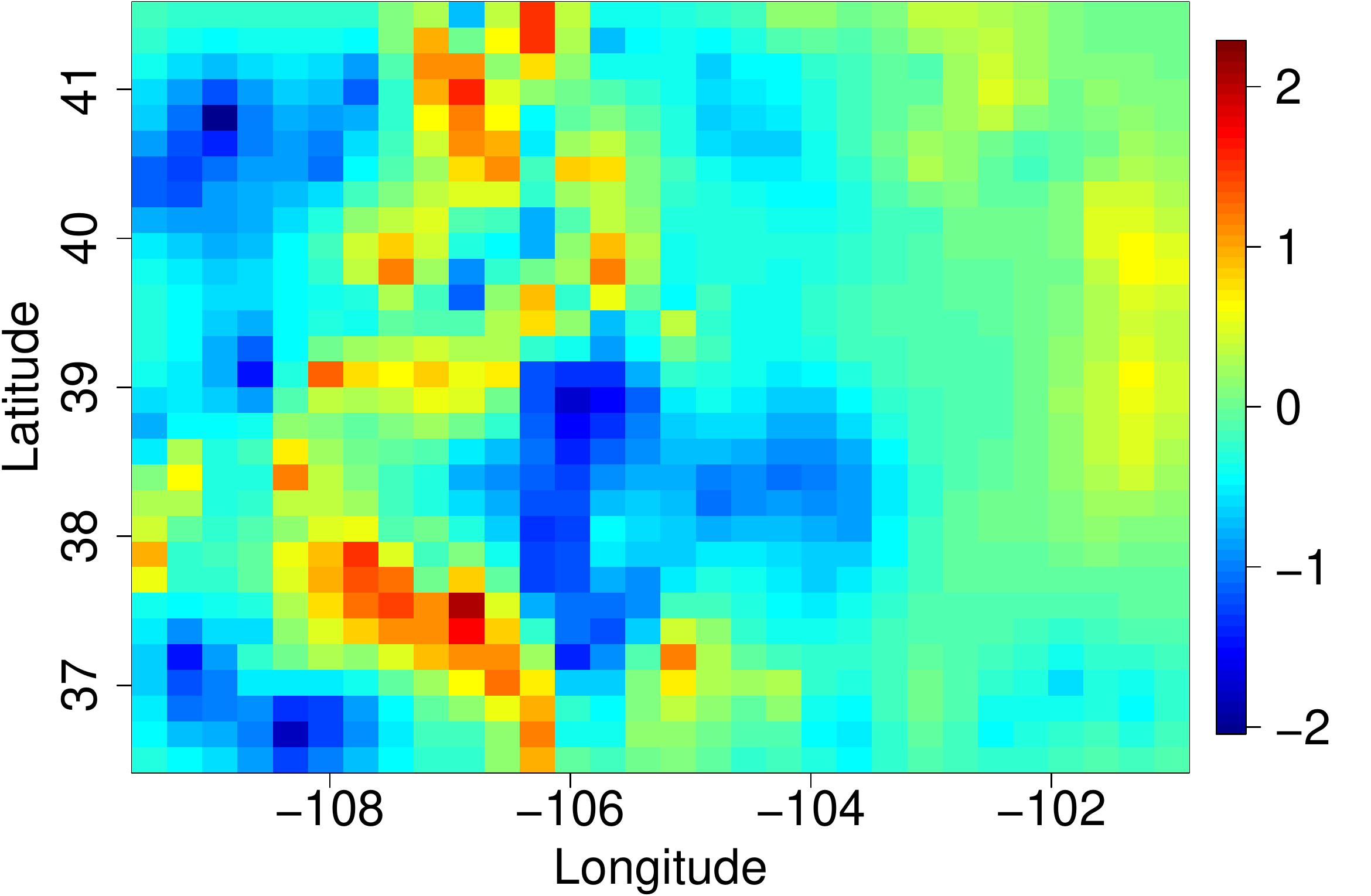}}
\subfigure[]{\label{fig:13b}\includegraphics[width=60mm]{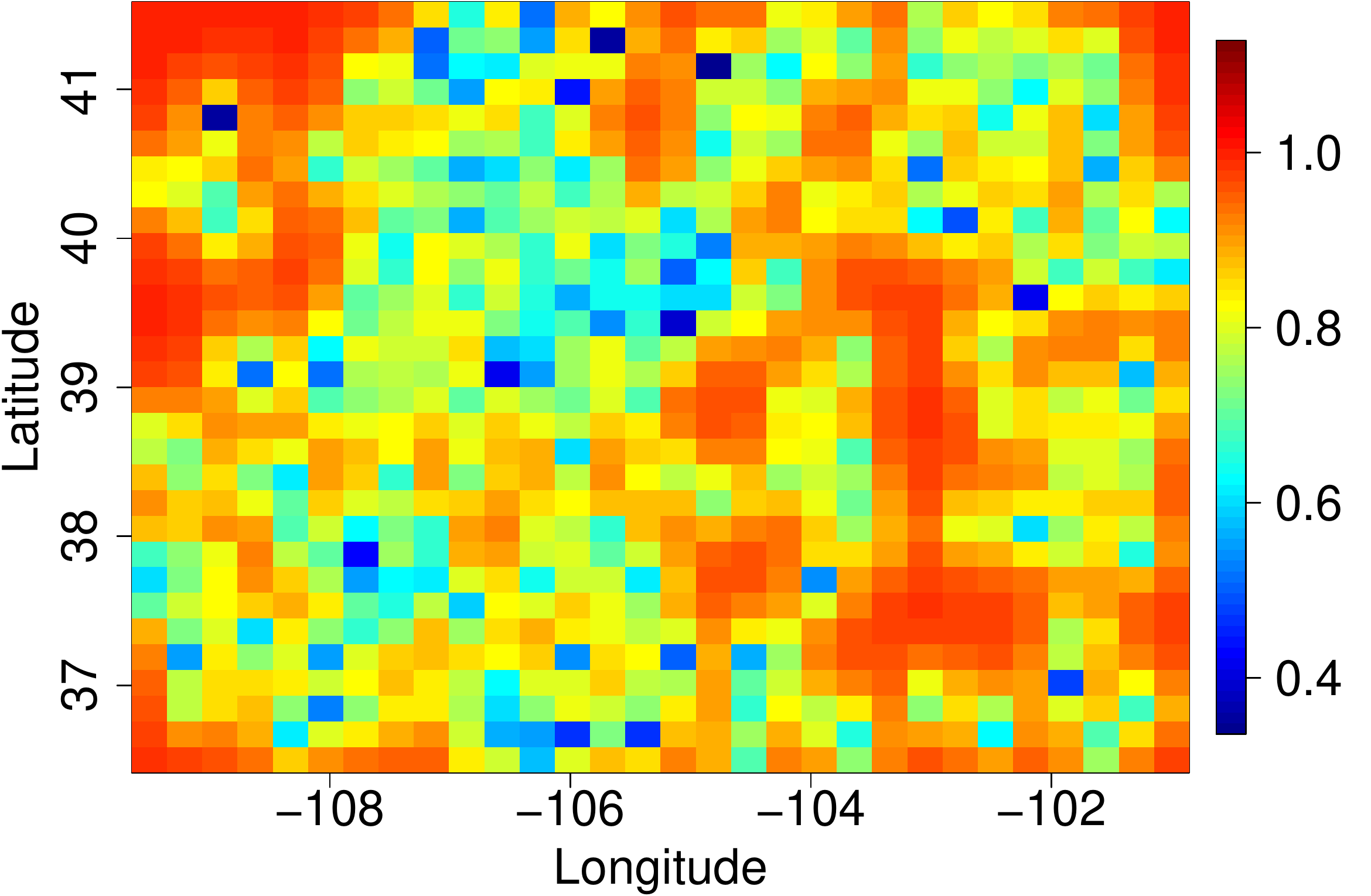}}\\
\subfigure[]{\label{fig:13c}\includegraphics[width=60mm]{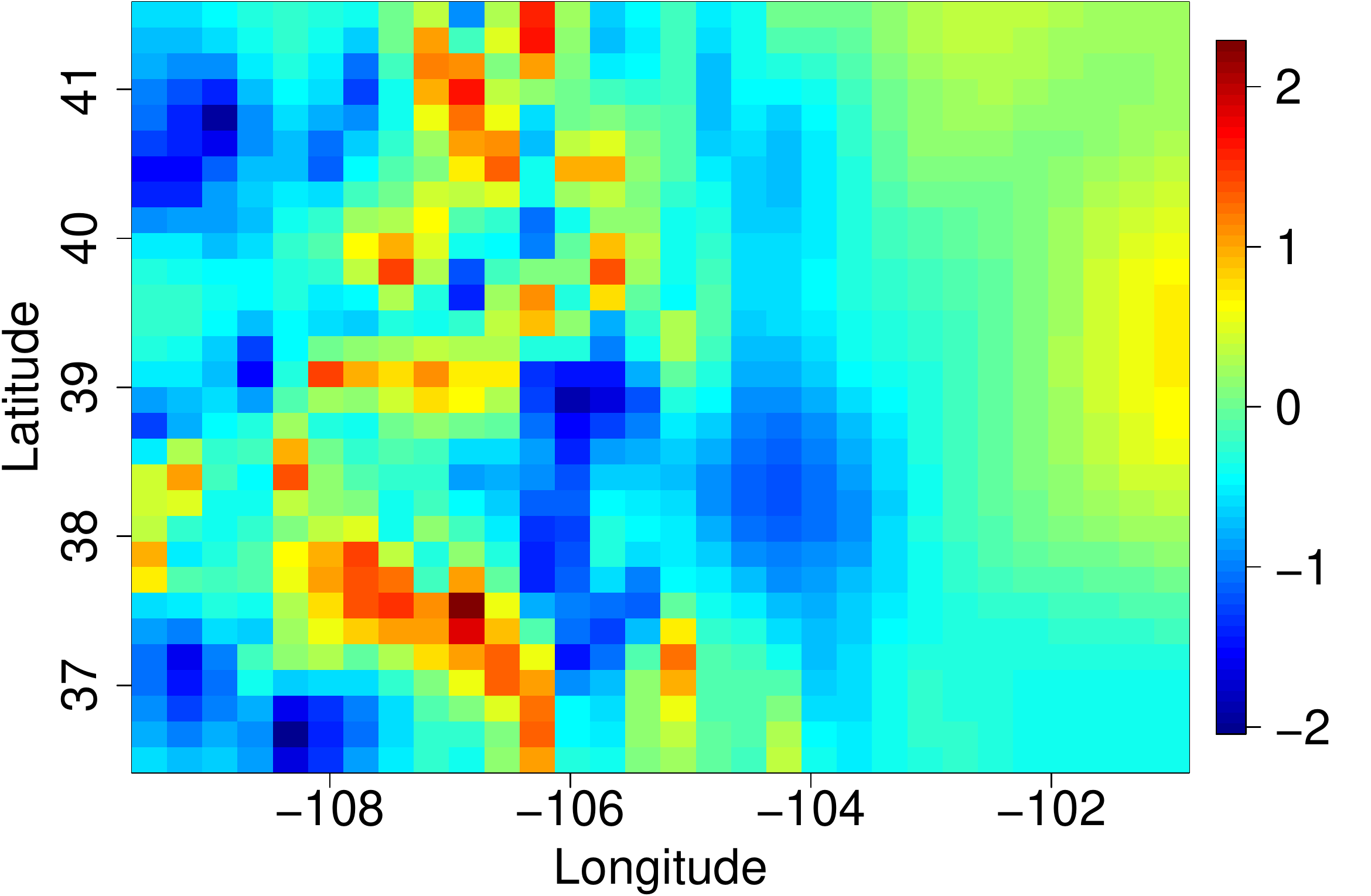}}
\subfigure[]{\label{fig:13d}\includegraphics[width=60mm]{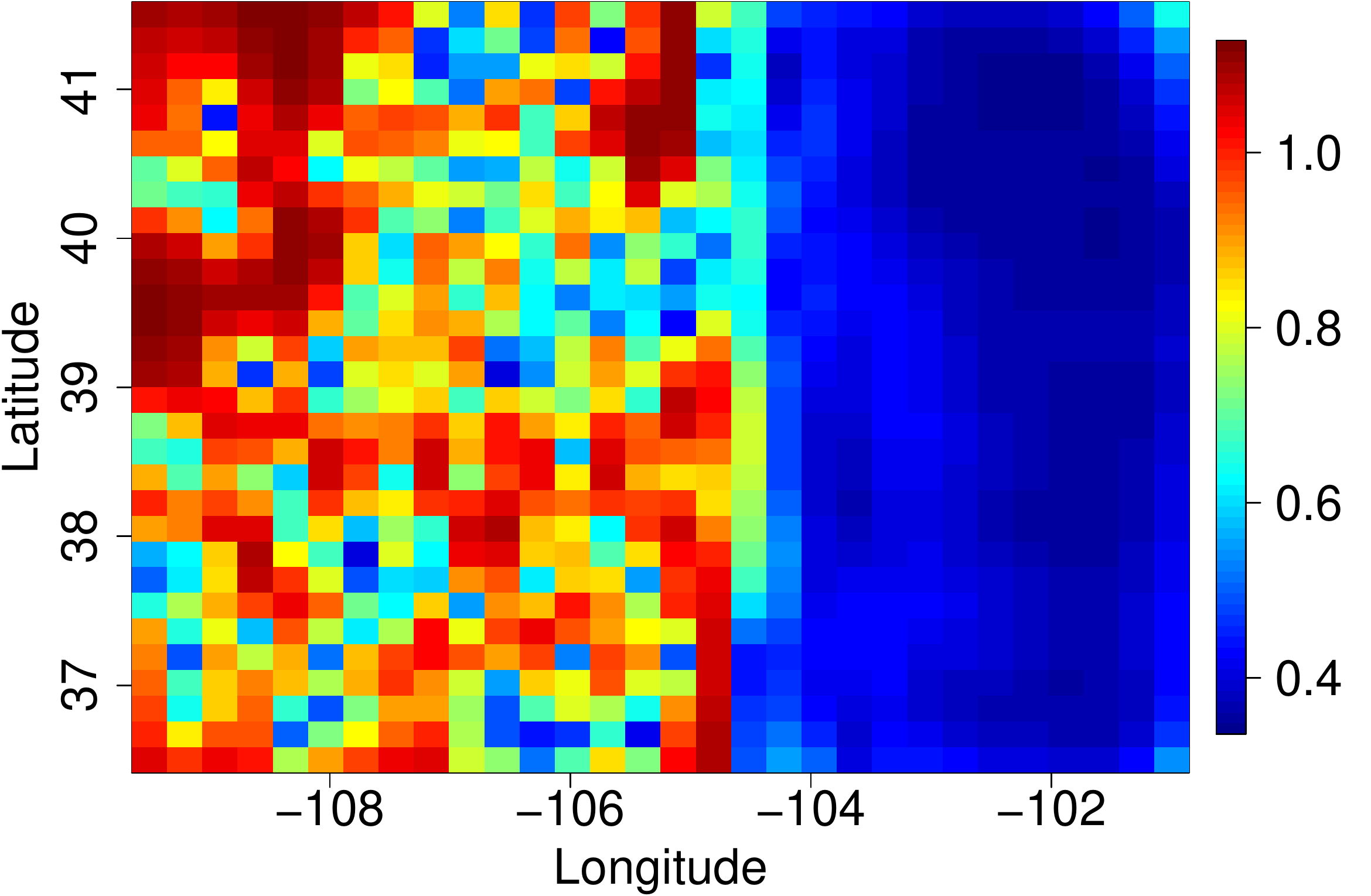}}\\

\caption{ Kriged values and kriging standard deviations for standardized log-transformed total annual precipitation data using the stationary approach ((a),(b)) and the nonstationary approach ((c),(d)). }
\label{fig:13}
\end{figure}

Based on the estimated stationary covariance model and the nonstationary covariance model, we perform kriging on the test set locations and additional grid locations. Figure~\ref{fig:13a} and Figure~\ref{fig:13c} show the kriged values for the stationary and nonstationary models, respectively. As the proposed method assumes the process to be regionally stationary, it takes into account the local features of each subregion, whereas the stationary model overlooks these local features causing the kriged values associated with the two models to look slightly different from each other. Kriged values in the Eastern subregion associated with the nonstationary model exhibit wider patches of highly correlated values than those based on the stationary model. We also observe remarkable differences in the kriging standard deviations, estimated from the stationary model (shown in Figure~\ref{fig:13b}) and the nonstationary model (shown in Figure~\ref{fig:13d}). Kriging standard deviations from the stationary model are nearly homogeneous throughout the entire domain, with more certain predictions in the Western subregion due to the availability of more dense observations. The stationary approach does not take into account the higher prediction uncertainty that arises due to higher topographical variability in the Western subregion as compared to the Eastern subregion. On the other hand, kriging standard deviations based on the nonstationary model are more realistic as they exhibits lower prediction uncertainty in the Eastern subregion and higher in the Western subregion; this can be attributed to the strong and weak spatial dependencies of these two subregions, respectively. Stronger spatial dependence provides more information for prediction, which leads to more certain predictions, and likewise, weaker spatial dependence leads to more uncertainty in prediction. In the context of deformation, a stronger spatial dependence is equivalent to a compressed subregion, where more observations are available in the neighborhood of a prediction location, leading to a reduction in prediction uncertainty. 

\begin{table}[t!] %***
\caption{Prediction scores for the stationary and nonstationary models.}
\label{table:1}\par
\vskip .2cm
\centerline{\tabcolsep=7truept\begin{tabular}{|lcc|} \hline %***5truept
Prediction score & Stationary model & Nonstationary model  \\\hline
MSPE & 0.44 & 0.37
 %\\NMSE &  0.731 & 0.730  
 \\MAE & 0.50 & 0.45 
 \\LogS  & 2.01 & 1.69 
 \\CRPS  &  0.36 & 0.33 \\ 
 \hline 
\end{tabular}}
\end{table}

We compute some commonly used prediction scores on a test set for quantification and comparison of the prediction performance for both methods. Table~\ref{table:1} summarizes the four prediction scores: mean squared prediction error (MSPE), mean absolute error (MAE), continuous ranked probability score (CRPS) and logarithmic score (LogS). Lower values for MSPE, MAE, CRPS and LogS are suggestive of better prediction performance. The percentage of reduction in (MSPE, MAE, LogS, CRPS) by using the proposed nonstationary model versus the stationary model is (15.91\%, 10\%, 15.92\%, 8.34\%), signifying superior prediction performance. Based on all prediction scores, and the visual inspection of kriging standard deviations, we conclude that our method provides improved point predictions with significantly better and realistic prediction uncertainties, in comparison with the stationary method. 
%\url{https://www.ncl.ucar.edu/Applications/Images/topo_8_lg.png} for topgraphic source.
\section{Discussion}\label{sec:disc}
In this article, we introduced a method to estimate a spatial deformation for modeling nonstationary spatial processes using functional data alignment of estimated regional variograms. To the best of our knowledge, this is the first study finding an intersection between Spatial Statistics and Elastic Functional Data Analysis. The proposed method allows for efficient estimation of the underlying nonstationary dependence structure by taking into account the stationary dependence structure of various subregions.

The proposed method allows for estimation of the deformation with a single realization of a spatial process. It also avoids the problem of folding of space, by allowing the deformation to be estimated in higher dimensions. Moreover, the estimated regional distance warping functions provide a neat exploratory tool that visualizes the degree of nonstationarity, and enable the identification of regions of low and high range spatial dependence.

As pointed out in Section~\ref{esti}, variogram alignment cannot account for the nonstationarity in sill and nugget. However, our method, in its entirety, can easily deal with this nonstationarity by allowing sill and nugget to spatially vary in the deformed space, i.e., by considering the following covariance function in the deformed space\[\text{Cov}(\textbf{s},\textbf{s}')=\sigma(\textbf{s})\sigma(\textbf{s}')\rho_\mathcal{D}\big(\|\theta(\textbf{s})-\theta(\textbf{s}')\|\big)+\tau(\textbf{s})\textbf{I}_{(\textbf{s}=\textbf{s}')}(\textbf{s},\textbf{s}')\text{\hspace{0.1cm} } \forall (\textbf{s},\textbf{s}')\in \mathcal{G}\times\mathcal{G},\] where $\rho_\mathcal{D}(\cdot)$ is any valid stationary and isotropic correlation function, $\sigma(\cdot)$ is a spatially varying standard deviation and $\tau(\cdot)$ is a spatially varying nugget. Variogram alignment attempts to homogenize the spatial range and smoothness, and therefore the  remaining heterogeneity in sill and nugget can be easily incorporated in the covariance function itself.

Our method is based on the subjective selection of subregions. Therefore, developing a method for the objective selection or an adaptive scheme for selection of subregions is one potential direction for future research. It would also be very interesting to develop a sophisticated method, based on covariate information, for objective partitioning of subregions. Although we have introduced a stepwise modeling approach, with separate steps for estimating the deformation and covariance function in the deformed space, it is still desirable to develop a scheme for joint estimation of deformation and covariance function, which can be another direction for the future.

%%%%%%%%%%%%%%%%%%%%%%%%%%%%%%%%%%%%%%%%%%%%%%%%%%%%%%%%%%%%%%%%%%%%%%%%%%%%%%%%%%%%%%%%%%%%%%%%%%%%%%%%%%%%%%%%%%%%%%%%%%%%
\vskip 14pt
\noindent {\large\bf Supplementary Materials}

Section S1 provides proofs for Properties 1 and 2. Section S2 provides discussion on the robustness of our method to different subdivisions of the spatial domain. Sections S3 provide extended results from the simulation study presented in Section \ref{sec:simulation}. Section S4 presents an additional simulation study. Section S5 gives a quantitative assessment of CMDS.

%\bibliographystyle{chicago}      % Chicago style, author-year citations
%\bibliography{main.bib}   % name your 

\newpage

\begin{center}
\large\textbf{Supplementary Materials}
\end{center}

\setcounter{section}{0}
\setcounter{equation}{0}
\def\theequation{S\arabic{section}.\arabic{equation}}
\def\thesection{S\arabic{section}}

\section{Proofs of Properties 1 and 2}
\noindent\textbf{Property 1:} For any $i\in \{1,2,...,k\}$, if $(\textbf{s},\textbf{s}') \in \mathcal{G}_i\times\mathcal{G}_i \implies \mathcal{L}(\textbf{s},\textbf{s}')=\{\mathcal{G}_i\}\implies \mathcal{P}(i,\textbf{s},\textbf{s}')=\|\textbf{s}-\textbf{s}'\|$ $\implies \mathcal{W}_i(\textbf{s},\textbf{s}')=\frac{\|\textbf{s}-\textbf{s}'\|}{\|\textbf{s}-\textbf{s}'\|}=1$ and $\mathcal{W}_j(\textbf{s},\textbf{s}')=0 \: \forall \: j\neq i$ $\implies \phi(\textbf{s},\textbf{s}')=\phi_i(\|\textbf{s}-\textbf{s}'\|)$.\\
\noindent\textbf{Property 2:} If the process $\{X(\textbf{s}):\textbf{s}\in \mathcal{G} \subset \mathbb{R}^{d^{\mathcal{G}}}\}$ is second-order stationary, then the regional distance warping functions are $identity\; functions$ (i.e., $\phi_i(\|\textbf{s}-\textbf{s}'\|)=\|\textbf{s}-\textbf{s}'\|$)\[\implies \phi(\textbf{s},\textbf{s}')=\sum_{\mathcal{G}_i\in \mathcal{L}(\textbf{s},\textbf{s}')}\mathcal{W}_i(\textbf{s},\textbf{s}')\|\textbf{s}-\textbf{s}'\|=\|\textbf{s}-\textbf{s}'\|\sum_{\mathcal{G}_i\in \mathcal{L}(\textbf{s},\textbf{s}')}\mathcal{W}_i(\textbf{s},\textbf{s}')=\|\textbf{s}-\textbf{s}'\|,\] as the sum of the weights $\mathcal{W}_i$  over ${\mathcal{G}_i\in \mathcal{L}(\textbf{s},\textbf{s}')}$ is guaranteed to be 1.
\par
\section{Robustness of Estimated Deformation to Domain Division}
\setcounter{equation}{0}
Let us consider a stochastic process $\{X(\textbf{s}),\;\textbf{s}\in\mathcal{G}\}$ with regional stationarity and global nonstationarity. Thus, it is possible to divide the entire domain $\mathcal{G}$ into subregions $\mathcal{G}_1,...,\mathcal{G}_k$ such that $\bigcup_{i=1}^k\mathcal{G}_i=\mathcal{G}$, and each of the $k$ regional processes $\{X(\textbf{s}),\; \textbf{s}\in\mathcal{G}_i\}$ are stationary processes admitting distinct isotropic variograms $\gamma_i(\|\cdot\|),\;i=1,2,...,k$. Here, let the partitioning $\{\mathcal{G}_1,...,\mathcal{G}_k\}$ to be referred as \textbf{True Partitioning} which is often unknown.

Let us assume that the regional variograms  $\{\gamma_i(\|\cdot\|),\;i=1,2,...,k\}$ are either known or can be estimated reasonably well enough. We align the $k$ distinct variograms $\{\gamma_i(\|\cdot\|),\;i=1,...,k\}$ to estimate $k$ distinct regional distance warping functions $\{\phi_i(\|\cdot\|),\;i=1,...,k\}$. Further, we define the global distance function for the \textbf{True Partitioning}  as
\begin{equation} \label{eqs:1}\phi(\textbf{s},\textbf{s}')=\sum_{\mathcal{G}_i\in \mathcal{L}(\textbf{s},\textbf{s}')}\mathcal{W}_i(\textbf{s},\textbf{s}')\phi_i(\|\textbf{s}-\textbf{s}'\|),\: \forall\: (\textbf{s},\textbf{s}') \in \mathcal{G}\times\mathcal{G},\end{equation}
where $\mathcal{L}(\textbf{s},\textbf{s}')$ is the set of subregions $\mathcal{G}_i$ such that the line segment joining the locations $\textbf{s}$ and $\textbf{s}'$ passes through all of the subregions in this set, and $\mathcal{W}_i(\textbf{s},\textbf{s}')$ are the location-dependent weights for the $i^{th}$ regional distance warping function. We define the weights as ${\mathcal{W}_i(\textbf{s},\textbf{s}')=\frac{\mathcal{P}(i,\textbf{s},\textbf{s}')}{\|\textbf{s}-\textbf{s}'\|}}$, where $\mathcal{P}(i,\textbf{s},\textbf{s}')$ is the length of the segment joining $\textbf{s}$ and $\textbf{s}'$ in $\mathcal{G}_i$.

Now, let us consider finer divisions of the domain where the $i^{th}$ subregion $\{\mathcal{G}_i,\;i=1,...,k\}$ is further divided into $m_i$ subregions $\mathcal{G}_{i1},...,\mathcal{G}_{im_i}$ such that \begin{equation}\label{eqs2}
  \bigcup_{j=1}^{m_i}\mathcal{G}_{ij}=\mathcal{G}_i,\;i=1,2,...,k\;\text{and}\;\;\bigcup_{i=1}^k\bigcup_{j=1}^{m_i}\mathcal{G}_{ij}=\mathcal{G}.
\end{equation}Here, we refer to the partitioning $\{\mathcal{G}_{ij},\;i=1,...,k,\;j=1,...,m_i\}$ as the \textbf{Guessed Partitioning}.
Then, the $\sum_{i=1}^km_i$ regional processes $\{X(\textbf{s}),\;\textbf{s}\in\mathcal{G}_{ij}\}$ $i=1,2,...,k,\;j=1,2,...,m_i$ are also stationary processes admitting the variograms $\gamma_{ij}(\|\cdot\|),\;i=1,2,...,k,\;\;j=1,2,...,m_i$, respectively. We align the $\sum_{i=1}^km_i$ regional variograms $\{\gamma_{ij}(\|\cdot\|),\;i=1,2...,k,\;j=1,2,...,m_i\}$ to obtain $\sum_{i=1}^km_i$ regional distance warping functions $\{\phi_{ij}(\|\cdot\|,\;i=1,2...,k,\;j=1,2,...,m_i\}$. The global distance function for the \textbf{Guessed Partitioning} is given by
\begin{equation} \label{eqs:3}\phi(\textbf{s},\textbf{s}')=\sum_{\mathcal{G}_{ij}\in \mathcal{L}(\textbf{s},\textbf{s}')}\mathcal{W}_{ij}(\textbf{s},\textbf{s}')\phi_{ij}(\|\textbf{s}-\textbf{s}'\|),\: \forall\: (\textbf{s},\textbf{s}') \in \mathcal{G}\times\mathcal{G},\end{equation}
where $\mathcal{L}(\textbf{s},\textbf{s}')$ is the set of $\textbf{guessed}$ subregions $\mathcal{G}_{ij}$ such that the line segment joining the locations $\textbf{s}$ and $\textbf{s}'$ passes through all of the subregions in this set, and $\mathcal{W}_{ij}(\textbf{s},\textbf{s}')$ are the location-dependent weights for the $(ij)^{th}$ regional distance warping function. We define the weights as ${\mathcal{W}_{ij}(\textbf{s},\textbf{s}')=\frac{\mathcal{P}(ij,\textbf{s},\textbf{s}')}{\|\textbf{s}-\textbf{s}'\|}}$, where $\mathcal{P}(ij,\textbf{s},\textbf{s}')$ is the length of the line segment joining $\textbf{s}$ and $\textbf{s}'$ that lies in subregion $\mathcal{G}_{ij}$.

\begin{figure}[t]
\begin{center}
\subfigure[]{\label{fig:1as}\includegraphics[scale=0.32]{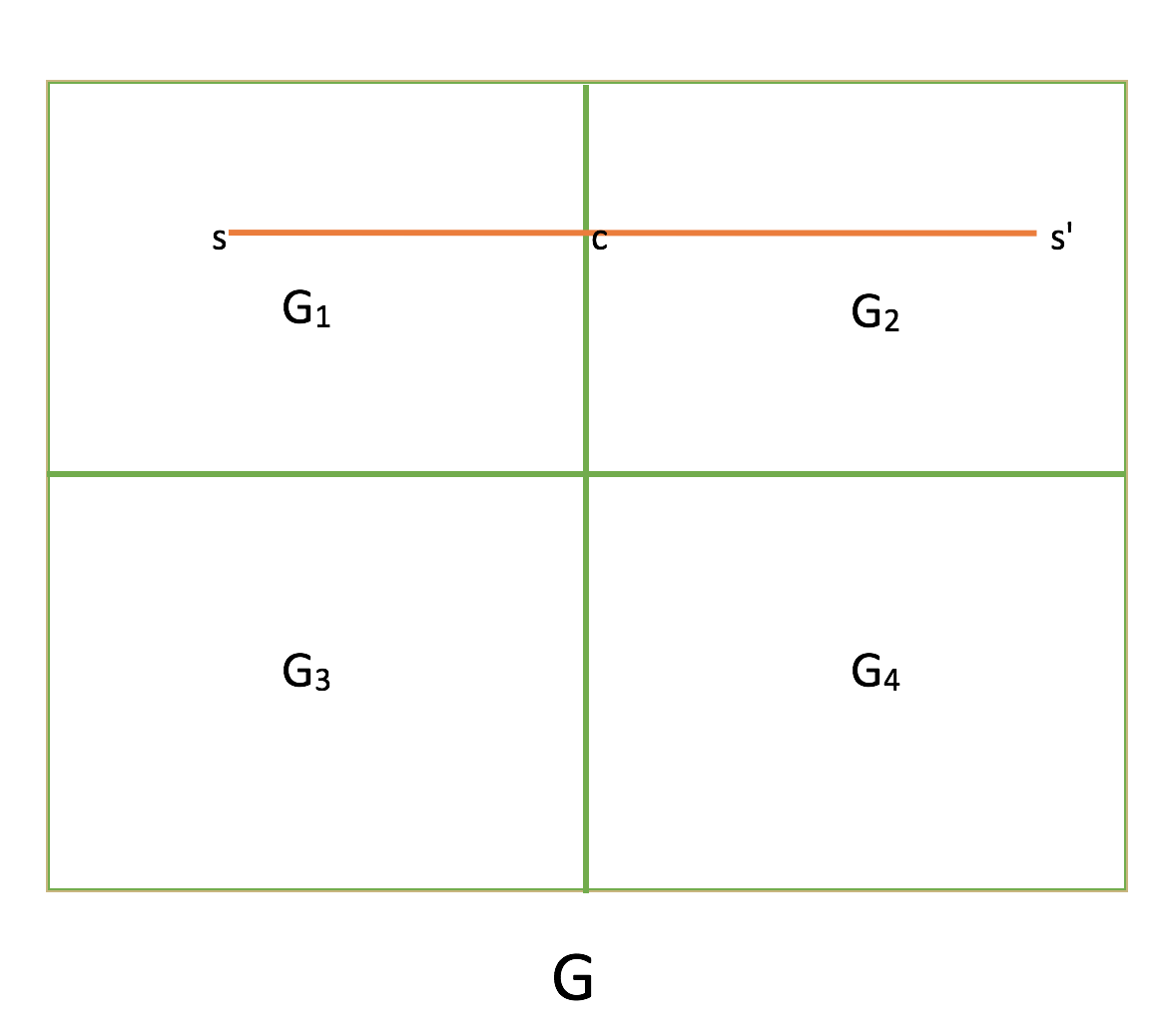}}
\subfigure[]{\label{fig:1bs}\includegraphics[scale=0.32]{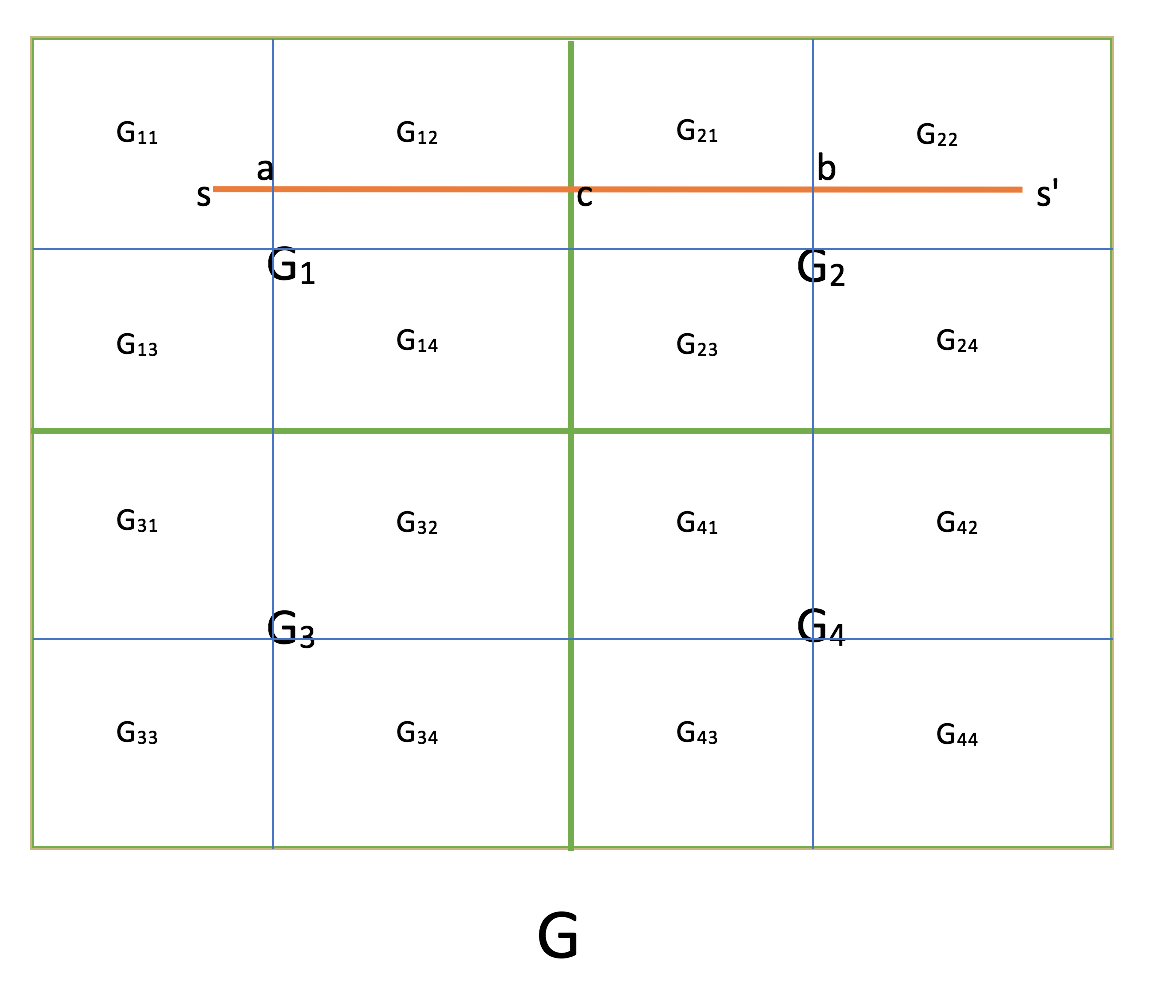}}
    \caption{(a) True Partitioning. (b) Guessed Partitioning.}
    \end{center}
\end{figure}

Note that $\gamma_{ij}(\|\cdot\|)=\gamma_i(\|\cdot\|),\;\forall\; j=1,...,m_i,\;i=1,2...,k$ because the sub-processes $\{X(\textbf{s}),\;\textbf{s}\in \mathcal{G}_{ij}\}$ will have the same spatial dependence structure as that of the parent stationary process $\{X(\textbf{s}),\;\textbf{s}\in \mathcal{G}_{i}\}$, and consequently, we will obtain many identical regional distance warping functions such that $\phi_{ij}(\|\cdot\|)=\phi_i(\|\cdot\|),\;\forall\; j=1,...,m_i,\;i=1,2...,k$. Therefore, because $\sum_{j=1}^{m_i}\mathcal{P}(ij,\textbf{s},\textbf{s'})=\mathcal{P}(i,\textbf{s},\textbf{s'}),\;\forall\; i=1,...,k$, Equation \ref{eqs:3} becomes \[ \phi(\textbf{s},\textbf{s}')=\sum_{\mathcal{G}_{i}\in \mathcal{L}(\textbf{s},\textbf{s}')}\sum_{j=1}^{m_i}\mathcal{W}_{ij}(\textbf{s},\textbf{s}')\phi_{i}(\|\textbf{s}-\textbf{s}'\|),\: \forall\: (\textbf{s},\textbf{s}') \in \mathcal{G}\times\mathcal{G},\]
\[=\sum_{\mathcal{G}_{i}\in \mathcal{L}(\textbf{s},\textbf{s}')}\sum_{j=1}^{m_i}\frac{\mathcal{P}(ij,\textbf{s},\textbf{s'})}{\|\textbf{s}-\textbf{s}'\|}\phi_{i}(\|\textbf{s}-\textbf{s}'\|),\: \forall\: (\textbf{s},\textbf{s}') \in \mathcal{G}\times\mathcal{G}\]
\[=\sum_{\mathcal{G}_{i}\in \mathcal{L}(\textbf{s},\textbf{s}')}\frac{\mathcal{P}(i,\textbf{s},\textbf{s'})}{\|\textbf{s}-\textbf{s}'\|}\phi_{i}(\|\textbf{s}-\textbf{s}'\|),\: \forall\: (\textbf{s},\textbf{s}') \in \mathcal{G}\times\mathcal{G}\]
\[=\sum_{\mathcal{G}_i\in \mathcal{L}(\textbf{s},\textbf{s}')}\mathcal{W}_i(\textbf{s},\textbf{s}')\phi_i(\|\textbf{s}-\textbf{s}'\|),\: \forall\: (\textbf{s},\textbf{s}') \in \mathcal{G}\times\mathcal{G},\]
leading to the same global distance function as given in Equation \ref{eqs:1}. This indicates robustness of our approach under finer subdivisions of the domain. Generally, the true partitioning $\{\mathcal{G}_1,...,\mathcal{G}_k\}$ is unknown. However, depending on the size of the data, we can divide the entire domain into very fine guessed subregions $\{\mathcal{G}_{ij},\;i=1,...,k,\;j=1,...,m_i\}$ such that Equation \ref{eqs2} approximately holds true.

The above result is further illustrated using a simple pictorial example. In Figure \ref{fig:1as}, the true partitioning is given by $\mathcal{G}=\bigcup_{i=1}^4\mathcal{G}_i$, $k=4$. We let $\textbf{s}\in\mathcal{G}_1$ and $\textbf{s'}\in \mathcal{G}_2$. Then, $\mathcal{L}(\textbf{s},\textbf{s'})=\{\mathcal{G}_1,\mathcal{G}_2\}$, $\mathcal{P}(1,\textbf{s},\textbf{s'})={\|\textbf{s}-\textbf{c}\|}$ and $\mathcal{P}(2,\textbf{s},\textbf{s'})={\|\textbf{c}-\textbf{s'}\|}$, resulting in weights $\mathcal{W}_1(\textbf{s},\textbf{s'})=\frac{\|\textbf{s}-\textbf{c}\|}{\|\textbf{s}-\textbf{s'}\|}$ and $\mathcal{W}_2(\textbf{s},\textbf{s'})=\frac{\|\textbf{c}-\textbf{s'}\|}{\|\textbf{s}-\textbf{s'}\|}$. This gives the global distance function $\phi(\textbf{s},\textbf{s'})=\frac{\|\textbf{s}-\textbf{c}\|\phi_1(\|\textbf{s}-\textbf{s'}\|)+\|\textbf{c}-\textbf{s'}\|\phi_2(\|\textbf{s}-\textbf{s'}\|)}{\|\textbf{s}-\textbf{s'}\|}$. Next, we compute the global distance function for a finer guessed partitioning of the domain as shown in Figure \ref{fig:1bs}: $\mathcal{G}=\bigcup_{i=1}^4\bigcup_{j=1}^{m_i}\mathcal{G}_{ij}$, $k=4,\ m_i=4,\; \forall\; i$. We let $\textbf{s}\in\mathcal{G}_{11}$ and $\textbf{s'}\in \mathcal{G}_{22}$. Then, $\mathcal{L}(\textbf{s},\textbf{s'})=\{\mathcal{G}_{11},\mathcal{G}_{12},\mathcal{G}_{21},\mathcal{G}_{22}\}$, $\mathcal{P}(11,\textbf{s},\textbf{s'})={\|\textbf{s}-\textbf{a}\|}$, $\mathcal{P}(12,\textbf{s},\textbf{s'})={\|\textbf{a}-\textbf{c}\|}$, $\mathcal{P}(21,\textbf{s},\textbf{s'})={\|\textbf{c}-\textbf{b}\|}$ and $\mathcal{P}(22,\textbf{s},\textbf{s'})={\|\textbf{b}-\textbf{s'}\|}$, resulting in weights $\mathcal{W}_{11}(\textbf{s},\textbf{s'})=\frac{\|\textbf{s}-\textbf{a}\|}{\|\textbf{s}-\textbf{s'}\|}$, $\mathcal{W}_{12}(\textbf{s},\textbf{s'})=\frac{\|\textbf{a}-\textbf{c}\|}{\|\textbf{s}-\textbf{s'}\|}$, $\mathcal{W}_{21}(\textbf{s},\textbf{s'})=\frac{\|\textbf{c}-\textbf{b}\|}{\|\textbf{s}-\textbf{s'}\|}$  and $\mathcal{W}_{22}(\textbf{s},\textbf{s'})=\frac{\|\textbf{b}-\textbf{s'}\|}{\|\textbf{s}-\textbf{s'}\|}$. This gives the global distance function $\phi(\textbf{s},\textbf{s'})=\frac{\|\textbf{s}-\textbf{a}\|\phi_{11}(\|\textbf{s}-\textbf{s'}\|)+\|\textbf{a}-\textbf{c}\|\phi_{12}(\|\textbf{s}-\textbf{s'}\|+\|\textbf{c}-\textbf{b}\|\phi_{21}(\|\textbf{s}-\textbf{s'}\|)+\|\textbf{b}-\textbf{s'}\|\phi_{22}(\|\textbf{s}-\textbf{s'}\|)}{\|\textbf{s}-\textbf{s'}\|}$. Applying $\phi_{ij}(\|\cdot\|)=\phi_i(\|\cdot\|), \forall\; i=1,...,k,\;j=1,...,m_i$, we obtain \begin{eqnarray}\nonumber
\phi(\textbf{s},\textbf{s'})=\frac{1}{\|\textbf{s}-\textbf{s'}\|}\times\{\|\textbf{s}-\textbf{a}\|\phi_{1}(\|\textbf{s}-\textbf{s'}\|)+\|\textbf{a}-\textbf{c}\|\phi_{1}(\|\textbf{s}-\textbf{s'}\|)\\\nonumber+\|\textbf{c}-\textbf{b}\|\phi_{2}(\|\textbf{s}-\textbf{s'}\|)+\|\textbf{b}-\textbf{s'}\|\phi_{2}(\|\textbf{s}-\textbf{s'}\|)\}\\\nonumber=\frac{1}{\|\textbf{s}-\textbf{s'}\|}\{\|\textbf{s}-\textbf{c}\|\phi_1(\|\textbf{s}-\textbf{s'}\|)+\|\textbf{c}-\textbf{s'}\|\phi_2(\|\textbf{s}-\textbf{s'}\|)\},
\end{eqnarray} which is the same as before.

However, in practice, we cannot estimate the variograms exactly even if the true processes share an identical spatial dependence. Therefore, while replacing $\gamma_{ij}(\|.\|),\;i=1,...,k, \;j=1,...,m_i$ with the estimated variograms $\hat{\gamma}_{ij}(\|.\|),\;i=1,...,k, \;j=1,...,m_i$, the exact equality $\gamma_{ij}(\|\cdot\|)=\gamma_i(\|\cdot\|),\;\forall\; j=1,...,m_i,\;i=1,2...,k$ becomes $\hat{\gamma}_{ij}(\|\cdot\|)\approx\hat{\gamma}_i(\|\cdot\|),\;\forall\; j=1,...,m_i,\;i=1,2...,k$. Consequently, the estimated regional distance warping functions follow $\hat{\phi}_{ij}(\|\cdot\|)\approx\hat{\phi}_i(\|\cdot\|),\;\forall\; j=1,...,m_i,\;i=1,2...,k$ instead of $\phi_{ij}(\|\cdot\|)=\phi_i(\|\cdot\|),\;\forall\; j=1,...,m_i,\;i=1,2...,k$. This approximation leads to slightly different deformations for the true partitioning and the guessed partitioning. Thus, accuracy of the estimation of regional variograms affects the robustness of our method under finer subdivisions of the domain.

There is a clear trade-off between the number of subregions and the amount of available data for estimation within each subregion. The guessed partitioning can be forced to consist of very small subregions such that Equation \ref{eqs2} approximately holds true. However, very small subregions lead to less data points per subregion and poor estimation of regional variograms. This is consequently reflected in the estimation of the regional distance warping functions and the final deformation. Therefore, the number of subregions must be driven by the extensiveness of available spatial data. This point is further demonstrated in the following simulation study.

\subsection{Simulation: Variogram Estimation Under Finer Subdivisions of the Domain}
We consider a zero-mean Gaussian process $X$ over a domain $\mathcal{G}=[0,8]^2$, with the following nonstationary Mat{\'e}rn covariance function \citep{Paciorek:2006aa}:
\begin{equation}\label{eqs:7}\text{C}^{NS}(\textbf{s}_i,\textbf{s}_j:\tilde{\eta})=\sigma(\textbf{s}_i)\sigma(\textbf{s}_j)\frac{|\Sigma(\textbf{s}_i)|^{1/4}|\Sigma(\textbf{s}_j)|^{1/4}}{2^{\nu-1}\Gamma(\nu)}\Big|\frac{\Sigma(\textbf{s}_i)+\Sigma(\textbf{s}_j)}{2}\Big|^{-1/2}(2\sqrt{\nu Q_{ij}})^{\nu}K_{\nu}(2\sqrt{\nu Q_{ij}}),\end{equation}where $\tilde{\eta}$ is the vector of parameters, $\sigma(\textbf{s})$ is a location-dependent standard deviation, $\nu$ is the smoothness parameter, $Q_{ij}$ is the Mahalanobis distance between two locations $\textbf{s}_i=(x_i,y_i)$ and $\textbf{s}_j=(x_j,y_j)$, $K_\nu$ is a modified Bessel function of the second order, and $\Sigma(\textbf{s})$ is a spatially varying kernel matrix that supervises the range and direction of spatial dependence.

\begin{figure}[h]
\centering     %%% not \center
\subfigure[]{\label{fig:3as}\includegraphics[width=60mm]{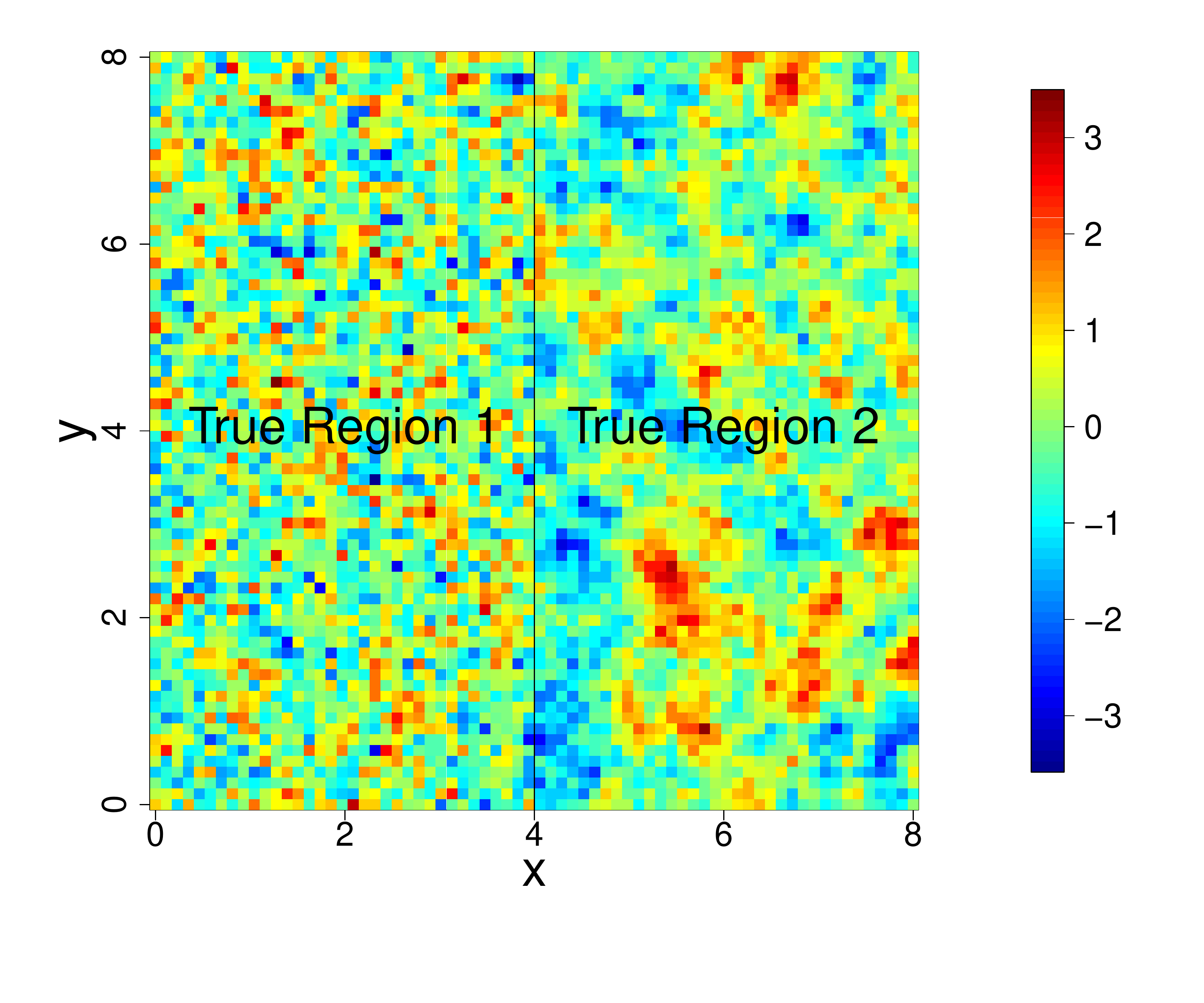}}
\subfigure[]{\label{fig:3bs}\includegraphics[width=60mm]{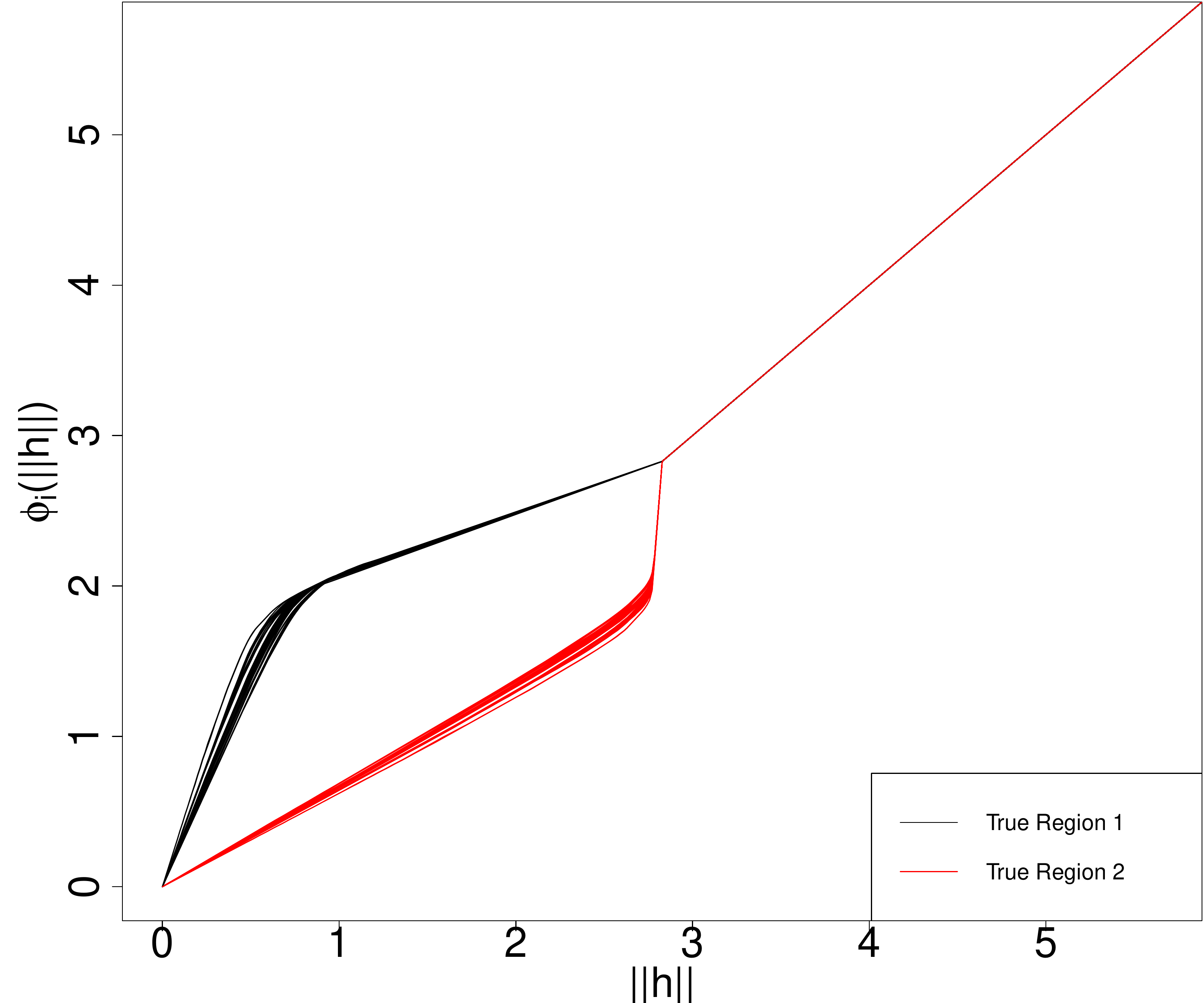}}

\caption{(a) One realization of the simulated process, with a solid black line depicting the true partitioning. (b) Estimated regional distance warping functions for the two subregions.}
\label{fig:3s}
\end{figure}

To obtain realizations from a regionally stationary process with nonstationarity only in the spatial range, we simulate 50 realizations of $X$ on a regular grid of $70\times 70$ points on $\mathcal{G}$, with $\nu=0.5$, $\{\sigma(\textbf{s})=1,\forall \textbf{s}\in \mathcal{G} \}$ and \[\Sigma(\textbf{s})_{2\times 2}=\begin{cases}\text{diag}(0.1^2,0.1^2), \text{\hspace{0.6cm} if } x\leq 4\\
\text{diag}(0.45^2,0.45^2), \text{\hspace{0.2cm} if } x>4.
\end{cases}\]For this simulation, the true partitioning consists of two disjoint regions, i.e.,  ``True Region 1"=$[0,4]\times[0,8]$ and ``True Region 2"=$(4,8]\times[0,8]$. We consider three cases for domain partitioning, starting with the true partitioning and progressing to finer subdivisions. For the true partitioning case we estimate the isotropic Mat{\'e}rn variogram for ``True Region 1" and ``True Region 2" for each of the 50 simulated realizations using Maximum Likelihood Estimation (MLE). During estimation, we fix the smoothness parameter $\nu=0.5$ to avoid identifiability issues \citep{zhang}. We then estimate the regional distance warping functions for ``True Region 1" and ``True Region 2" by aligning their respective estimated regional variograms. The estimated regional variograms become numerically constant for distances greater than $\sqrt{8}$, and therefore, we set the value $\|\textbf{h}_{t}\|=\sqrt{8}$ and assign identity regional distance warping functions for $\|\textbf{h}\|>\sqrt{8}$. Figure \ref{fig:3as} shows one realization of the simulated process, with a solid black line indicating the true partitioning. The estimated regional distance warping functions are shown in Figure \ref{fig:3bs}. The 50 pairs of regional distance warping functions are nearly identical for each run indicating that enough data is available per subregion to efficiently estimate the regional variograms.

\begin{figure}[p]
\centering     %%% not \center
\subfigure[]{\label{fig:4as}\includegraphics[width=60mm]{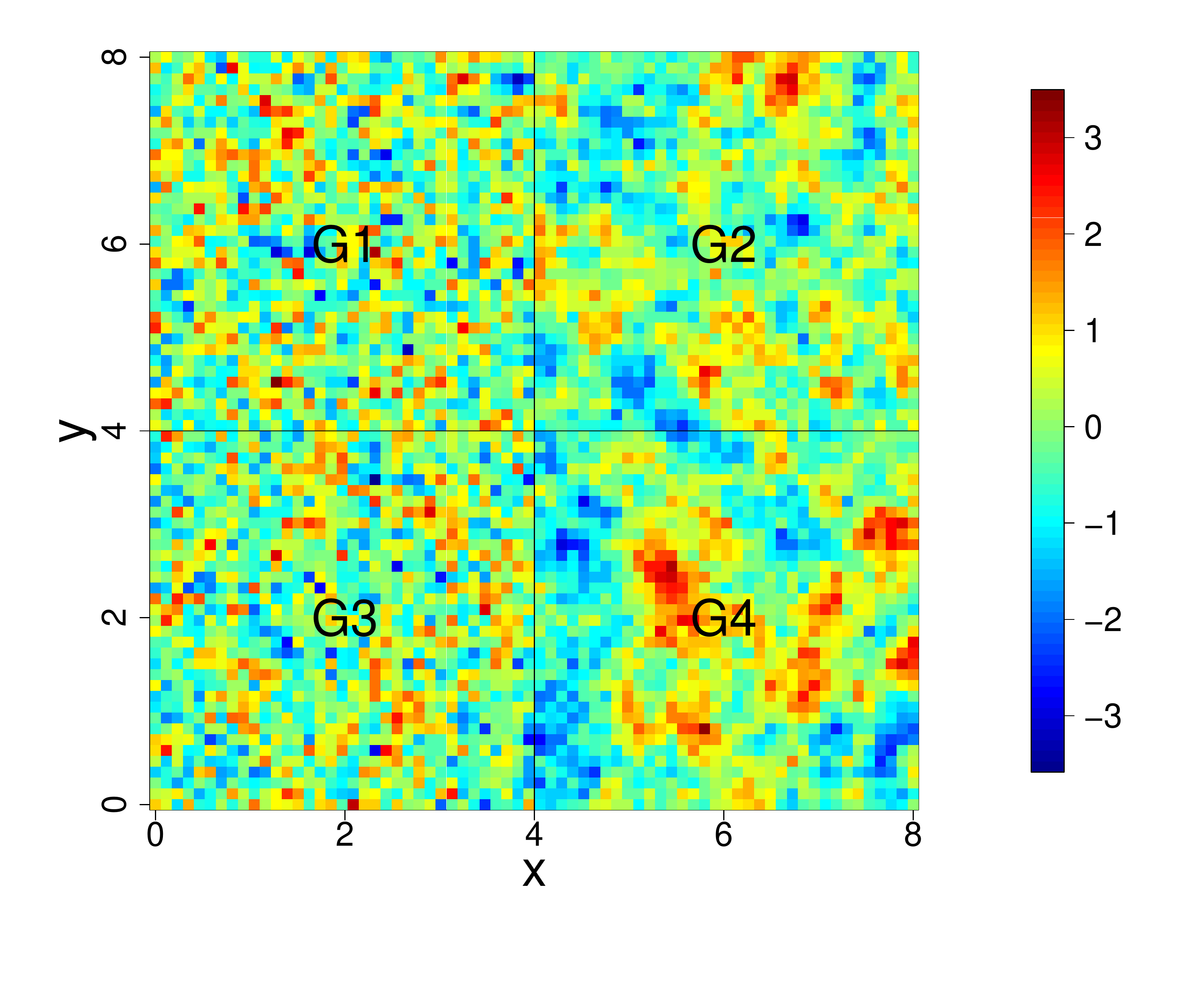}}
\subfigure[]{\label{fig:4bs}\includegraphics[width=60mm]{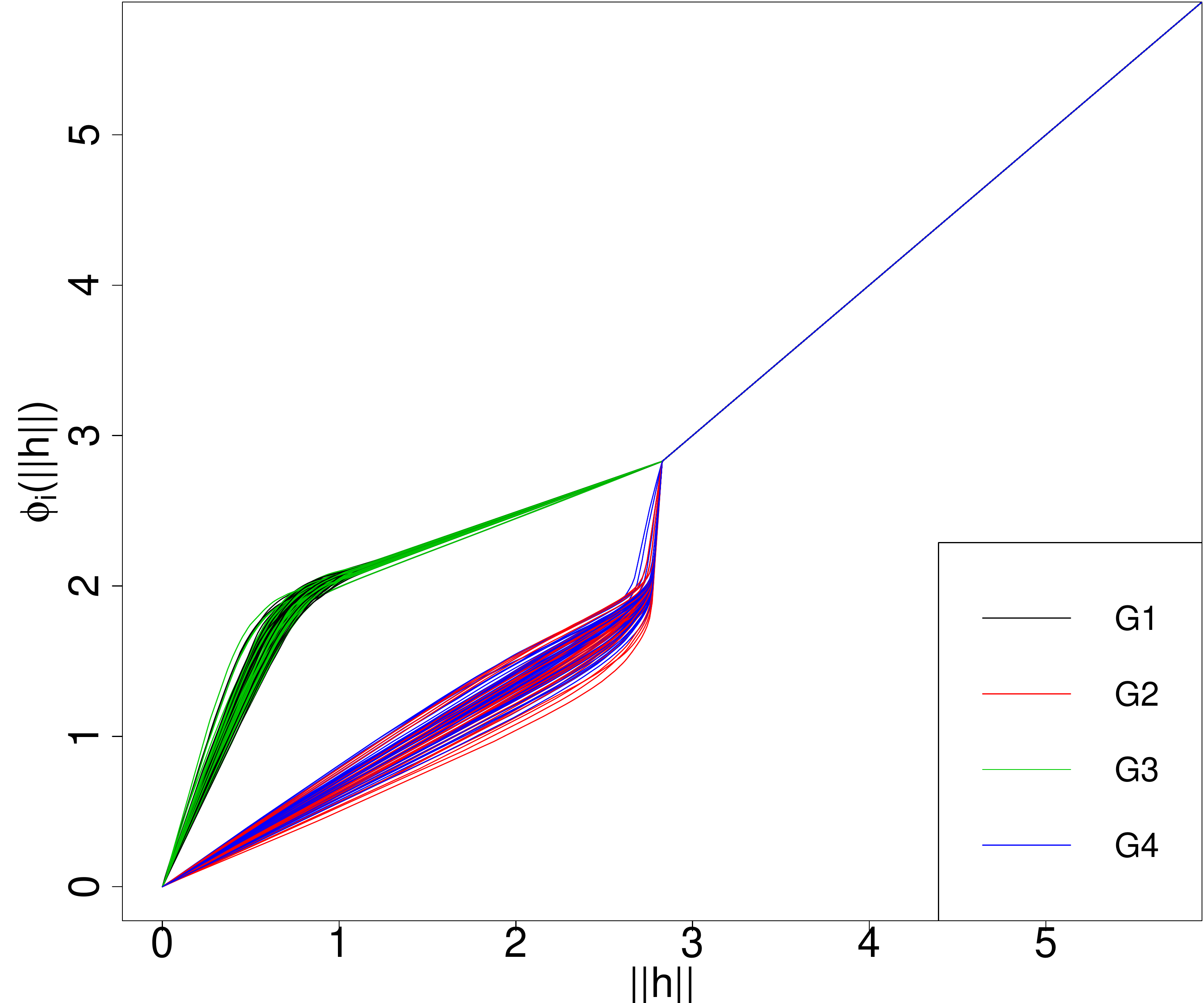}}\\
\subfigure[]{\label{fig:4cs}\includegraphics[width=60mm]{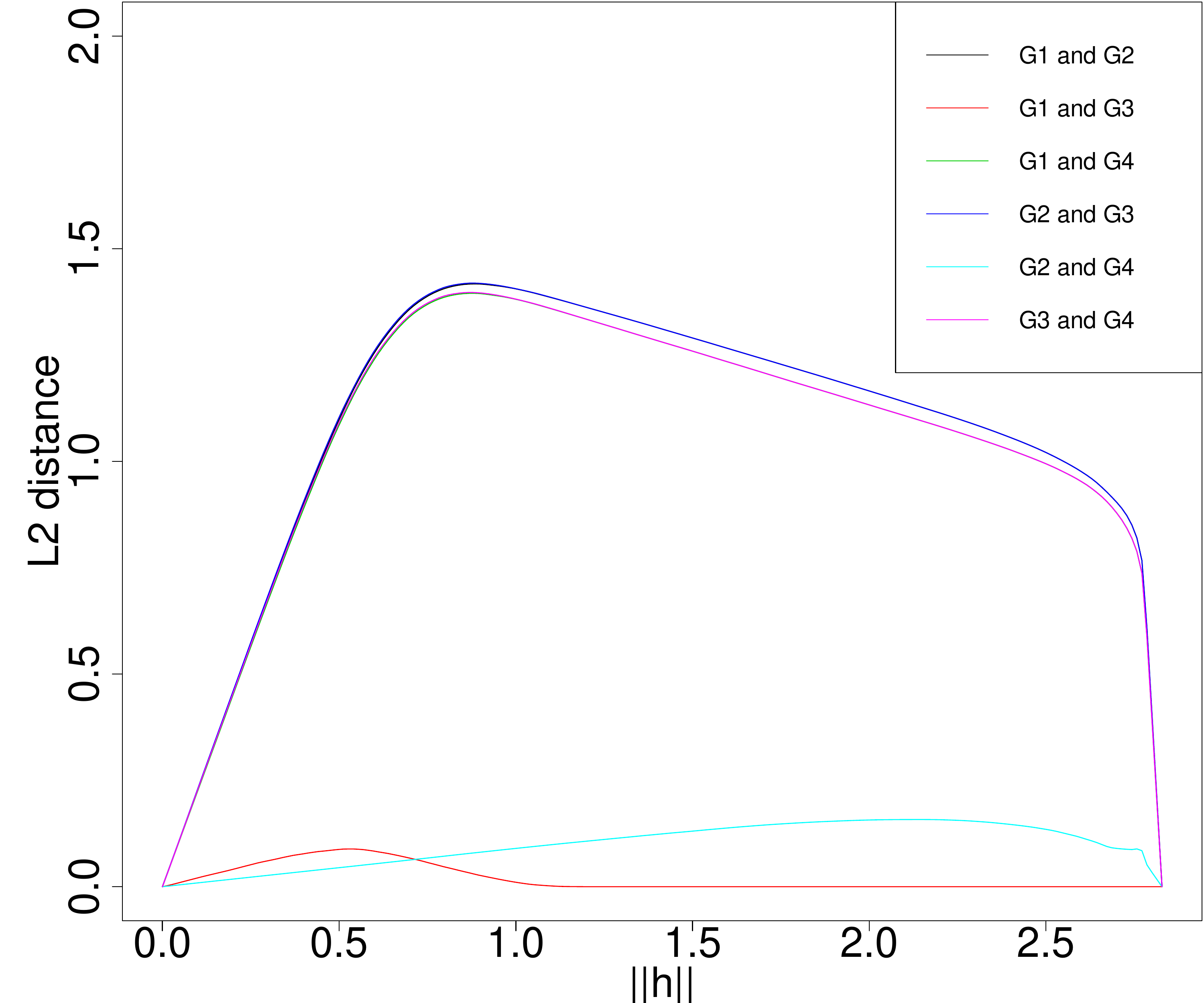}}
\subfigure[]{\label{fig:4ds}\includegraphics[width=60mm]{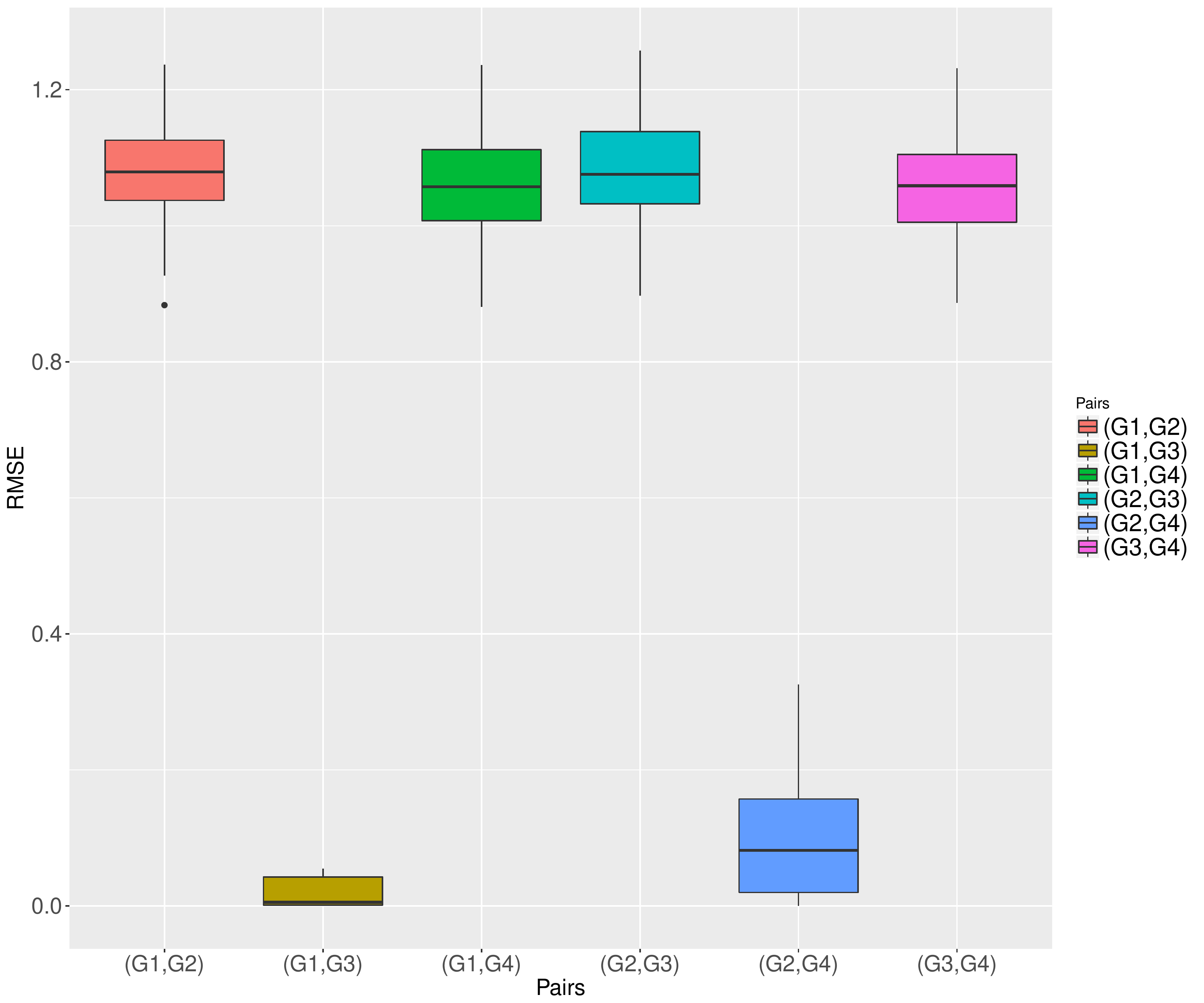}}
\caption{(a) One realization of the simulated process, with solid black lines depicting the partitioning. (b) Estimated regional distance warping functions. (c) Pointwise $\mathbb{L}^2$ distances between regional distance warping functions. (d) Boxplots of pairwise RMSE values.}
\label{fig:4s}
\end{figure}
For the second case, we consider a finer subdivision of the domain into four equal subregions $\{\mathcal{G}_1,\mathcal{G}_2,\mathcal{G}_3,\mathcal{G}_4\}$. Following the same estimation procedure as described for the true partitioning, we estimate the isotropic Mat{\'e}rn regional variograms and regional distance warping functions for the four subregions. Figure \ref{fig:4as} shows one realization of the simulated process, with solid black lines depicting the partitioning. The estimated regional distance warping functions are shown in Figure \ref{fig:4bs} for each of the 50 runs; they show very similar behavior within subregions $(\mathcal{G}_1,\mathcal{G}_3)$ and $(\mathcal{G}_4,\mathcal{G}_4)$, which is expected because of the common true underlying variograms for the pair of sub-processes ($\{X(\textbf{s}),s\in\mathcal{G}_1\},\{X(\textbf{s}),s\in\mathcal{G}_3\})$ and ($\{X(\textbf{s}),s\in\mathcal{G}_2\},\{X(\textbf{s}),s\in\mathcal{G}_4\})$. The very similar pairs of regional distance warping functions in Figure \ref{fig:4bs}, as well as the closeness of Figure \ref{fig:3bs} and Figure \ref{fig:4bs} in terms of their shapes, demonstrate that the finer partitioning and the true partitioning lead to nearly identical deformations.

Figure \ref{fig:4cs} shows the pointwise $\mathbb{L}^2$ distance between the regional distance warping functions for every pair of subregions in the finer partitioning, averaged over 50 runs and evaluated up to $\|\textbf{h}_{t}\|$. Figure \ref{fig:4ds} shows the boxplot for the root mean squared error (RMSE) between every pair of the regional distance warping functions evaluated up to $\|\textbf{h}_{t}\|$. The $\mathbb{L}^2$ distances and the boxplots of RMSE values for the pairs $(\mathcal{G}_1,\mathcal{G}_3)$ and $(\mathcal{G}_2,\mathcal{G}_4)$ are concentrated near 0, which provides quantitative validation of the pairwise proximity of the corresponding regional distance warping functions.

\begin{figure}[!t]
\centering     %%% not \center
\subfigure[]{\label{fig:5as}\includegraphics[width=60mm]{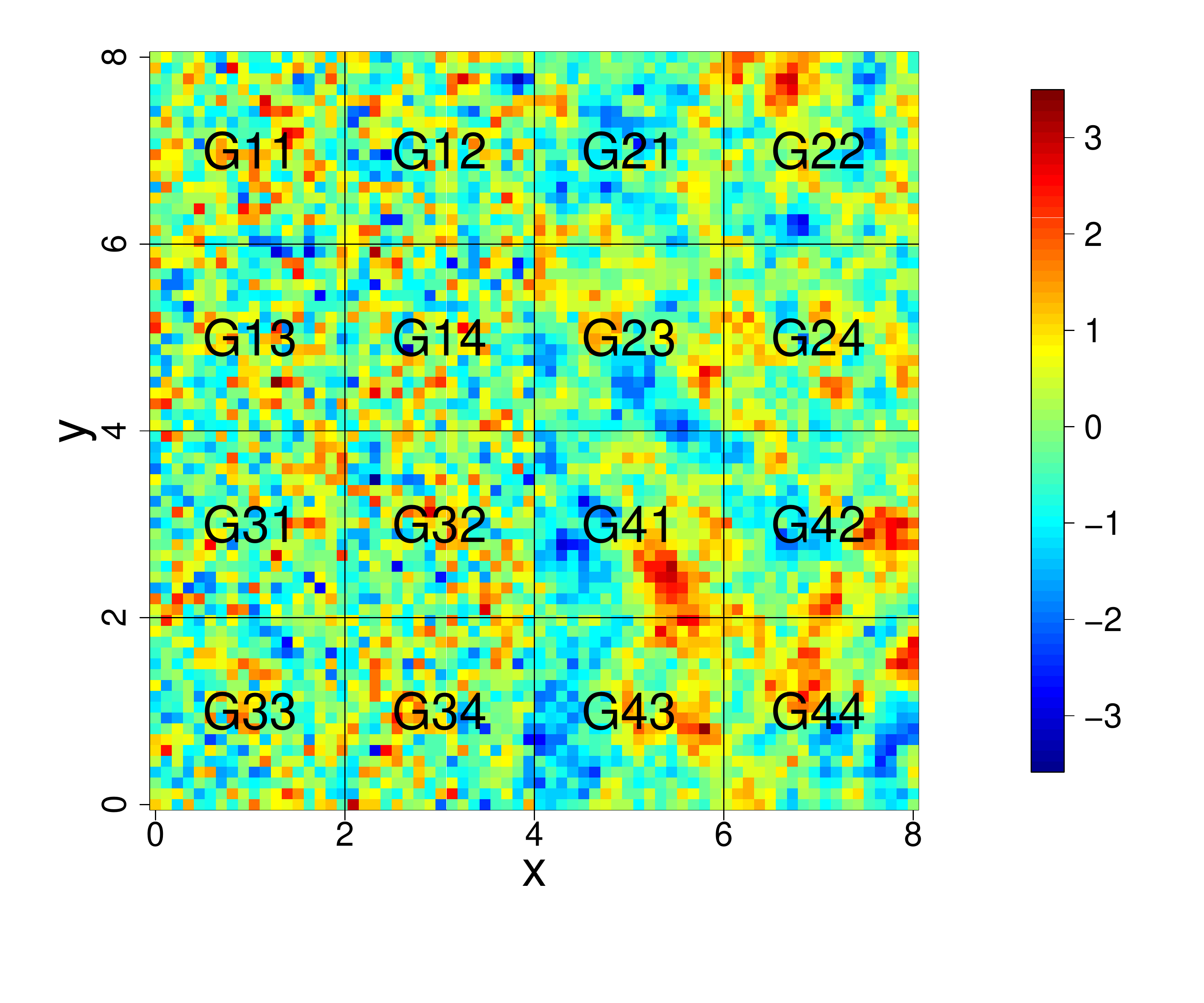}}
\subfigure[]{\label{fig:5bs}\includegraphics[width=60mm]{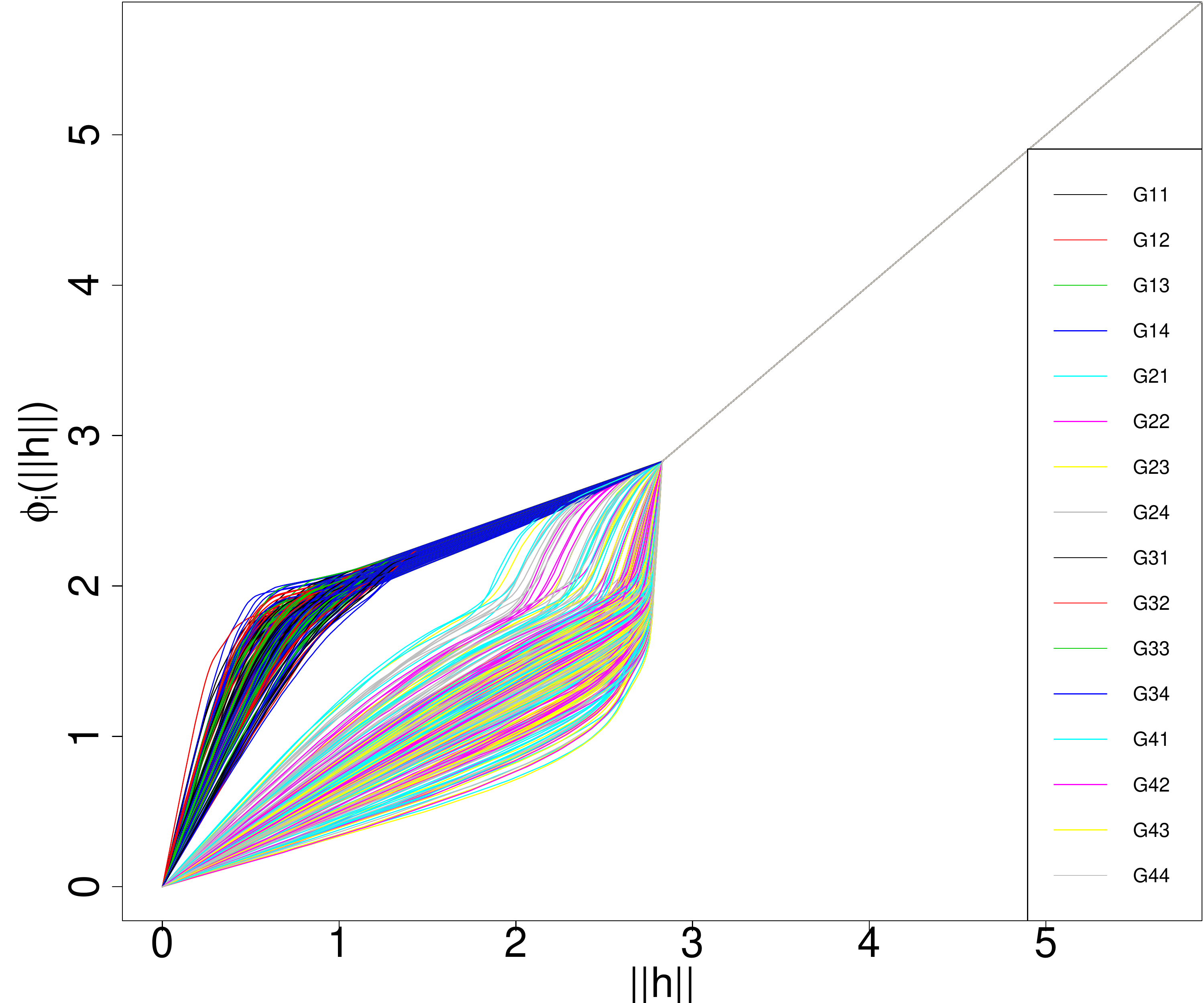}}
\caption{(a) One realization of the simulated process, with solid black lines depicting partitioning. (b) Estimated regional distance warping functions.}
\label{fig:5s}
\end{figure}
For the third case, we consider an even finer subdivision of the domain into sixteen subregions $\{\mathcal{G}_{ij},\;i,j=1,...,4.\}$. Figure \ref{fig:5as} shows one realization from the simulated process, with solid black lines depicting this partitioning. Similarly to the other two cases considered previously, we estimate the regional distance warping functions and show them in Figure \ref{fig:5bs}. The estimated regional distance warping functions show more variability in this case due to fewer data points in each subregion, resulting in an inefficient estimation of the regional variograms. The sub-processes $\{X(\textbf{s}),\textbf{s}\in\mathcal{G}_{ij},\;j=1,...,4\}$ have a common true underlying variogram for each $i=1,\dots,4$, but their poor estimation results in regional distance warping functions that differ from those displayed in Figure \ref{fig:3bs} and Figure \ref{fig:4bs}. Consequently, the estimated deformations in this case will look different than in the previous two cases. This shows the trade-off between the number of subregions chosen to partition the original domain and the accuracy of estimation of the true regional variograms.
\par

\section{Sensitivity Analysis: Based on the Simulation Study from the Main Manuscript}
\setcounter{equation}{0}

For identification of subregions, we can start with a small number of subregions such that Equation \ref{eqs2} approximately holds. However, there can be additional misspecification of the boundary of the two subregions. In this simulation, we consider sensitivity of our method to such a misspecification. We use the same simulated dataset as in Section 3 in the manuscript. Here, we estimate the deformed space in 2-D with two slightly misspecified subregions, and compare the result to the true partitioning one. The true partitioning in this case is $\{\mathcal{G}_1,\mathcal{G}_2\}$, where $\mathcal{G}_1=[0,1]\times[0,2]$ and $\mathcal{G}_2=(1,2]\times[0,2]$. We consider two cases: (1) $\mathcal{G}_1=[0,0.8]\times[0,2]$ and $\mathcal{G}_2=(0.8,2]\times[0,2]$; and (2) $\mathcal{G}_1=[0,1.2]\times[0,2]$ and $\mathcal{G}_2=(1.2,2]\times[0,2]$.
\begin{figure}[t]
\centering     %%% not \center
\subfigure[]{\label{fig:7sa}\includegraphics[width=44mm]{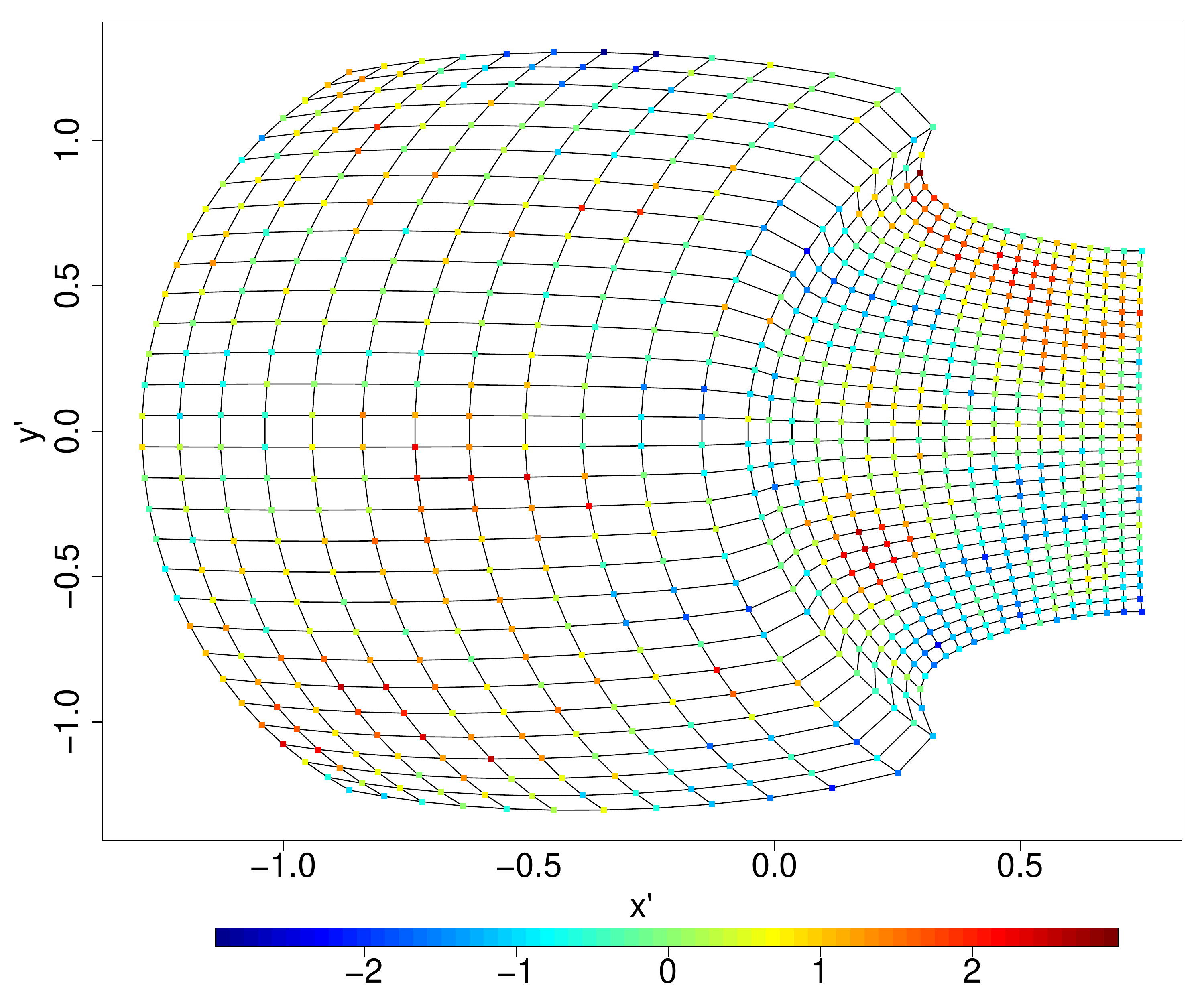}}
\subfigure[]{\label{fig:7sb}\includegraphics[width=44mm]{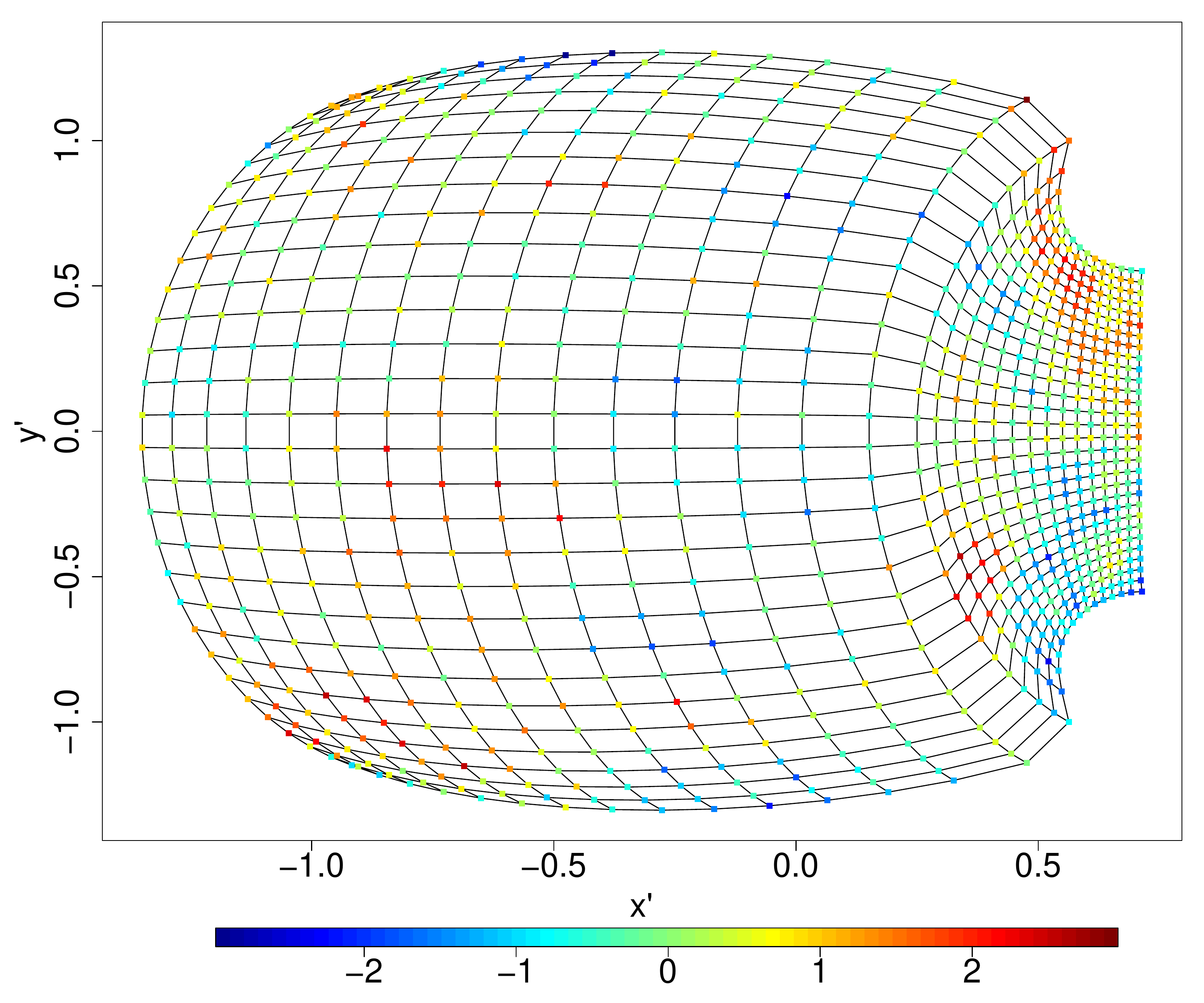}}
\subfigure[]{\label{fig:7sc}\includegraphics[width=44mm]{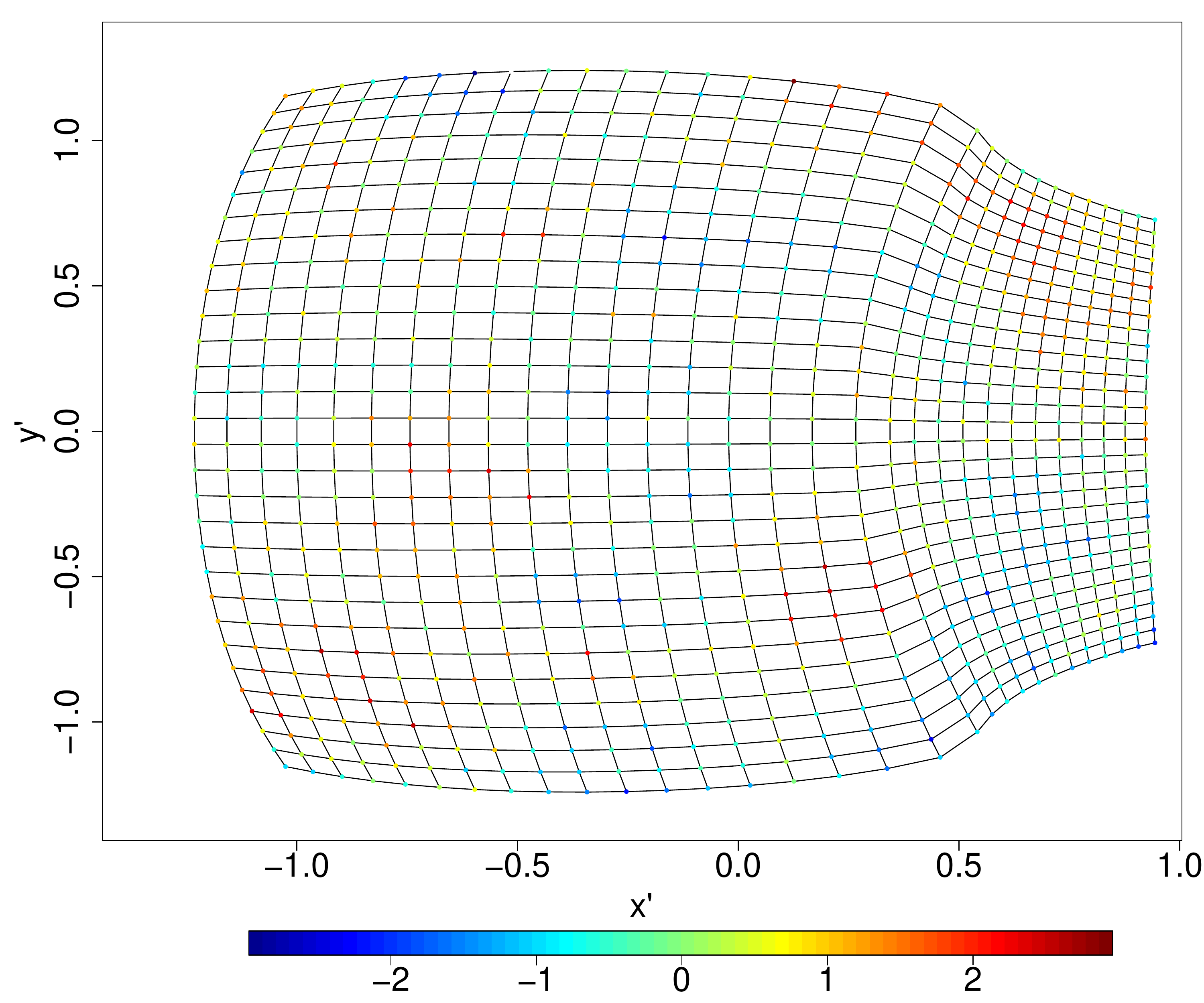}}
\caption{Estimated deformed space in 2-D with (a) misspecified partition (1), (b) true partitioning and (c)  misspecified partition (2).}
\label{fig:7s}
\end{figure}

The estimated deformed space in 2-D for all three cases (true partitioning, (1) and (2)) are shown in Figure \ref{fig:7s}. Figure \ref{fig:7sb} shows the estimated deformed space under the true partitioning. Figure \ref{fig:7sa} shows the estimated deformed space for (1) while Figure \ref{fig:7sc} shows the estimated deformed space for (2). The change point from the stretched subregion to the squeezed subregion is shifted slightly to the left for (1) and slightly to the right for (2) as compared to the true partitioning deformed space. However, for all three cases, the overall pattern of deformation is very similar with a stretched left subregion and a squeezed right subregion, allowing homogeneity in the spatial range throughout the entire domain.

\section{Additional Simulation Study: Second-order Stationary Process}\label{sub:stat}
\setcounter{equation}{0}
In this section, we present an additional simulation study based on a second-order stationary Gaussian process. 
For spatially non-varying parameters, Equation~\ref{eqs:7} is reduced to a stationary covariance function. We set smoothness $\nu=0.6$, standard deviation $\{\sigma(\textbf{s})=1,\forall \textbf{s}\in \mathcal{G}\}$ and kernel matrix to a constant $\{\Sigma(\textbf{s})_{2\times 2}=\text{diag}(0.1849,0.1849),\;\forall \textbf{s}\in \mathcal{G}\}$, to simulate a zero-mean second-order stationary Gaussian process $X$, at $30\times 30$ points on a regular grid, on the domain $\mathcal{G}=[0,2]^2$.

\begin{figure}[p]
\centering     %%% not \center
\subfigure[]{\label{fig:7a2s}\includegraphics[width=60mm]{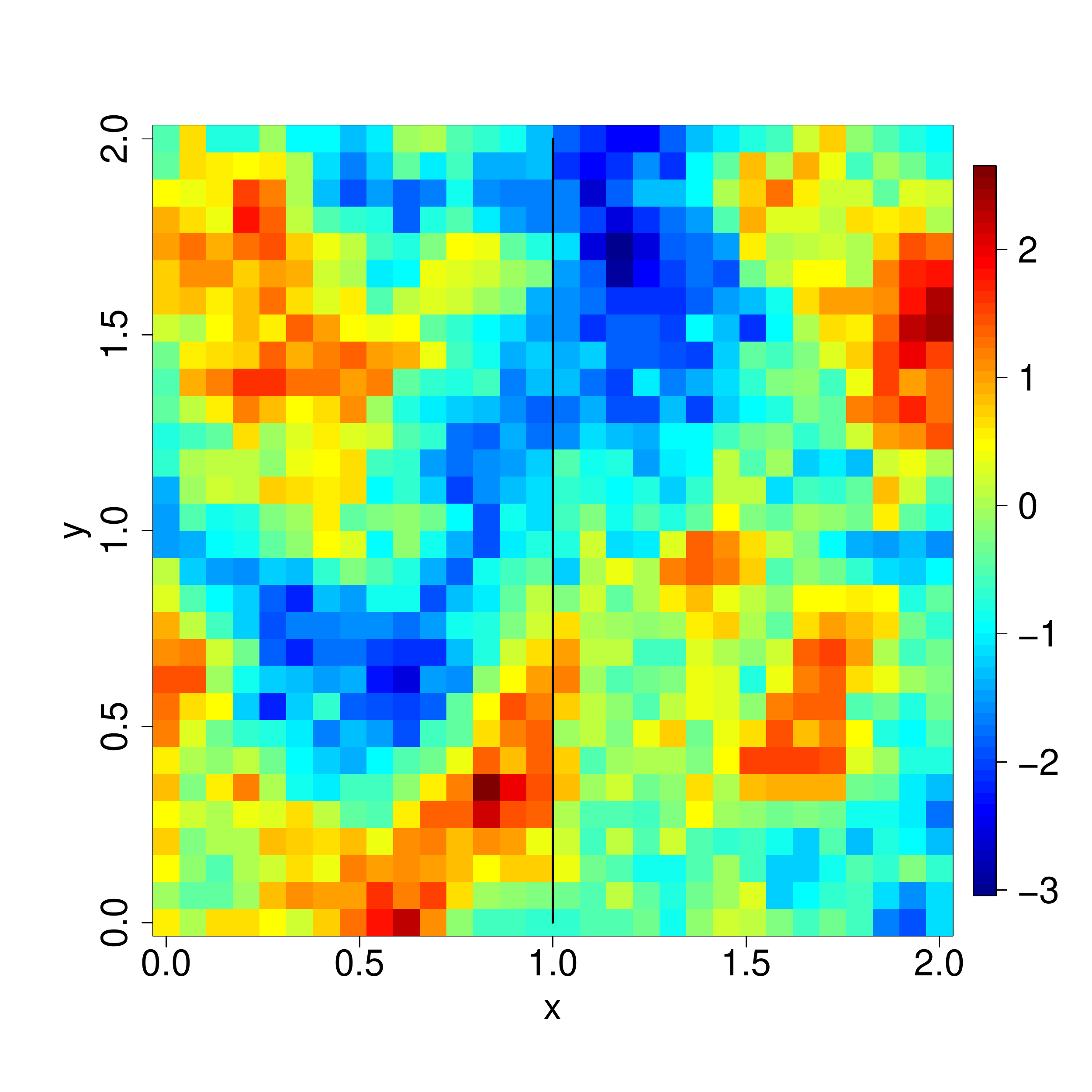}}
\subfigure[]{\label{fig:7b2s}\includegraphics[width=60mm]{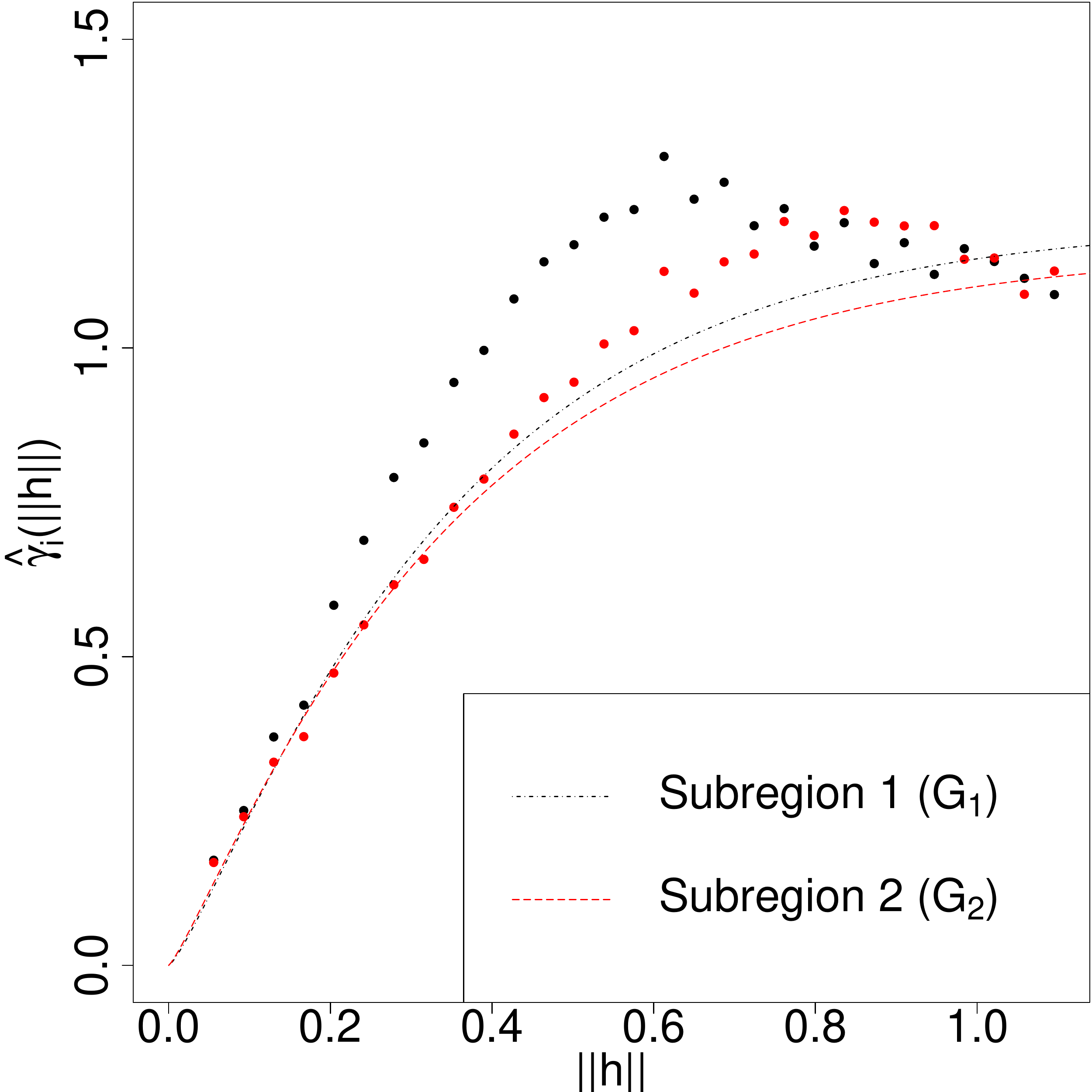}}
\subfigure[]{\label{fig:7c2s}\includegraphics[width=60mm]{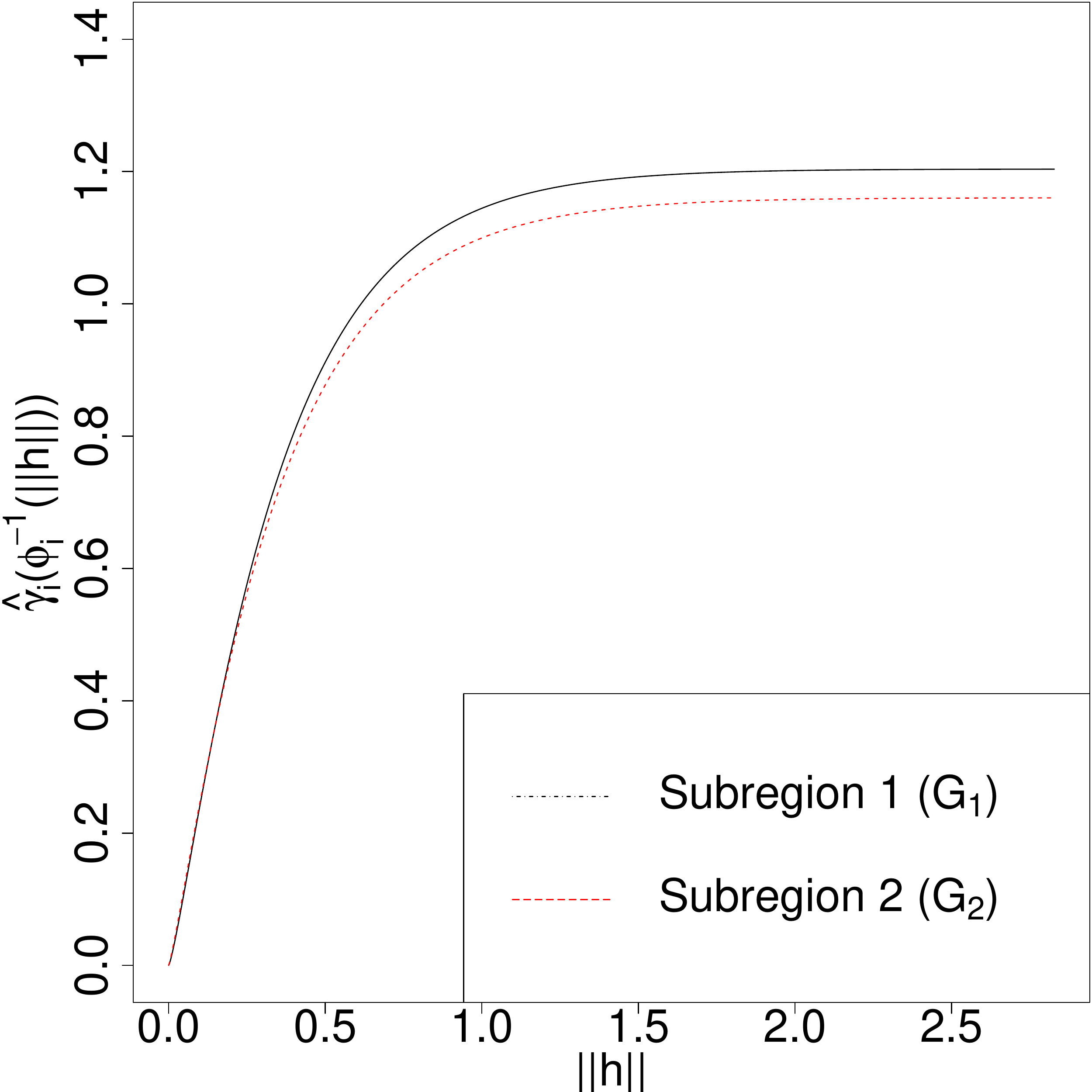}}
\subfigure[]{\label{fig:7d2s}\includegraphics[width=60mm]{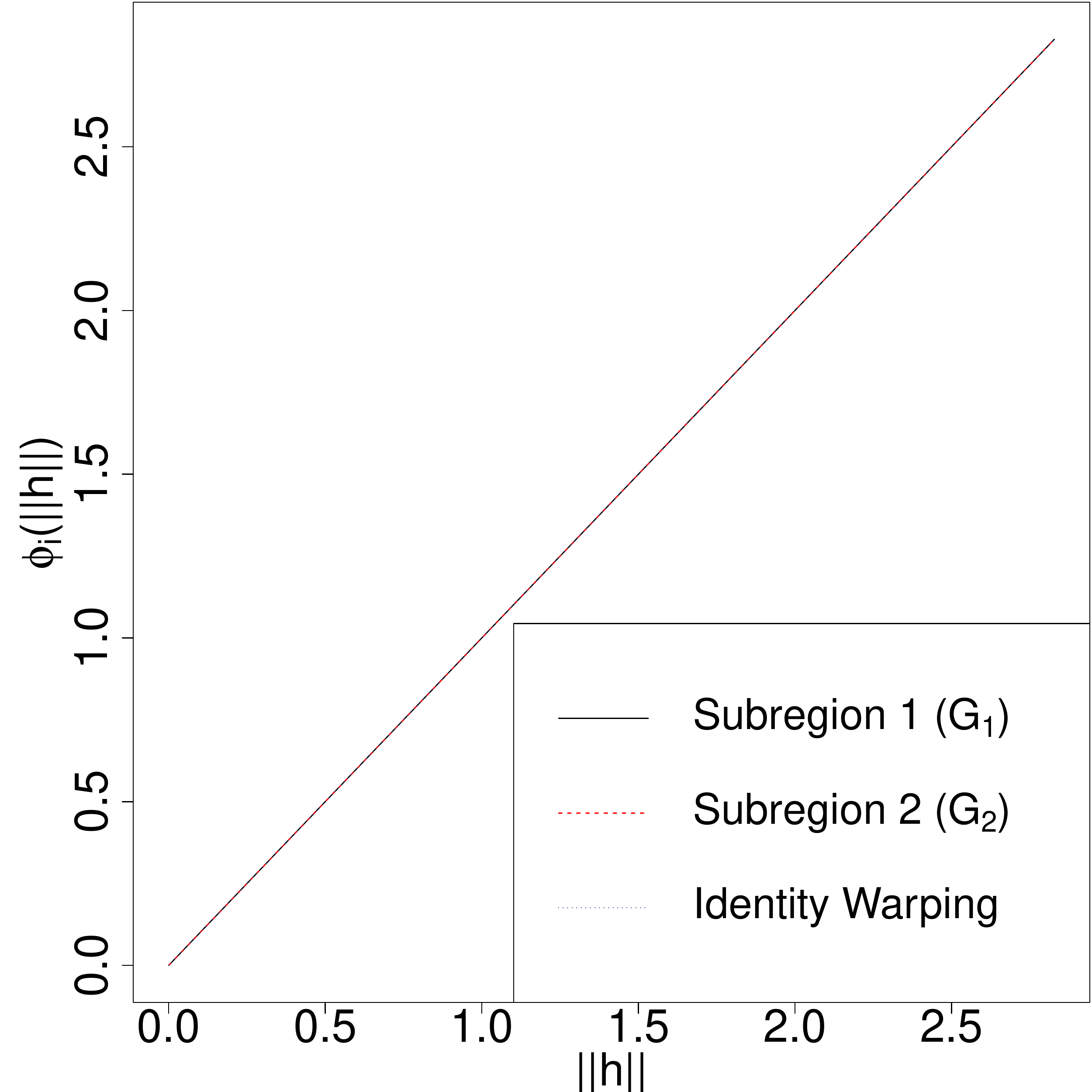}}
\caption{(a) Example of a realization of the zero mean second-order stationary Gaussian process with solid black line indicating the partitioning of geographic space. (b) Estimated regional variograms overlaid on the regional empirical variograms. (c) Registered variograms. (d) Regional distance warping functions. } 
\label{fig:72}
\end{figure} 

In this example, the dependence structure is already known to be homogeneous over the entire domain $\mathcal{G}=[0,2]^2$, and therefore, we can arbitrarily divide $\mathcal{G}$ into subregions to test whether the estimate of the deformed space provided by our method is identical to the geographic space. We proceed with the same partitioning as in Section 3 of the main manuscript, i.e., we divide $\mathcal{G}$ into two subregions $\mathcal{G}_1=[0,1]\times[0,2]$ and $\mathcal{G}_2=(1,2]\times[0,2]$, and fit the isotropic Mat{\'e}rn variogram models for each subregion.

\begin{figure}[t]
\centering     %%% not \center
\subfigure[]{\label{fig:81as}\includegraphics[width=60mm]{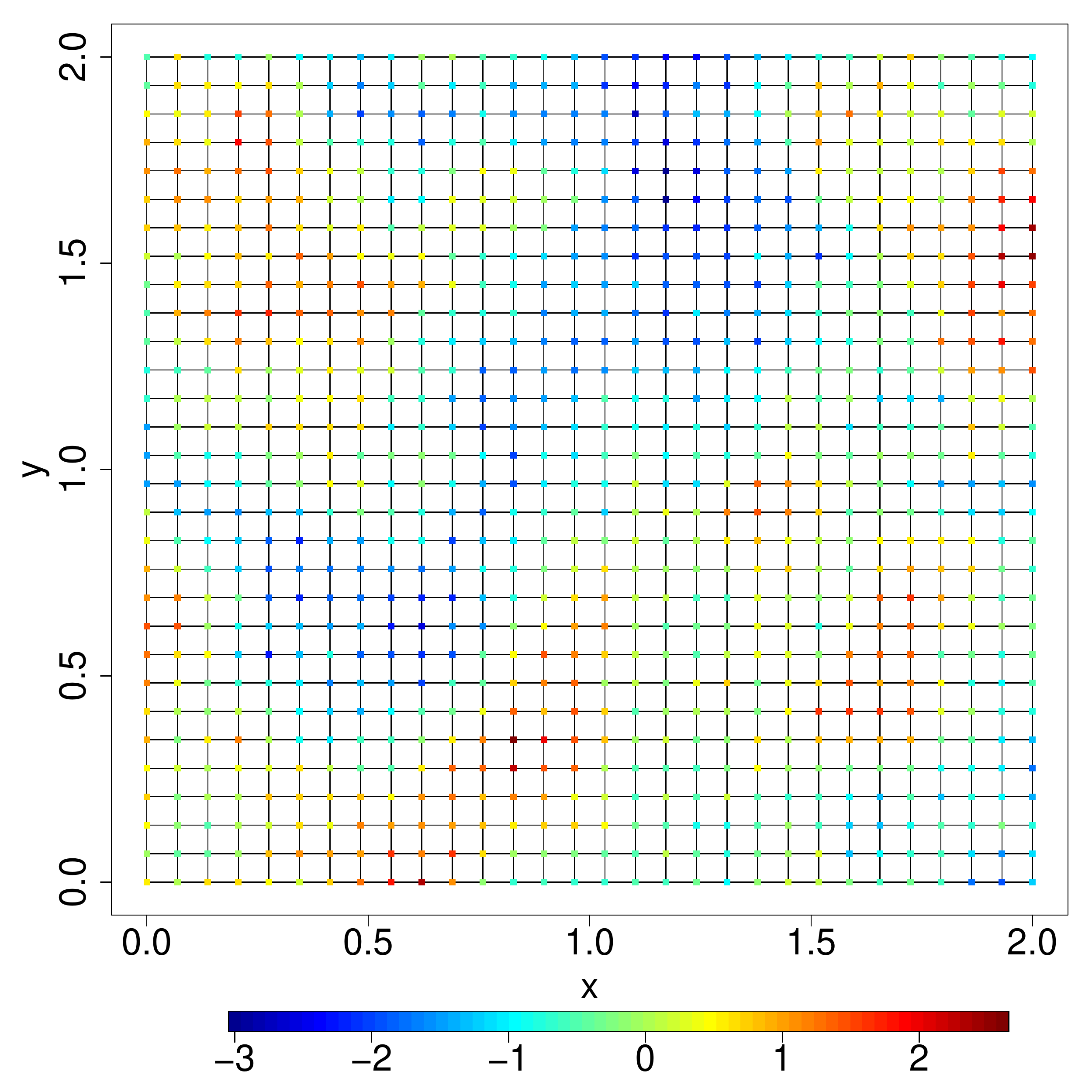}}
\subfigure[]{\label{fig:81bs}\includegraphics[width=60mm]{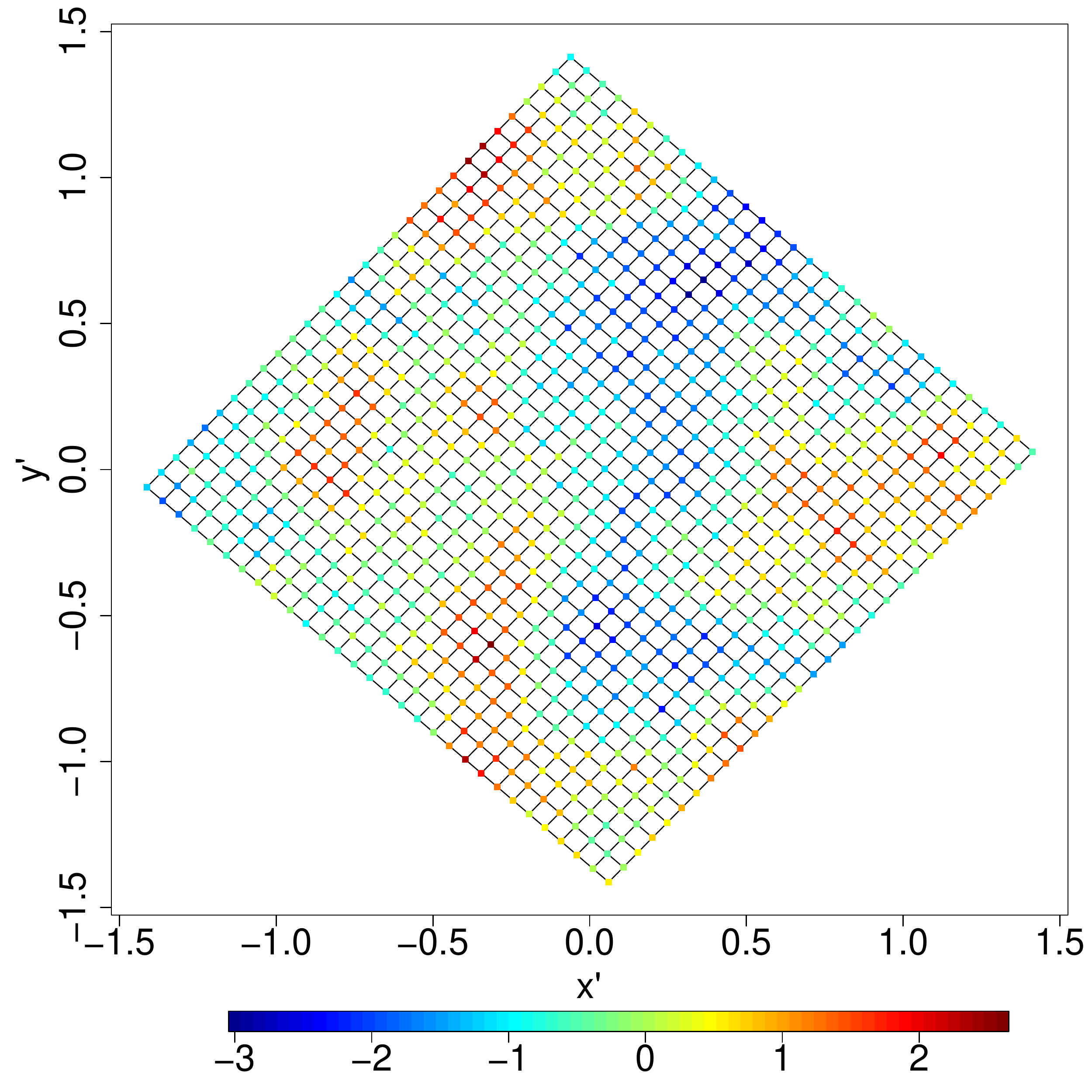}}
\caption{(a) Geographic space. (b) Estimated deformed space. }
\label{fig:81}  
\end{figure}

Figure~\ref{fig:72} shows the results of the registration step for the simulated data. A realization from the second-order stationary process in the geographic space is shown in Figure~\ref{fig:7a2s}, with the solid black line indicating the chosen partitioning. The two regional Mat{\'e}rn variograms (Figure~\ref{fig:7b2s}) show negligible variation in spatial range, which matches the settings of this simulation. This negligible variation results in a very small phase variation captured by the estimated regional distance warping functions shown in Figure~\ref{fig:7d2s}, which are almost identical to the identity warping, further validating the second-order stationarity of the process. The estimated deformed space for $\psi=0$ (2-D) (Figure~\ref{fig:81bs}) is extremely similar to the geographic space shown in Figure~\ref{fig:81as}, in terms of inter-point distances. This means that no deformation is required to achieve stationarity. Note that the fitting of regional variograms is a crucial step in estimating the deformed space, and if the fit is inadequate, the proposed method may estimate a slight deformation even if the original process is second-order stationary.

\section{Effectiveness of Classical Multidimensional Scaling}
\setcounter{equation}{0}

\begin{figure}[!t]
    \centering
    \includegraphics[scale=0.2]{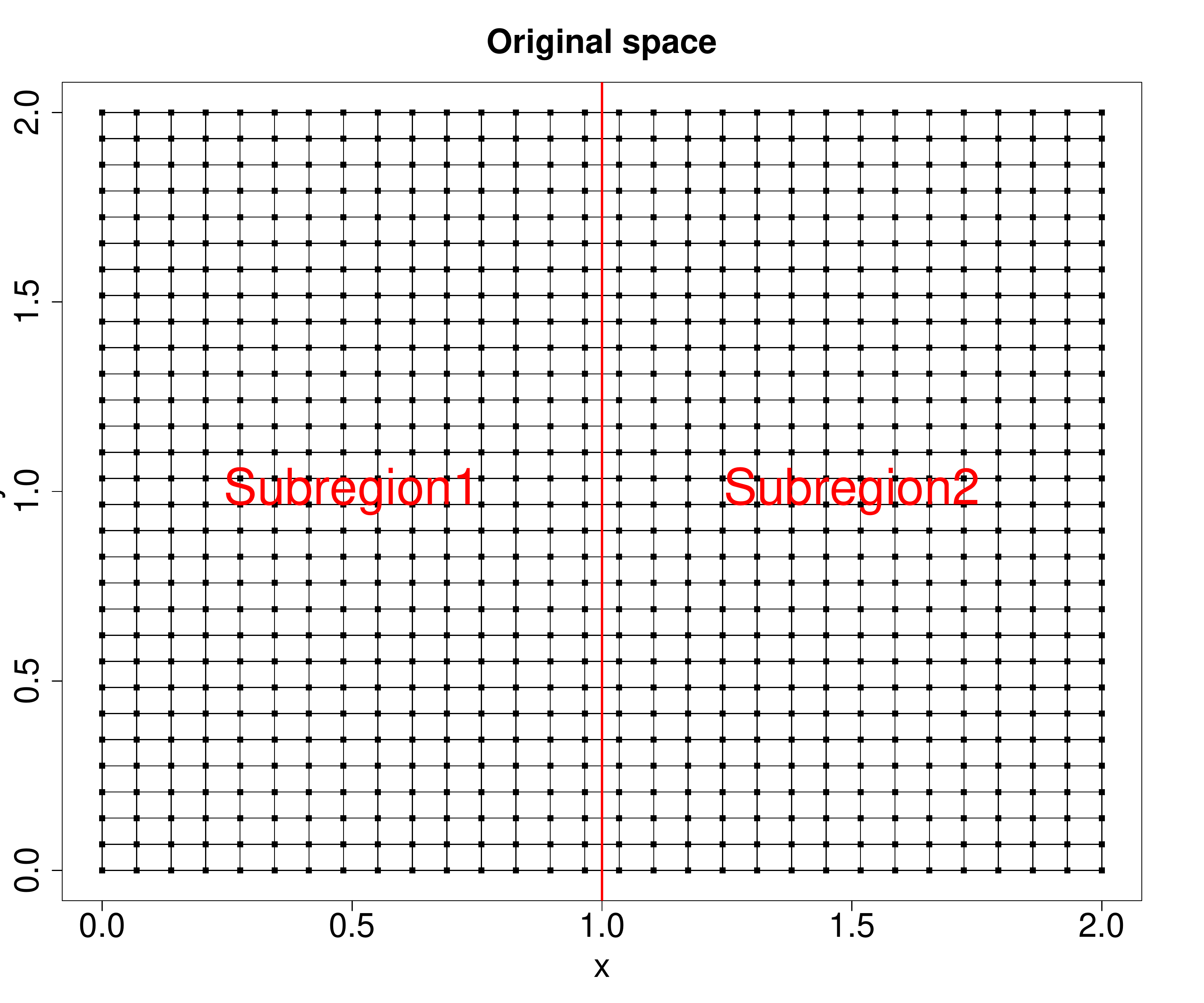}
    \caption{Original space with two subregions.}
    \label{fig:8s}
\end{figure}

% \begin{figure}[!t]
%\centering     %%% not \center
%\subfigure[]{\label{fig:9a}\includegraphics[width=55mm]{1a.pdf}}
%\subfigure[]{\label{fig:9b}\includegraphics[width=55mm]{c1def2d.pdf}}\\
%\subfigure[]{\label{fig:9c}\includegraphics[width=55mm]{c1def3d.pdf}}
%\subfigure[]{\label{fig:9d}\includegraphics[width=55mm]{cmds1.pdf}}
%
%\caption{Case 1 (1): (a) Regional distance warping function (black: warping for subregion 1, red: warping for subregion 2). (b) Deformed space in 2d. (c) Deformed space in 3d. (d) Plot of NMSE vs. dimension of deformed space.}
%\label{fig:91}
%\end{figure}
Our method relies on mapping the transformed distance matrix $\Delta_{n\times n}=\{\hat{\phi}(\textbf{s}_i,\textbf{s}_j)\}_{i,j=1}^{n}$ to the deformed coordinates $\hat{\theta}(\textbf{s}_i),\;i=1,...,n,$ (in some finite dimensional space), which we achieve using classical multidimensional scaling (CMDS). In this simulation, we investigate the effectiveness of CMDS in mapping the transformed distance matrix to deformed coordinates. We consider $30\times30$ regularly spaced points in the domain $[0,2]^2$, shown in Figure \ref{fig:8s}, and divide it into two subregions using the line $x=1$. We consider two different types of parametric regional distance warping functions:
\begin{enumerate}
    \item Case 1:  \[\phi_1(\|\textbf{h}\|)=\begin{cases}\sqrt{8}\frac{e^{a\|\textbf{h}\|/\sqrt{8}}-1}{e^a-1}, \text{\hspace{0.2cm} if } a\neq0\\
  \|\textbf{h}\|, \text{\hspace{0.2cm} if } a=0
  \end{cases}\]\[ \phi_2(\|\textbf{h}\|)=\begin{cases}\sqrt{8}\frac{e^{-a\|\textbf{h}\|/\sqrt{8}}-1}{e^{-a}-1}, \text{\hspace{0.2cm} if } a\neq0\\
  \|\textbf{h}\|, \text{\hspace{0.2cm} if } a=0
  \end{cases}\]
  \item Case 2:\[\phi_1(\|\textbf{h}\|)=\sqrt{8}\textbf{B}^C(\|\textbf{h}\|/\sqrt{8}|\alpha_1,\beta_1)\quad \phi_2(\|\textbf{h}\|)=\sqrt{8}\textbf{B}^C(\|\textbf{h}\|/\sqrt{8}|\alpha_2,\beta_2),\]
      
\end{enumerate}
where $\textbf{B}^C(\cdot|\alpha,\beta)$ is the cumulative distribution function of the Beta distribution for given shape parameters $\alpha$ and $\beta$. The parameter $|a|$ controls the intensity of warping in Case 1, whereas the parameters $(\alpha_1,\beta_1,\alpha_2,\beta_2)$ control the shape of the regional distance warping functions in Case 2. We consider four settings for these parameters in each case: Case 1 $|a|=$ (1) $0$ (identity warping), (2) $0.5$, (3) $1.5$, (4) $2.5$; Case 2 $(\alpha_1,\beta_1,\alpha_2,\beta_2)=$ (1) $(0.7,1.5,1,1/2.5)$, (2) $(1,1.4,1,1/2.5)$, (3) $(0.25,1.4,8,2)$, (4) $(0.25,1.4,2,1)$. For each of the parameter settings, we compute the global distance function $\phi(\textbf{s},\textbf{s'})$ using the proposed method, and estimate the deformed space in dimensions $d^\mathcal{D}=2,....,30$. We then compute the normalized mean squared error (NMSE) between the transformed distance matrix $\Delta_{900\times 900}=\{{\phi}(\textbf{s}_i,\textbf{s}_j)\}_{i,j=1}^{900}$ and the distance matrix of the estimated deformed space under different values of $d^\mathcal{D}$. A value of NMSE equal to 1 indicates perfect mapping of distances to the coordinates using CMDS. 

% \begin{figure}[!t]
%\centering     %%% not \center
%\subfigure[]{\label{fig:11a}\includegraphics[width=55mm]{1c.pdf}}
%\subfigure[]{\label{fig:11b}\includegraphics[width=55mm]{c4def2d.pdf}}\\
%\subfigure[]{\label{fig:11c}\includegraphics[width=55mm]{c4def3d.pdf}}
%\subfigure[]{\label{fig:11d}\includegraphics[width=55mm]{cmds4.pdf}}
%
%\caption{Case 1 (3): (a)-(d) Same as in Figure \ref{fig:91}.}
%\label{fig:111}
%\end{figure}
\begin{figure}[!t]
\centering     %%% not \center
\subfigure[]{\label{fig:10as}\includegraphics[width=80mm]{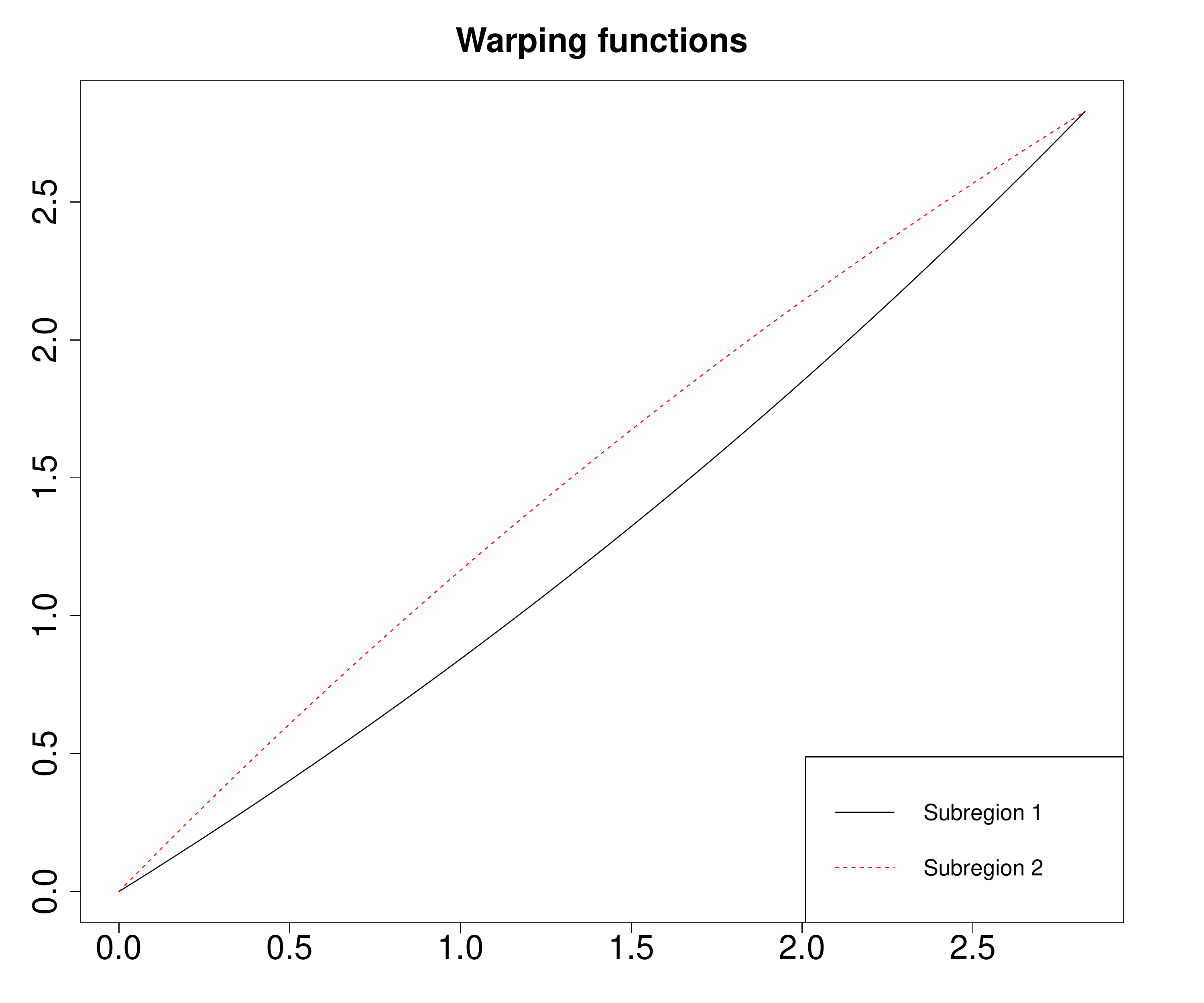}}
\subfigure[]{\label{fig:10bs}\includegraphics[width=80mm]{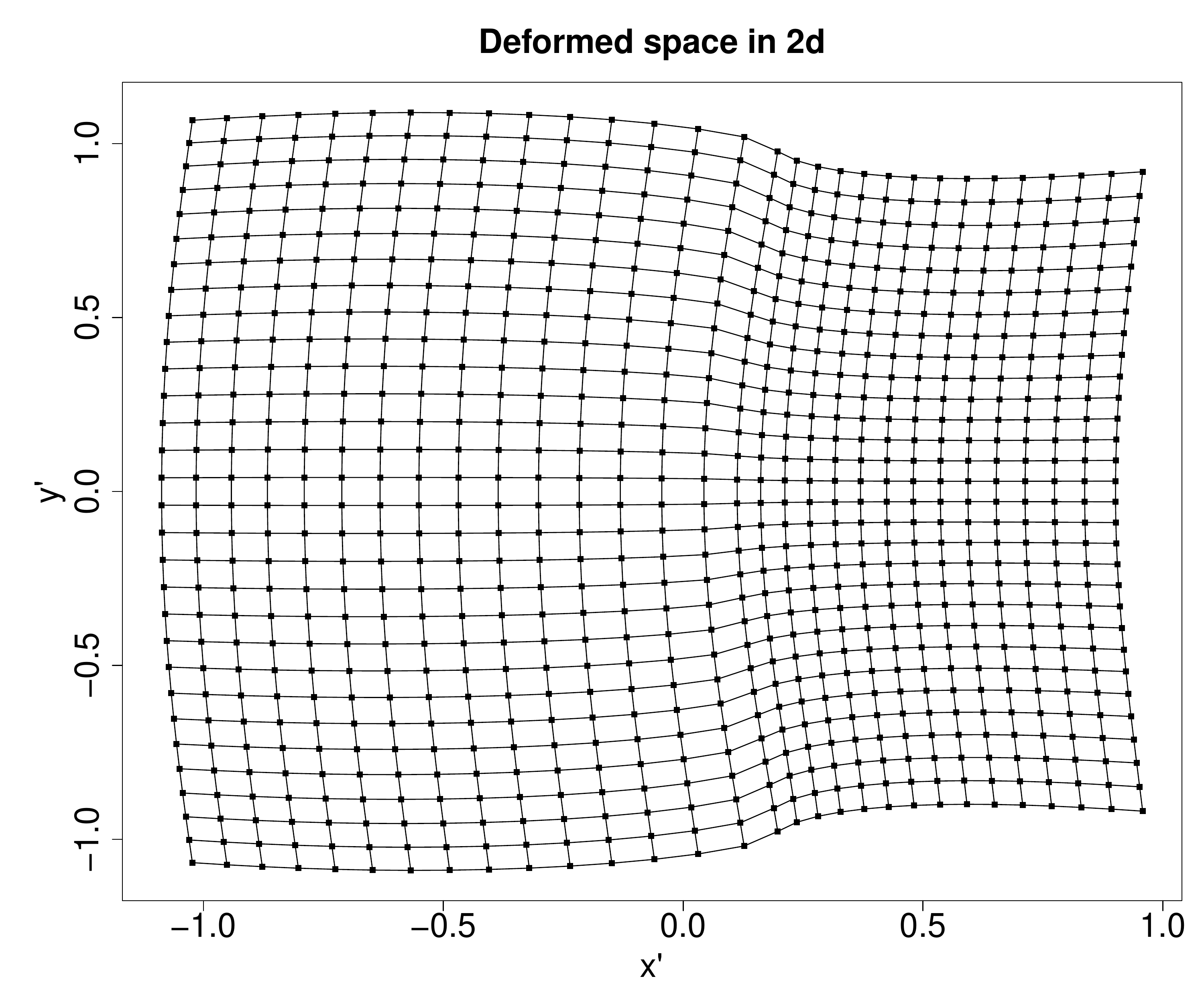}}\\
\subfigure[]{\label{fig:10cs}\includegraphics[width=80mm]{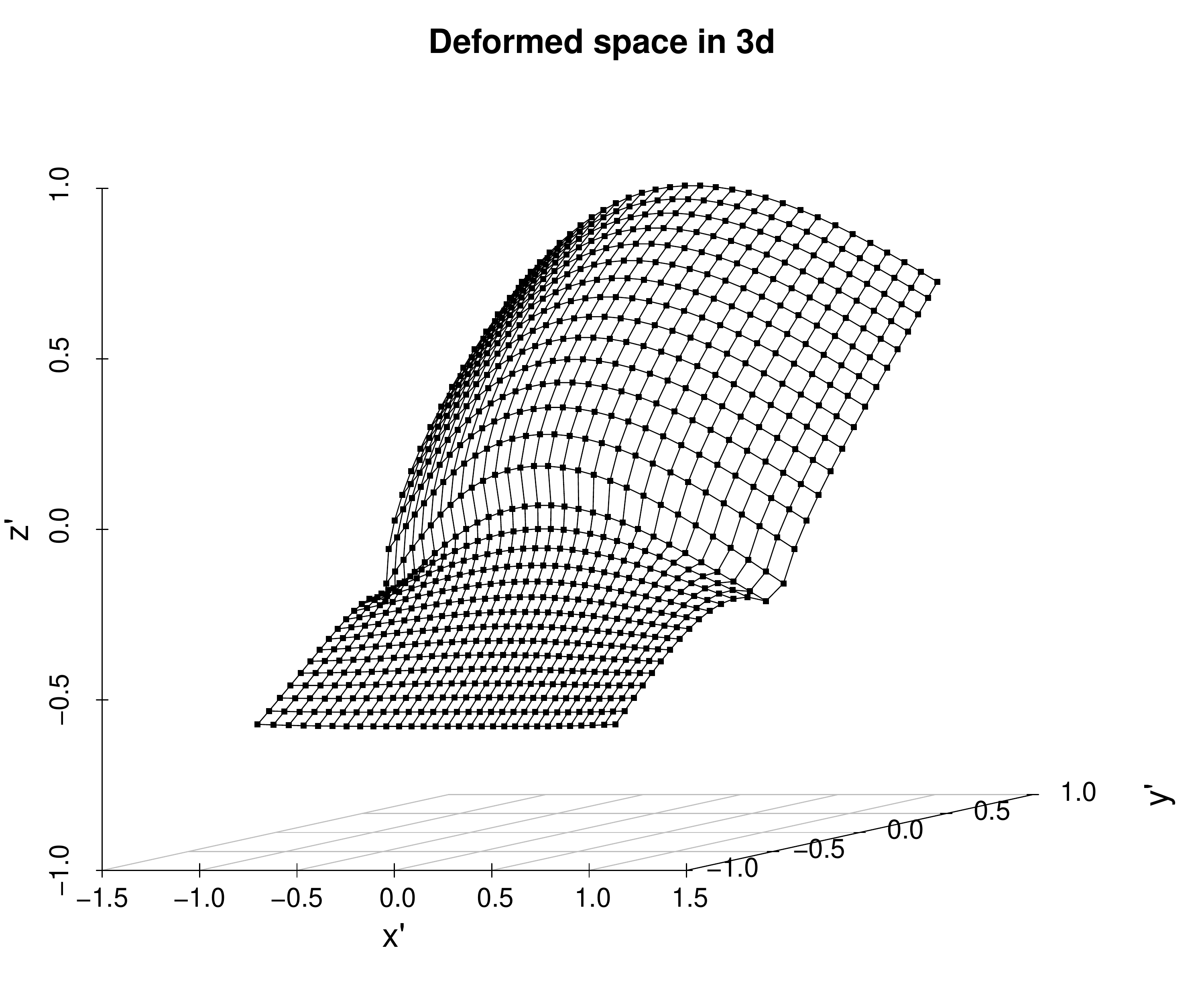}}
\subfigure[]{\label{fig:10ds}\includegraphics[width=80mm]{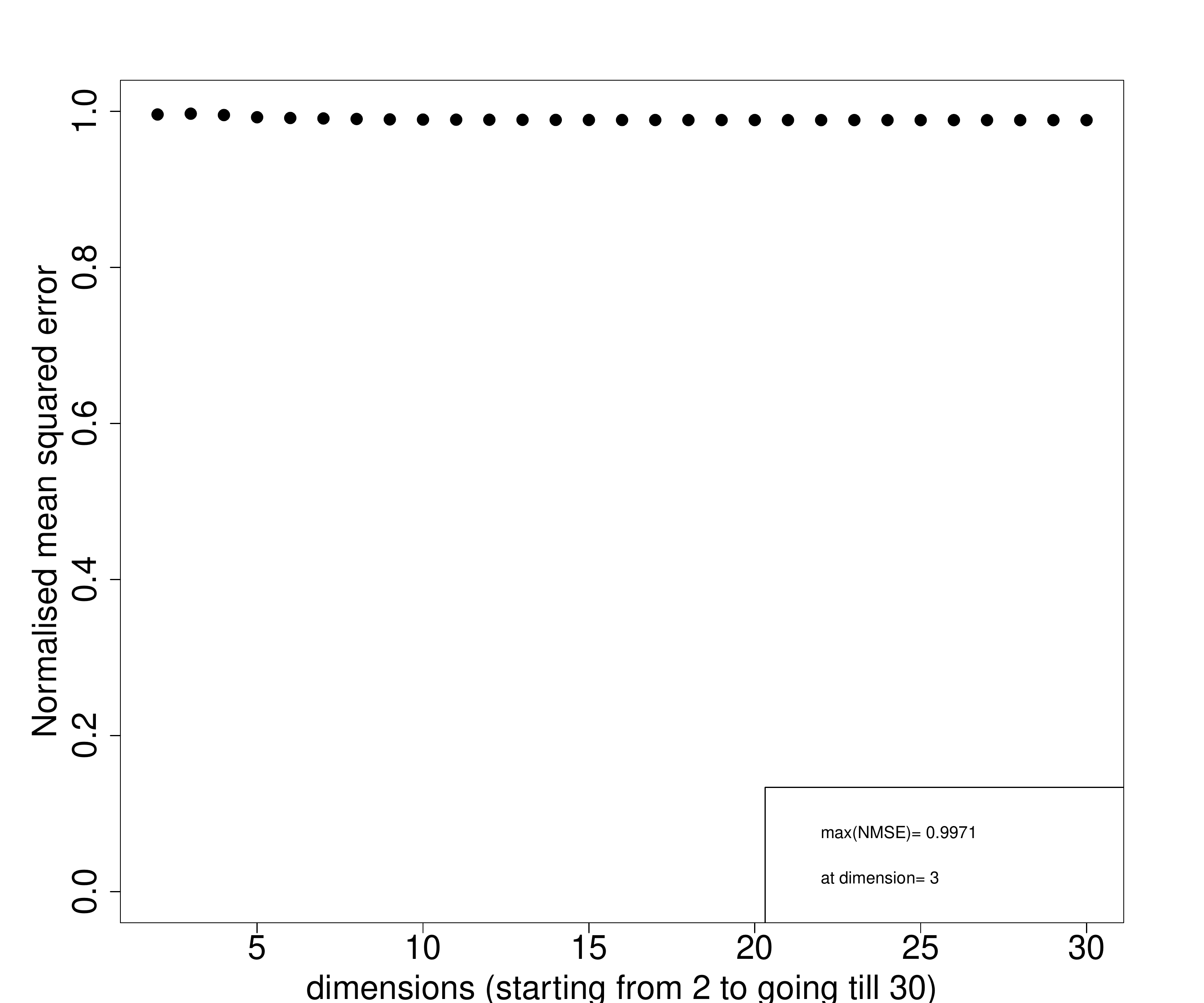}}

\caption{Case 1 (2): (a) Regional distance warping function (black: warping for subregion 1, red: warping for subregion 2). (b) Deformed space in 2d. (c) Deformed space in 3d. (d) Plot of NMSE vs. dimension of deformed space.}
\label{fig:91}
\end{figure}
\begin{table}[!t]
\centering
 \begin{tabular}{|c c c c|}
 \hline
Case & Parameter Setting & Max NMSE & Dimension for Max NMSE\\ [0.5ex]
 \hline
 1 & (1) & 1 &  2 \\
 \hline
 1 &  (2) & 0.9971 & 3 \\
 \hline
 1 & (3) & 0.9791 & 3 \\
 \hline
 1  & (4) & 0.946 & 3  \\
 \hline
 2 & (1) & 0.9428 & 3 \\
 \hline
 2 &  (2) & 0.9682 & 2 \\
 \hline
 2 & (3) & 0.681 & 30 \\
 \hline
 2  & (4) & 0.8258 & 30  \\
 \hline
\end{tabular}
\caption{Quantitative assessment of the estimated deformed space based on CMDS for different types of regional distance warping functions.}
\label{table:s1}
\end{table}
\begin{figure}[p]
\centering     %%% not \center
\subfigure[]{\label{fig:12as}\includegraphics[width=80mm]{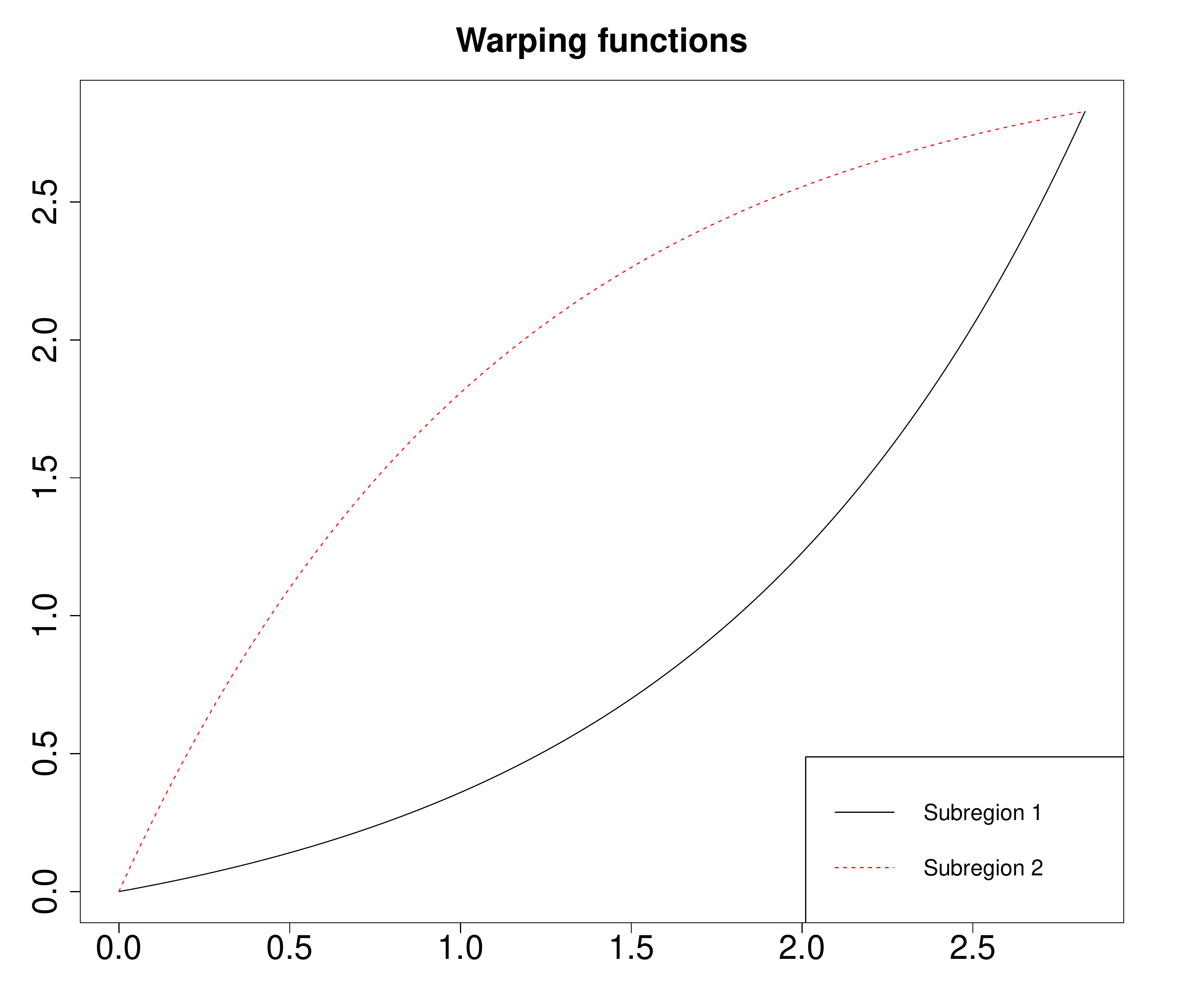}}
\subfigure[]{\label{fig:12bs}\includegraphics[width=80mm]{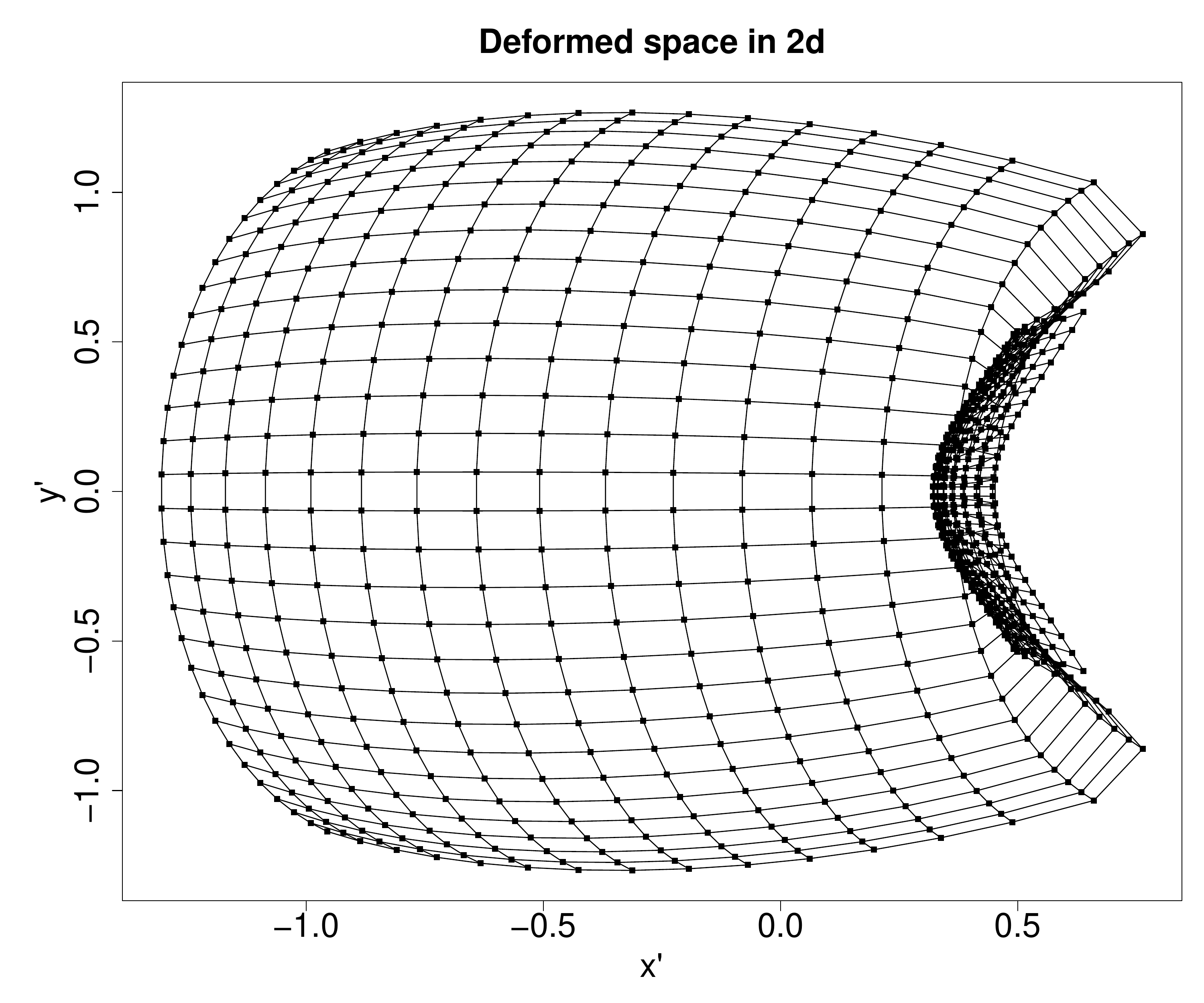}}\\
\subfigure[]{\label{fig:12cs}\includegraphics[width=80mm]{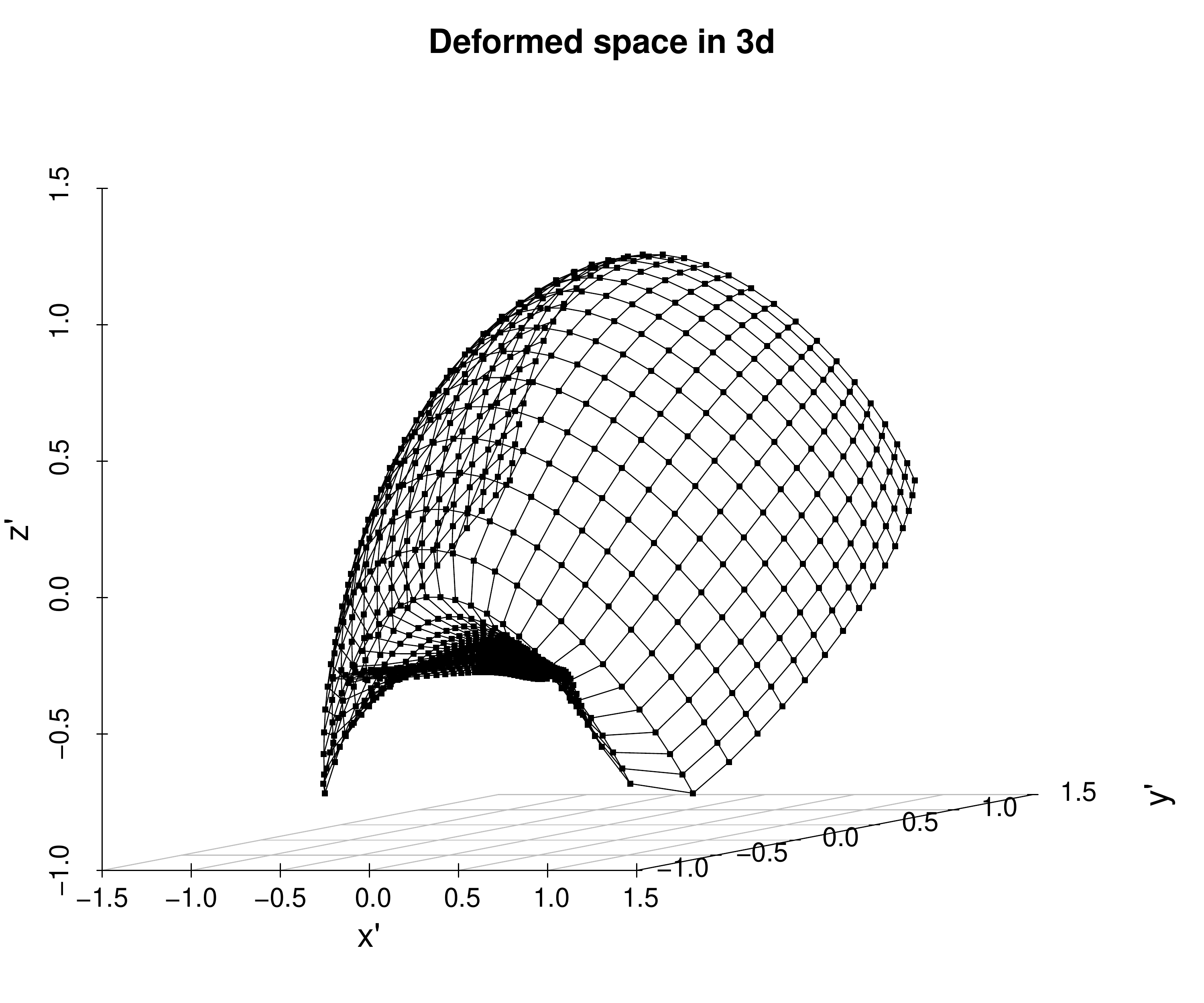}}
\subfigure[]{\label{fig:12ds}\includegraphics[width=80mm]{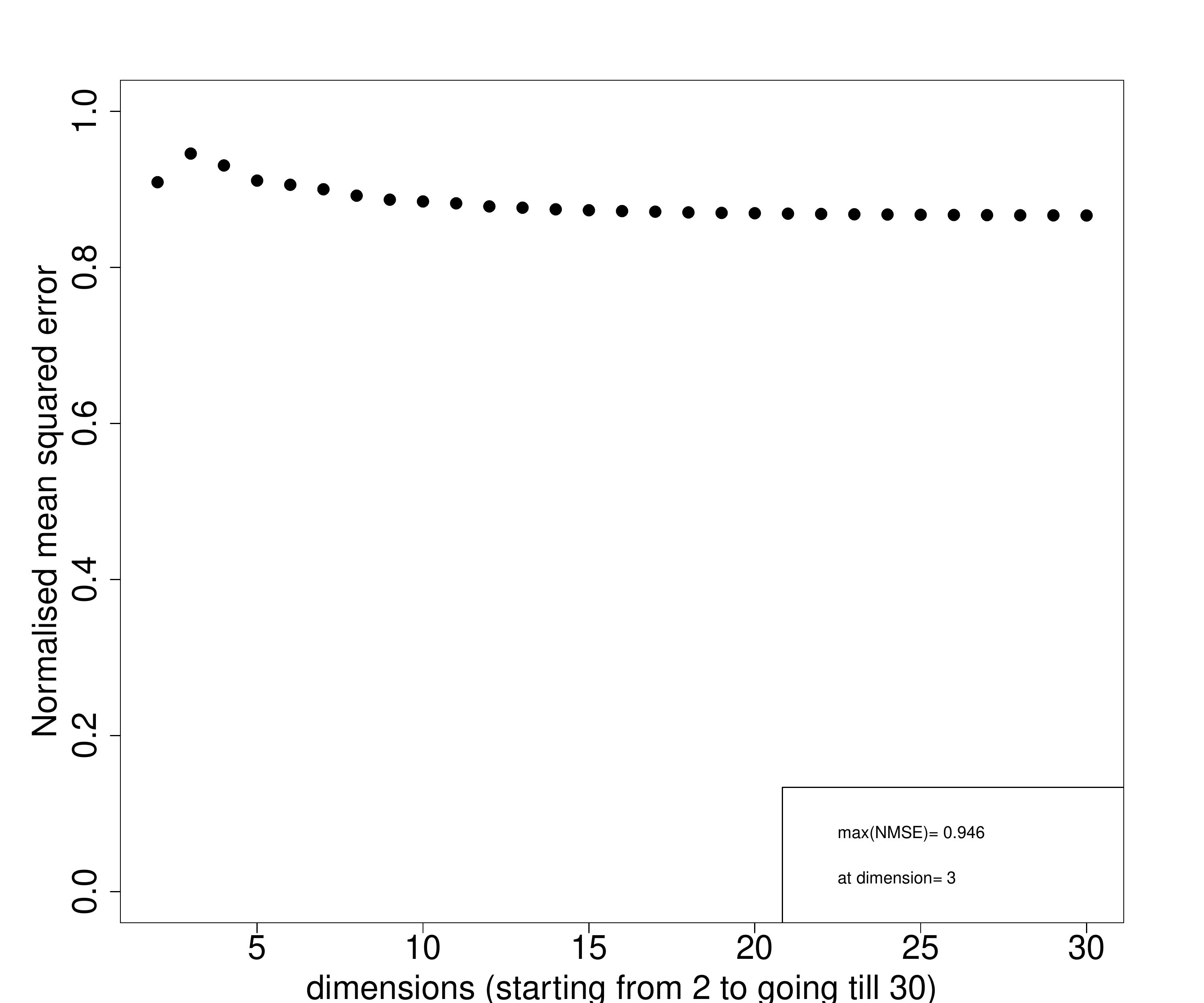}}

\caption{Case 1 (4): (a)-(d) Same as in Figure \ref{fig:91}.}
\label{fig:100}
\end{figure}

\begin{figure}[p]
\centering     %%% not \center
\subfigure[]{\label{fig:13as}\includegraphics[width=80mm]{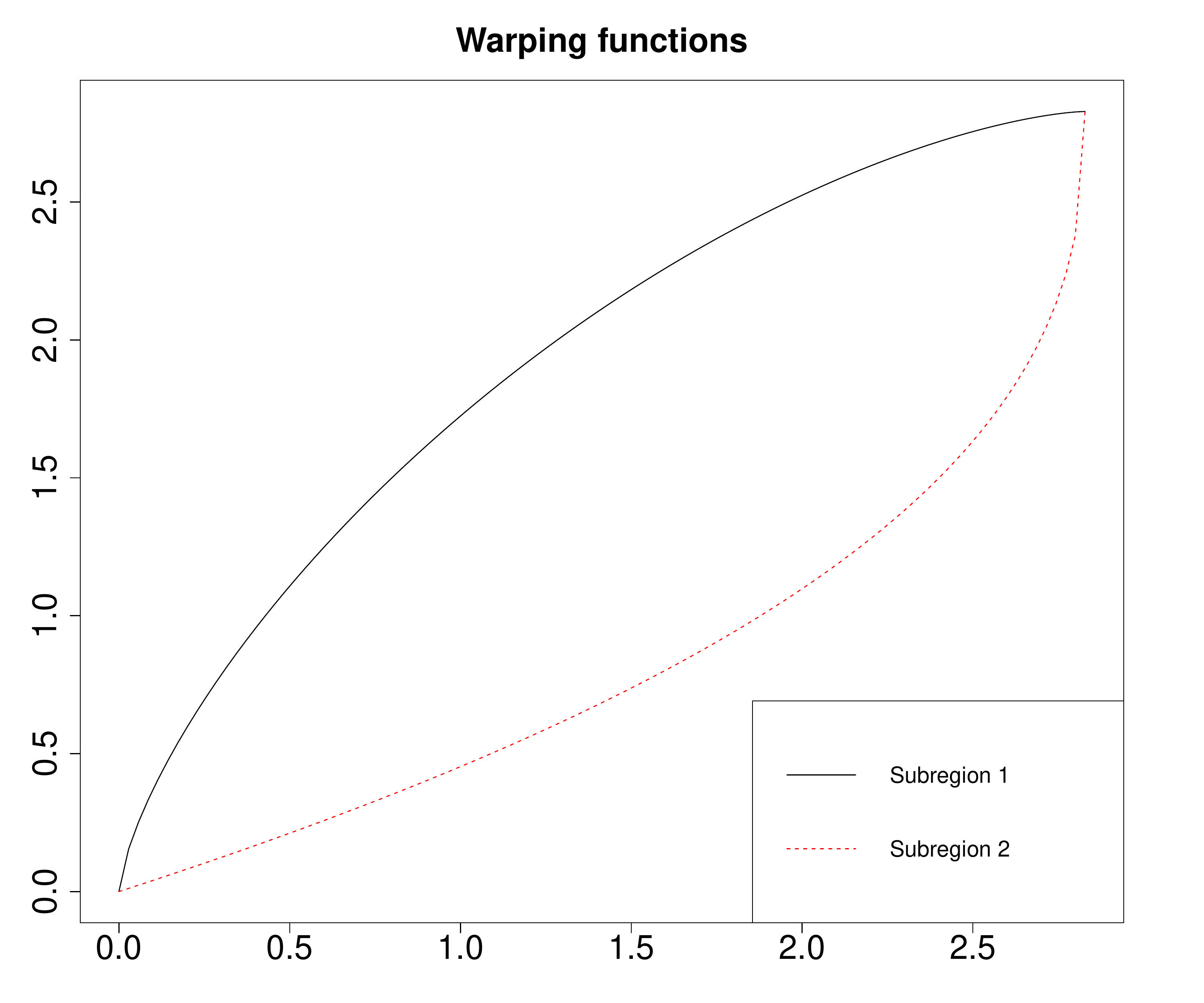}}
\subfigure[]{\label{fig:13bs}\includegraphics[width=80mm]{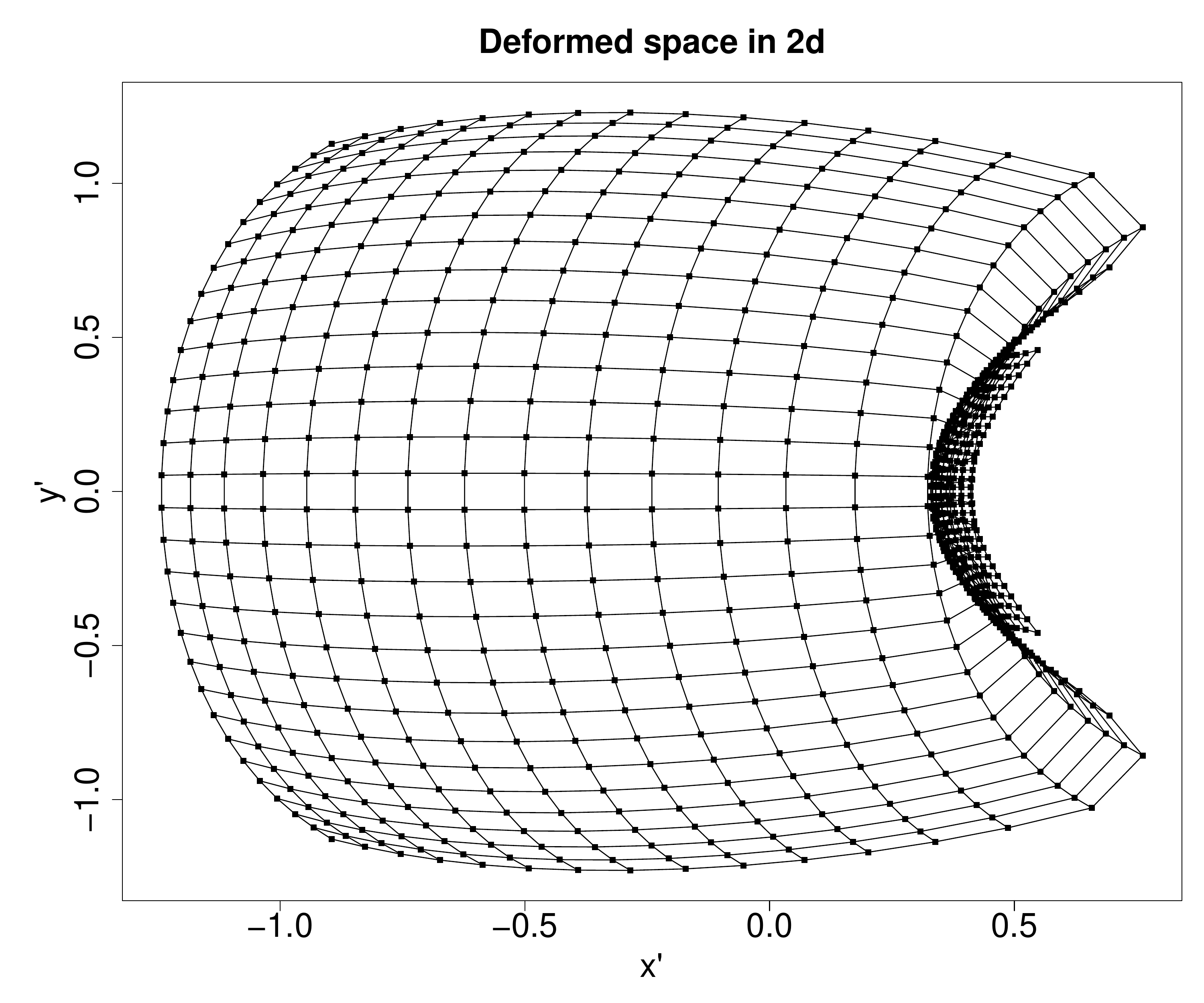}}\\
\subfigure[]{\label{fig:13cs}\includegraphics[width=80mm]{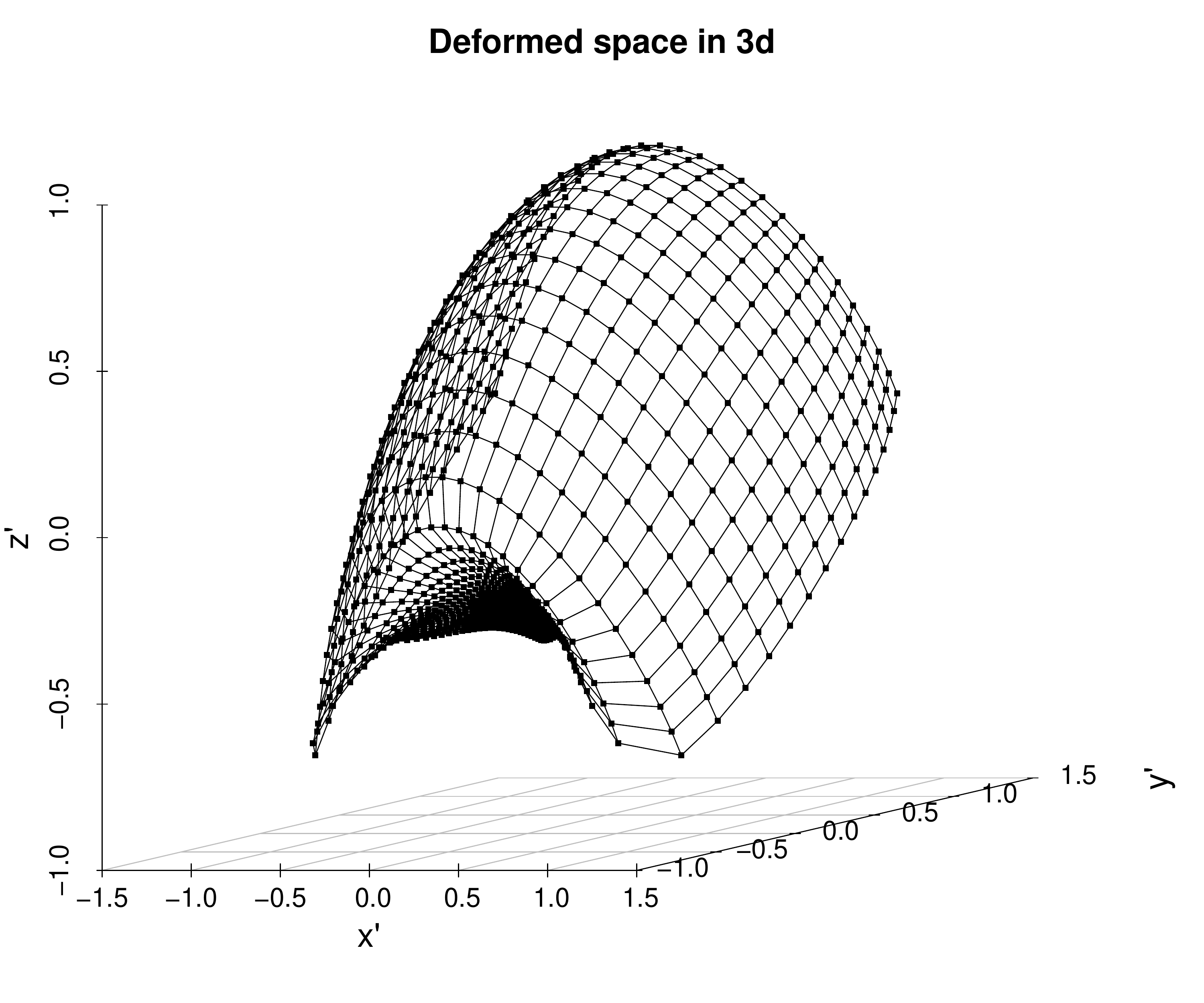}}
\subfigure[]{\label{fig:13ds}\includegraphics[width=80mm]{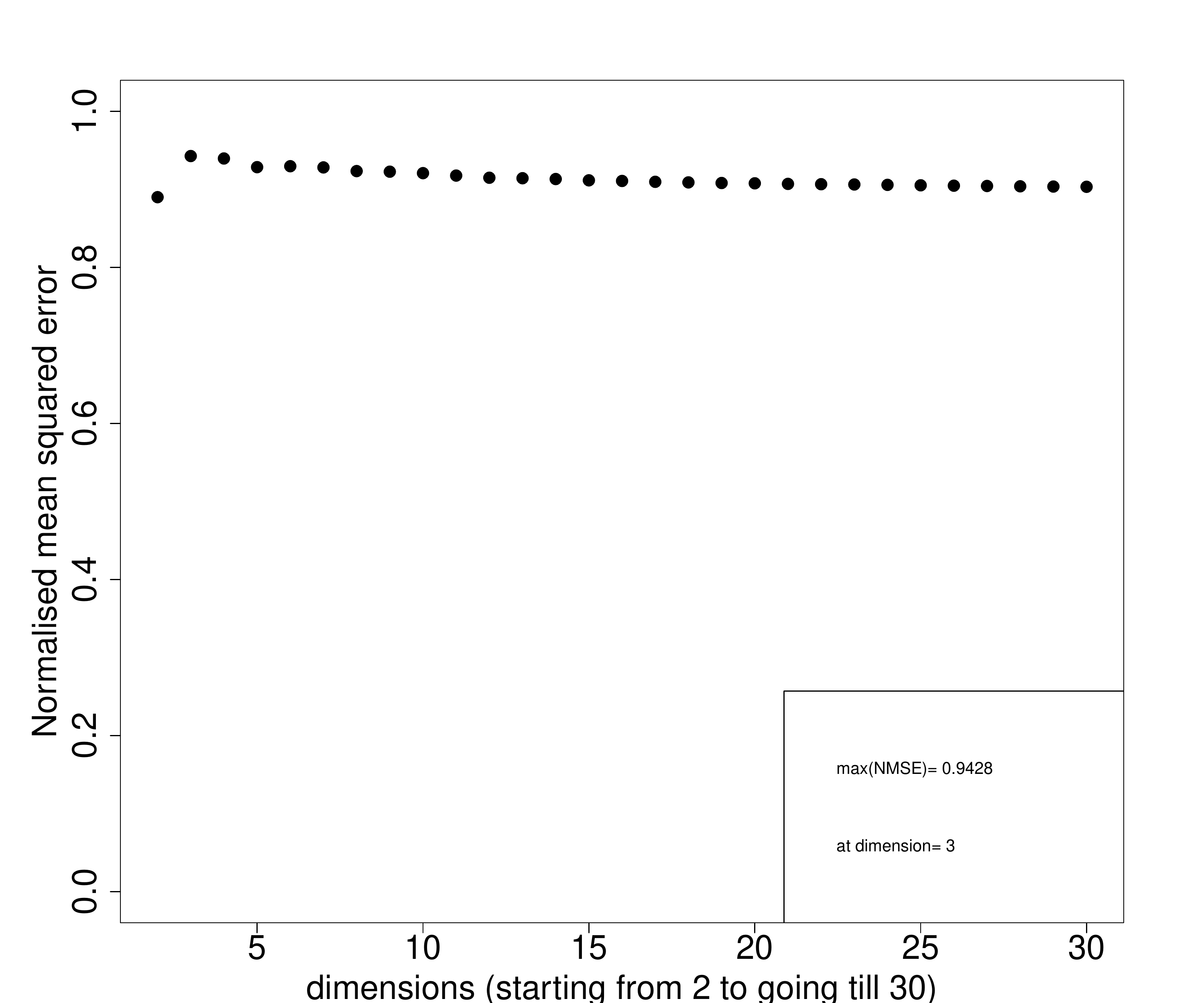}}

\caption{Case 2 (1): (a)-(d) Same as in Figure \ref{fig:91}.}
\label{fig:111}
\end{figure}

%\begin{figure}[!t]
%\centering     %%% not \center
%\subfigure[]{\label{fig:14a}\includegraphics[width=55mm]{2b.pdf}}
%\subfigure[]{\label{fig:14b}\includegraphics[width=55mm]{bc22ddef.pdf}}\\
%\subfigure[]{\label{fig:14c}\includegraphics[width=55mm]{bc23ddef.pdf}}
%\subfigure[]{\label{fig:14d}\includegraphics[width=55mm]{bc2fplot.pdf}}
%
%\caption{Case 2 (2): (a)-(d) Same as in Figure \ref{fig:91}.}
%\label{fig:14}
%\end{figure}

\begin{figure}[p]
\centering     %%% not \center
\subfigure[]{\label{fig:15as}\includegraphics[width=80mm]{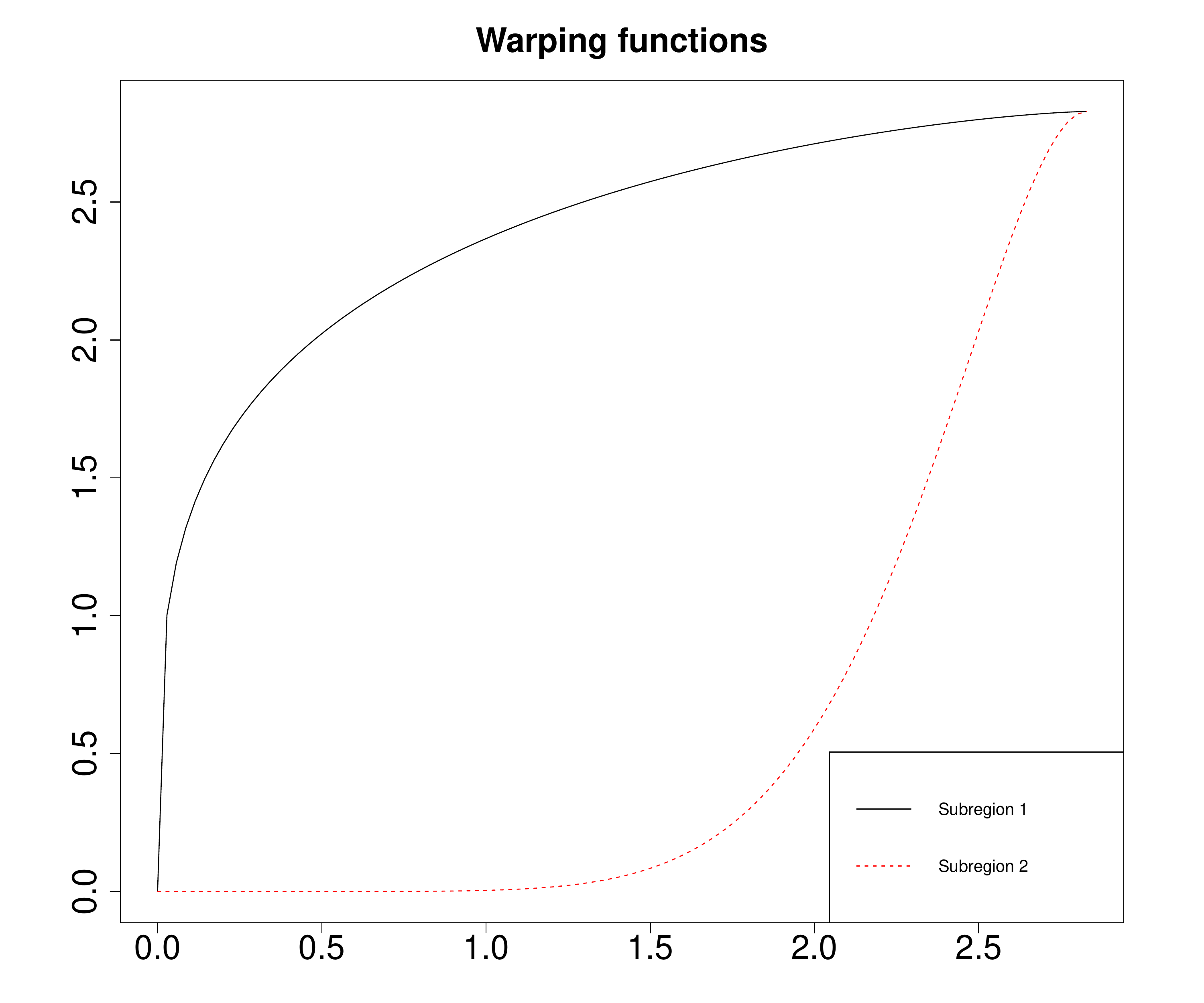}}
\subfigure[]{\label{fig:15bs}\includegraphics[width=80mm]{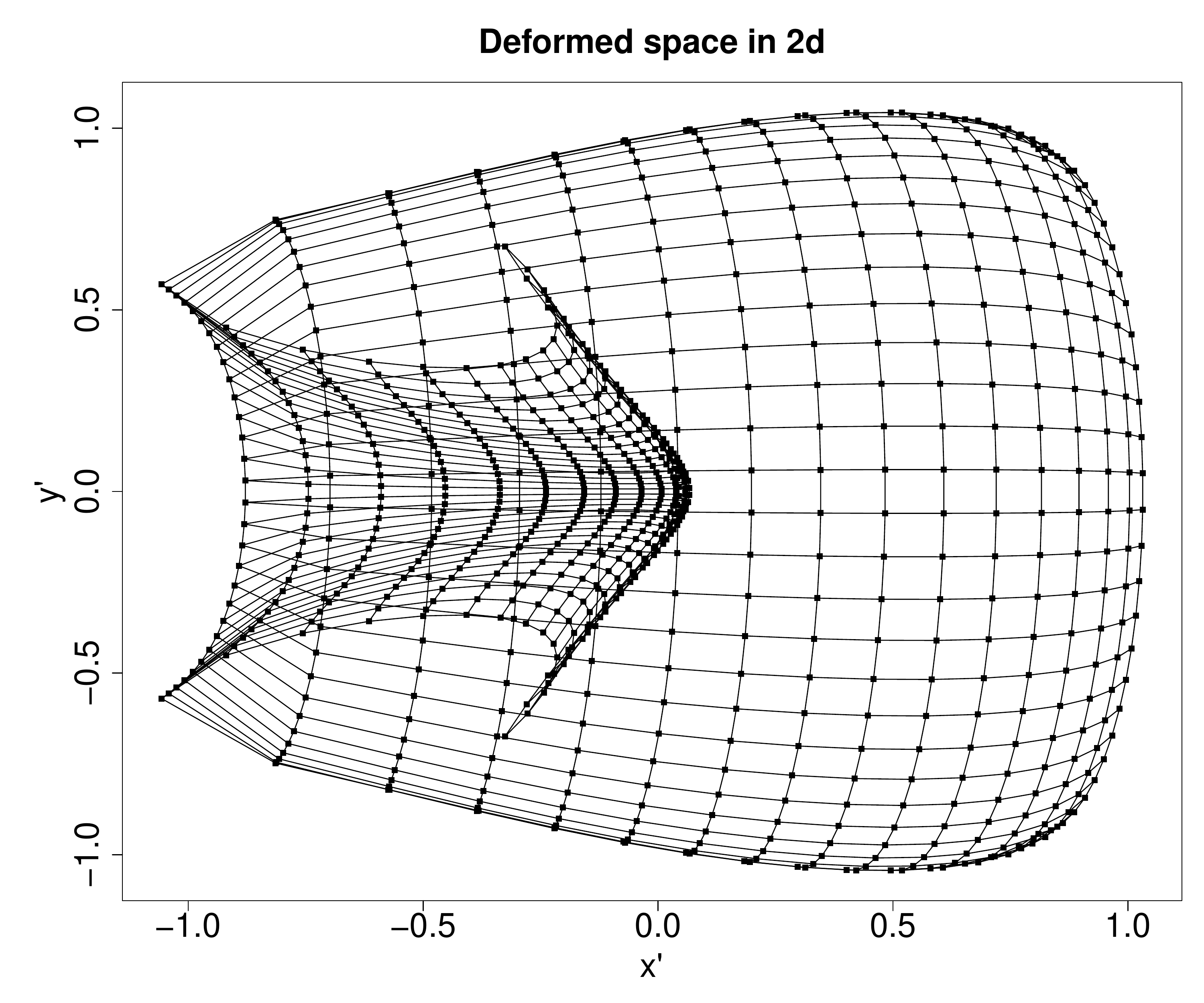}}\\
\subfigure[]{\label{fig:15cs}\includegraphics[width=80mm]{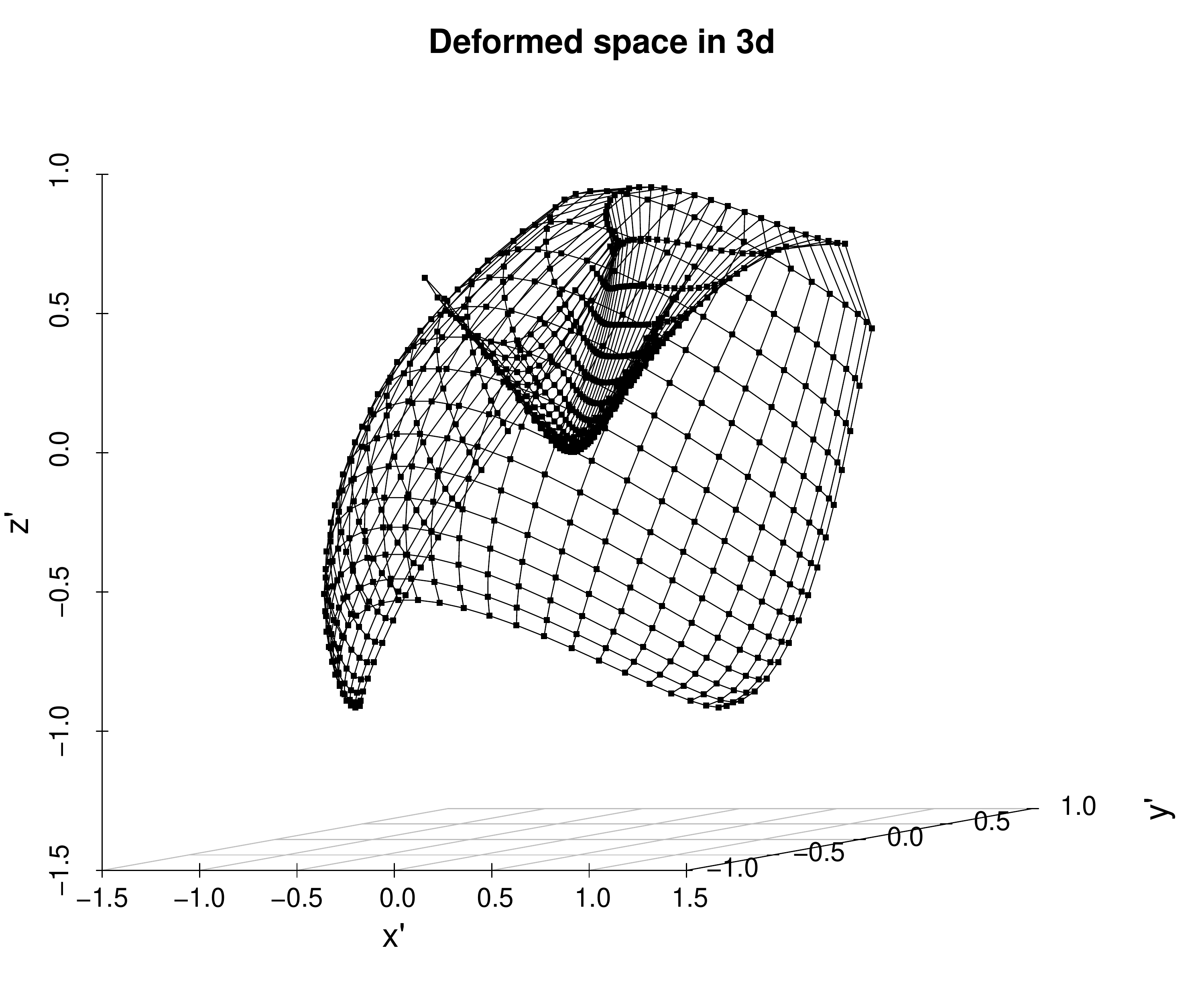}}
\subfigure[]{\label{fig:15ds}\includegraphics[width=80mm]{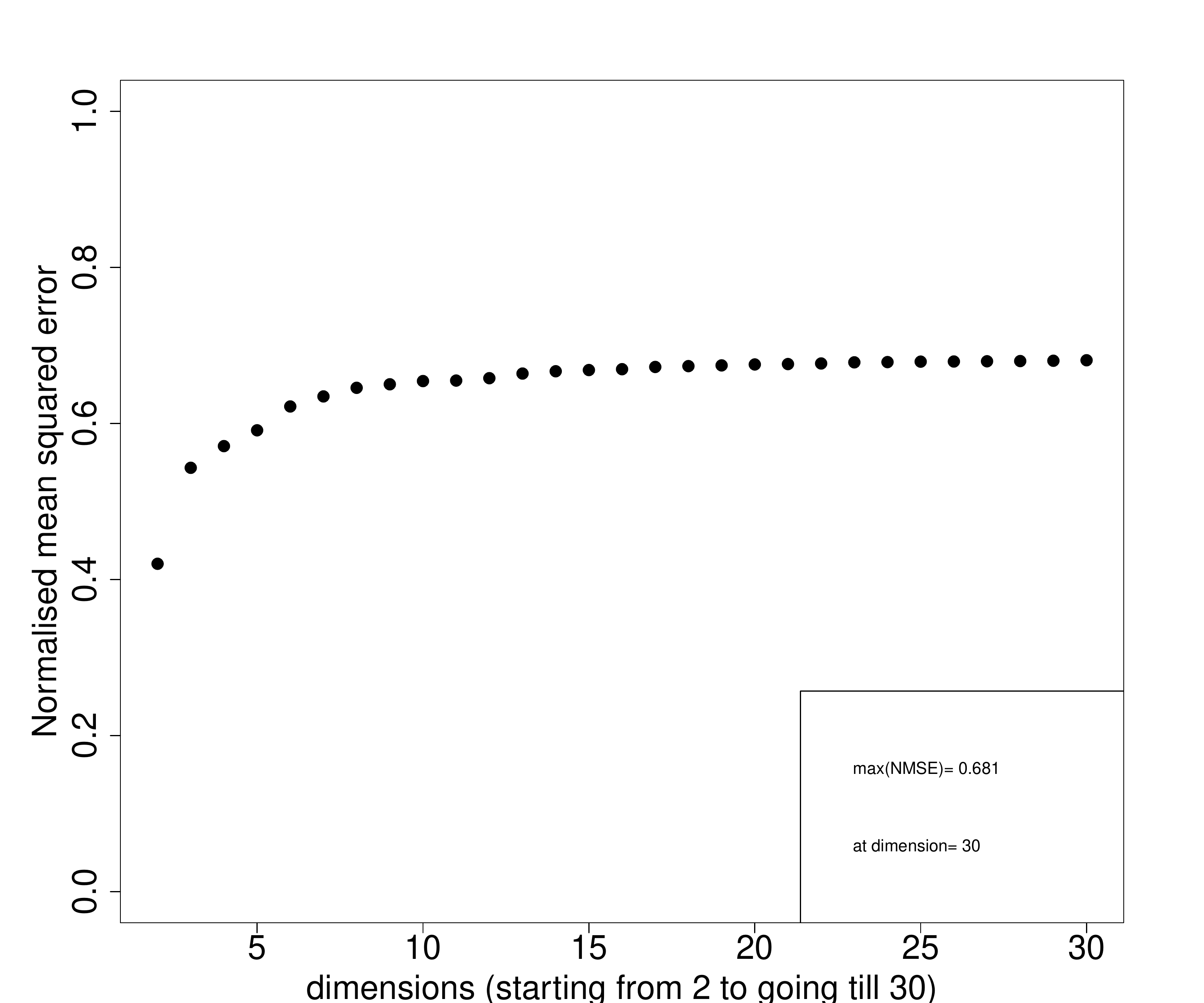}}

\caption{Case 2 (3): (a)-(d) Same as in Figure \ref{fig:91}.}
\label{fig:122}
\end{figure}

Table \ref{table:s1} reports the maximum NMSE and the dimension $d^\mathcal{D}$ at which the maximum NMSE is attained for all of the eight simulated situations. Figures \ref{fig:91}-\ref{fig:122} show the summary of results for Case 1 (2) and (4) and Case 2 (1) and (3), respectively (we do not show figures for the other parameter settings for brevity). We observe that, for Case 1, as we increase the value of $|a|$ (i.e., the intensity of regional distance warping functions), the performance of CMDS decreases. However, even in Case 1 (4) with considerable amount of warping, the NMSE is 0.946 for a deformed space of dimension 3, a value very close to 1 indicating a very good approximation. For Case 2 (1) and (2), CMDS performs  very well and produces NMSE values close to 1. However, for Case 2 (3) and (4) with extreme regional warping functions (in parts of the subdomain the warping functions become nearly vertical or horizontal), CMDS does not perform well. These settings correspond to extreme deformations of the geographic space, which we do not expect in realistic applications.

\end{document}